\documentclass[modern,times]{aastex63}
\pdfoutput=1
\usepackage[utf8]{inputenc}
\usepackage{amsmath}
\usepackage{amssymb}
\submitjournal{ApJS}
\usepackage{soul}
\usepackage{import}
\usepackage{hyperref}
\usepackage{calculator}
\usepackage{gensymb}

\usepackage{multirow}
\usepackage{bm}
\usepackage[section]{placeins}

\usepackage{printlen}
\makeatletter
\newcommand\footnoteref[1]{\protected@xdef\@thefnmark{\ref{#1}}\@footnotemark}
\newcommand{\as}{$^{\prime\prime}$}

\newcommand{\siml}{$\sim$}

\newcommand{\newref}[2]{\hyperref[#1]{\ref*{#1}(#2)}}

\edef\writeimagedesc#1,#2,#3,#4.#5{The VLA #1 band #2 config contours are overlaid on the X-ray image binned to a size of #3 and smoothed with a Gaussian kernel of FWHM=#4.#5\as. }
\newcommand{\Radiocontours}[1]{The radio contours are given by #1 mJy/beam. }
\usepackage[maxfloats=256]{morefloats}

\newcommand{\comment}[1]{}

\makeatother

\begin{document}
\title{X-Ray-to-Radio Offset Inference from Low-Count X-Ray Jets}
\shorttitle{Offsets from low-count images}

\author[0000-0001-9018-9553]{Karthik Reddy}
\affil{Department of Physics, University of Maryland Baltimore County, 1000 Hilltop Circle, Baltimore, MD 21250, USA}
\email{karthik@umbc.edu}
\author[0000-0002-2040-8666]{Markos Georganopoulos}
\affil{Department of Physics, University of Maryland Baltimore County, 1000 Hilltop Circle, Baltimore, MD 21250, USA}
\affil{NASA Goddard Space Flight Center, Code 663, Greenbelt, MD 20771, USA}
\author[0000-0002-7676-9962]{Eileen T. Meyer}
\affil{Department of Physics, University of Maryland Baltimore County, 1000 Hilltop Circle, Baltimore, MD 21250, USA}

\shortauthors{Reddy, Georganopoulos, Meyer}
\received{2020 Nov 01}
\revised{2020 Dec 30}
\accepted{2021 Jan 03}

\begin{abstract}
Observations of positional offsets between the location of X-ray and radio features in many resolved, extragalactic jets indicates that the emitting regions are not co-spatial, an important piece of evidence in the debate over the origin of the X-ray emission on kpc scales. 
The existing literature is nearly exclusively focused on jets with sufficiently deep \emph{Chandra} observations to yield accurate positions for X-ray features, but most of the known X-ray jets are detected with tens of counts or fewer, making detailed morphological comparisons difficult. Here we report the detection of X-ray-to-radio positional offsets in 15 extragalactic jets from an analysis of 22 sources with low-count \textit{Chandra} observations, where we utilized the Low-count Image Reconstruction Algorithm (LIRA). This algorithm has allowed us to account for effects such as Poisson background fluctuations and nearby point sources which have previously made the detection of offsets difficult in shallow observations. Using this method, we find that in 55 \% of knots with detectable offsets, the X-rays peak upstream of the radio, questioning the applicability of one-zone models, including the IC/CMB model for explaining the X-ray emission. We also report the non-detection of two previously claimed X-ray jets. Many, but not all, of our sources follow a loose trend of increasing offset between the X-ray and radio emission, as well as a decreasing X-ray to radio flux ratio along the jet.

\end{abstract}

\section{Introduction\label{sec:introduction}}

About 10\% of the Active Galactic Nuclei (AGN) produce two-sided relativistic jets that transport energy and momentum
from the central parsec-scale region out to kiloparsec and often to megaparsec scales \citep{Padovani_2017}. There is growing evidence that such jets play a crucial role in the evolution of galaxies and clusters \citep[e.g.,][]{croton2006many,fabian2012}. 
Even after half a century of study, much about jets remains uncertain, including their composition, kiloparsec-scale speeds, and high-energy emission mechanism(s) \citep[for reviews, see~][]{harris2006x,worrall2009x,blandford2019relativistic}. These uncertainties translate to large differences in the estimates of the total jet power, with obvious implications for understanding the impact of jets on their hosts and larger environment \citep[e.g.][]{harris2006x}.

Radio telescopes such as the Very Large Array (VLA) have %
detected thousands of resolved, kiloparsec-scale extragalactic jets, and it is now well-established that the radio emission is synchrotron radiation from energetic electrons spiraling in a highly magnetized plasma \citep[e.g.,][]{bridle1984extragalactic}. In many cases this synchrotron emission can extend up to optical frequencies or even higher \citep[e.g.,][]{sambruna2002survey}. Since its launch in 1999, the \textit{Chandra} X-ray observatory has detected X-rays from over 150 kpc-scale radio jets  \citep[e.g,.][]{sambruna2004survey,sambruna2006deep,marshall2011x,Marshall_2018,Massaro_2018}. %
Interestingly, many of these exhibit very high X-ray flux and/or hard X-ray spectral index relative to radio  \citep[e.g.,][]{sambruna2004survey}, implying a second or even third \citep[e.g.,][]{meyer2018detection} spectral component in the jet spectral energy distributions (SED), separate from the radio-synchrotron component. Following \citet{breiding2017fermi}, we will refer to such jets as MSC (multi-spectral component) jets for rest of the paper. Predominantly, the X-ray emission from MSC jets has been
attributed to the inverse-Compton scattering of the cosmic microwave background \citep[IC/CMB; ][]{tavecchio2000x,celotti2001large} by a jet that remains, throughout its extent, highly relativistic and closely aligned to our line of sight \citep[e.g.,][]{sambruna2004survey,2011ApJ...730...92H,Perlman_2011,2012ApJ...748...81K,Marshall_2018}. 

However, the IC/CMB model is discordant with many observations \cite[for a review, see~][]{worrall2009x}.
Specifically, it requires low-energy electrons (lower by $\sim 100$ than what the radio spectrum requires) that upscatter the CMB photons to X-ray wavelengths. The longer lifetimes of these X-ray emitting electrons would cause the X-ray emission to persist downstream of the radio knot \citep{2007RMxAC..27..188H}. However, the X-ray emission, when observed at similar angular resolution, peaks before the radio in the knots of many MSC jets
 \citep[e.g.,][]{2007ApJ...657..145S,kataoka2008chandra,Clautice:2016zai,Marchenko_2017}, while the opposite is almost never seen. Even the presence of sharply defined `knots' (compact, bright features) is difficult to reconcile with the very long radiative lifetimes of the electrons which are supposed to produce X-rays under IC/CMB. \added{Furthermore, an IC/CMB model for the X-ray emission in these jets unavoidably predicts a very high and steady flux in the GeV band \citep{georganopoulos2006quasar}; this emission has been strongly ruled out in many cases using the \emph{Fermi}/LAT \citep{meyer2015ruling,breiding2017fermi}.}

As an alternative to IC/CMB, the synchrotron mechanism  has been invoked to explain the X-ray emission from MSC jets \citep[e.g.,][]{ostrowski2002radiation,stawarz2004multiwavelength,2005MNRAS.360..926W,jester2006new,kataoka2008chandra,Clautice:2016zai}. In this scenario, there are separate high-energy and low-energy electron populations that produce X-rays and radio, respectively. Although co-spatiality between radio and X-rays is not required in the second-synchrotron case, whether it exists or not (subject to limited resolution) is an important constraint on both emission mechanisms and theoretical jet models.   
Under the synchrotron interpretation, a few possibilities to explain cases of X-ray-before-radio knots have been explored, such as a strong shock followed by a weaker shock downstream \citep[both produced by a bend in the jet;][]{2005MNRAS.360..926W}, synchrotron time lags associated  with downstream advection and radiative losses \citep{bai2003radio}, and moving knots producing forward and reverse shocks \citep{stawarz2004multiwavelength,kataoka2008chandra}.

Quantifying the positional offsets between radio and X-ray features in jets present an important way to constrain the X-ray emission mechanism and also the models of knot formation. A proper study of the offsets should first establish the frequency of their occurrence and any systematic trends that may emerge. This has not been done previously, in part because of the limited number of deep X-ray observations from which to infer offsets. %
Indeed, the majority of \textit{Chandra} detected X-ray jets have been made with shallow "snapshot" observations \citep[ and the references therein]{kataoka2005x,2011ApJS..197...24M,Massaro_2018,Stuardi_2018} where the faint knots are detected with only a few counts. These low-counts observations have not generally been analyzed for radio/X-ray offsets. 

In this paper, we present a new analysis of low-count X-ray jets and present our method for detecting radio/X-ray positional offsets using LIRA for these cases. These measurements will be included in a larger study of the entire X-ray jet population and their properties (including offsets) in a forthcoming publication (Reddy et al.,).  Section \ref{sec:data_desc} describes our sample and data reduction procedures including a description of LIRA and our method to infer offsets. Section \ref{sec:results} describes the offsets that are inferred using our method and Section \ref{sec:discussion} summarizes our work and presents a short discussion.
We adopt the cosmology \added{from \citet{2014A&A...571A...1P}} in which $H_0=67.8~\text{km}~\text{s}^{-1}~\text{Mpc}^{-1}$, $\Omega_m=0.308$ and $\Omega_\Lambda=0.692$. Spectral index $\alpha$ is given by the flux density, $F_\nu=\nu^{-\alpha}$ and photon index by $\Gamma=\alpha+1$.

\section{Methods\label{sec:data_desc}}
\subsection{Sample Selection}
We have utilized the XJET\footnote{\url{https://hea-www.harvard.edu/XJET/}} database (last updated in 2015) as well as a thorough search of the literature to compile an initial list of all known/claimed resolved X-ray emitting radio jets, a total of 189 sources. In this paper, we focus only on measuring offsets in low-count jet observations, because of the challenge of making observations in this regime, which requires special techniques not required in deep/high-counts imaging. A full analysis of offsets for the entire sample of resolved X-ray jets (including those introduced here) will be presented in a subsequent publication. We first identified 69 jets with fewer than 20 counts in at least one individual jet feature or knot. We were not able to analyze about 2/3 of these jets for the following reasons:
 ({\sl i)} X-ray imaging revealed underexposed cores (15 sources; $<35$ counts) that did not permit their accurate localization and/or spectral analysis. ({\sl ii)} They lacked any point or point-like components (e.g., knots or hotspots) in the radio where X-ray emission was reported (8 sources).  ({\sl iii}) They did not have a detectable core in the radio (18 sources), likely due to being highly misaligned (i.e., the core is de-beamed). Furthermore, we have only examined X-ray knots that do not have another X-ray component within two native ACIS-S pixels (0.492\as). This selection ensured that emission from one component did not induce a false offset in the other while also allowing an ROI that covered sufficient area to compute a centroid; 6 sources were excluded based on this criteria. With these exclusions our final analysis sample comprised 37 features in 22 jets.

In the appendix, Table \ref{table:sample} lists  all 69 low count X-ray jets, which  includes common names, IAU names, J2000 coordinates, redshifts, angular scales (kpc/arcsec), class of the source (e.g., FR-I, FR-II, quasar) and references to the corresponding publications. In the case of 4C+55.15, which was previously classified as a flat-spectrum radio quasar, \citet{McConville_2011} find unusually bright and steady $\gamma$-ray emission and suggest that this source might be a compact symmetric object (CSO). Hence we note its class as quasar/CSO. 

\subsection{Radio Data}
We have primarily used archival VLA observations to produce one or more high-resolution radio images of our sources. In a few cases, we imaged archival data from the  Australian Telescope Compact Array (ATCA), or used images available from the NASA Extragalactic Database (NED)\footnote{\url{https://ned.ipac.caltech.edu/}}. The radio data were chosen to have a resolution ($\approx$0.2\as-0.4\as)
to match the resolution of the X-rays (\siml0.25\as), except for Centaurus B and B3 1428+422, where the only available imaging was of resolution 1.3\as.  Table \ref{table:radio_obs} summarizes the details of the radio observations used in our analysis. For data reduction and imaging we followed standard procedures outlined in the handbook of Common Astronomy Software Applications \citep[CASA;][]{2007ASPC..376..127M}. For the VLA data, either 3C 286 or 3C 48 were used as flux calibrators and mostly the sources themselves as phase calibrators. After initial amplitude and phase calibration, a few rounds of self-calibration and imaging were performed  using the CASA tasks \texttt{gaincal} and \texttt{tclean}. The CLEAN algorithm was used for de-convolution with Briggs weighting (robust=0.5). For ATCA data, the initial pre-processing was done using MIRIAD \citep{1995ASPC...77..433S} and were imported in to CASA using the \texttt{importuvfits} task. For all the ATCA sources, 1934-638 was used as the flux calibrator and the sources themselves as the phase calibrators. The images were produced following the same procedure as used for the VLA data. Radio fluxes for knots were estimated by fitting an elliptical Gaussian fit using the CASA interface. For knots with poor fits, the pixel values inside their respective regions were summed to obtain the flux.

\begin{deluxetable*}{lllcCl}
  \tablecaption{Details of Radio Observations for 22 Analyzed Low-Counts Jets \label{table:radio_obs}}
  \tabletypesize{\scriptsize}
  \tablehead{
    \colhead{Name} &
    \colhead{Program}&
    \colhead{Date}&
    \multicolumn{1}{p{1.5cm}}{\centering Frequency\\GHz} &
    \multicolumn{1}{p{2cm}}{\centering Beam size\\\arcsec$\times$\arcsec}&
    \multicolumn{1}{p{1cm}}{\centering RMS\\$10^{-5}$Jy/beam}
  }
  \startdata
 3C 6.1 & AP0380 & 1999 Aug 02 & 8.4 & 0.25\times0.24 & 3.3\\
3C 17 & AS0179 & 1985 Mar 08 & 4.8 & 0.48\times0.39 & 101.0\\
3C 133 & AL0164 & 1987 Oct 09 & 4.8 & 0.39\times0.36 & 27.4\\
0529+075 & AH0824 & 2003 Aug 12 & 4.8 & 0.42\times0.35 & 10.5\\
3C 179 & AA0149 & 1992 Nov 10 & 8.4 & 0.29\times0.19 & 3.3\\
4C+25.21 & AK0353 & 1994 Mar 21 & 8.4 & 0.30\times0.21 & 5.7\\
TXS 0833+585 & AL0164 & 1987 Oct 09 & 4.8 & 0.43\times0.38 & 1.0\\
3C 213.1 & AK0403 & 1995 Jul 27 & 8.4 & 0.31\times0.24 & 6.9\\
3C 220.2 & AK0403 & 1995 Jul 27 & 8.4 & 0.33\times0.23 & 5.9\\
4C+55.17 & AM0672 & 2000 Nov 05 & 8.4 & 0.40\times0.21 & 26.7\\
PKS 1046-409 & C890 (ATCA) & 2004 May 12 & 19 & 0.73\times0.48 & 10.9\\
3C 275.1 & AG0247 & 1987 Sep 14 & 4.8 & 0.42\times0.40 & 3.5\\
3C 280.1 & AK0180 & 1987 Jul 26 & 4.8 & 0.47\times0.41 & 7.6\\
PKS 1311-270 & AK0353 & 1994 Mar 21 & 8.4 & 0.44\times0.23 & 6.8\\
4C+11.45 & AK0353 & 1994 Mar 21 & 8.4 & 0.30\times0.23 & 5.7\\
Centaurus B & C890 (ATCA) & 2002 Feb 02 & 8.6 & 1.35\times1.28 & 94.8\\
PKS 1402+044 & AG0670 & 2009 Oct 04 & 4.8 & 0.48\times0.37 & 8.4\\
B3 1428+422 & AC0755 & 2004 Dec 06 & 1.4 & 1.82\times1.39 & 8.0\\
3C 334 & AB0369 & 1986 May 05 & 4.8 & 0.40\times0.38 & 5.0\\
3C 327.1 & AM0548 & 1996 Oct 26 & 4.8 & 0.41\times0.38 & 6.8\\
3C 418 & AF0376 & 2000 Dec 10 & 8.4 & 0.24\times0.23 & 23.5\\
2123-463 & C890 (ATCA) & 2004 May 10 & 17.7 & 0.50\times0.50 & 32.4\\
\enddata
\end{deluxetable*}

\clearpage
\subsection{HST data}
We retrieved optical and IR data from the HST archive\footnote{\url{https://archive.stsci.edu/hst/}} and re-calibrated it by following the HST handbooks for the respective instruments. We utilized the latest version of \textit{Astrodrizzle} from the Drizzlepac software \citep{2015ASPC..495..281A} to create a final cosmic-ray corrected and combined image for each source. For sources with multi-epoch observations, the \texttt{tweakreg} command was used to align the images. The count rate was  measured using the APPHOT package from PyRAF \citep{2012ascl.soft07011S} and was then converted to flux density using the PHOTOFLAM keyword from the header of each image. The flux was then extinction-corrected using the values of nearest matching band reported in NED.

\subsection{Chandra data}
We retrieved archival observations from the \textit{Chandra} archive and  re-processed  %
the data using the standard methods outlined in Chandra Interactive Analysis of Observations (CIAO) threads\footnote{\url{http://cxc.harvard.edu/ciao/guides/index.html}}. We utilized CIAO v4.11 \citep{2006SPIE.6270E..1VF} and calibration information from CALDB v4.8.3. We checked the light curves for any background flares and excluded those time intervals accordingly. We only considered events falling in the energy range of 0.5-7 keV and in the final binned image, we excluded counts from pixels which had less than 2\% of the total exposure time. To ensure consistency across our analysis, we binned all the event files on half-pixel boundaries (bin size: 0.5) of the ACIS-S instrument (see sec.~\ref{subsec:lira_data_prep}). We set the image size to even powers of two (64$\times$64 pixels for smaller images and 128$\times$128 for larger ones) to conform to the LIRA input requirements. All the selected sources except 3C 6.1 have only one \textit{Chandra} observation. For 3C 6.1, which has two individual observations, we chose the observation with the longest exposure (ObsID: 3009; 36.5 ks exposure) and ignored the other (obsID: 4363; 19.9 ks exposure), which only reduced the total number of counts in the analyzed regions by 4. Table \ref{table:xray_obs} provides the details of the X-ray  observations utilized in our work. 
\startlongtable
\begin{deluxetable*}{llclcl}
  \tablecaption{Details of \textit{Chandra} X-ray Observations for 22 sources \label{table:xray_obs}}
  \tabletypesize{\scriptsize}
  \tablehead{
    \colhead{Name} &
    \multicolumn{1}{p{1cm}}{\centering Chandra\\ObsID} &
    \multicolumn{1}{p{2cm}}{\centering Effective\\exposure (ks)\textsuperscript{a}} &
    \colhead{Mode}&
    \colhead{counts/s}&
    \colhead{\texttt{blur}\textsuperscript{b}}
  }
  \startdata
3C 6.1 & 3009 & 33.7 & VFAINT & 0.048 & 0.3\\
3C 17 & 9292 & 7.9 & VFAINT & 0.206 & 0.3\\
3C 133 & 9300 & 7 & VFAINT & 0.083 & 0.25\\
0529+075 & 9289 & 8.7 & VFAINT & 0.077 & 0.3\\
3C 179 & 2123 & 8.6 & FAINT & 0.201 & 0.3\\
4C+25.21 & 10307 & 19.8 & FAINT & 0.011 & 0.25\\
TXS 0833+585 & 7870 & 3.8 & FAINT & 0.158 & 0.28\\
3C 213.1 & 9307 & 7.6 & VFAINT & 0.008 & 0.28\\
3C 220.2 & 18098 & 11.4 & VFAINT & 0.059 & 0.3\\
4C+55.17 & 4842 & 28 & FAINT & 0.114 & 0.25\\
PKS 1046-409 & 3116 & 3.8 & FAINT & 0.227 & 0.25\\
3C 275.1 & 2096 & 21 & FAINT & 0.174 & 0.2\\
3C 280.1 & 21398 & 15.8 & VFAINT & 0.043 & 0.25\\
PKS 1311-270 & 10306 & 17.9 & FAINT & 0.049 & 0.3\\
4C+11.45 & 10310 & 17.8 & FAINT & 0.029 & 0.27\\
Centaurus B & 3120 & 3.7 & FAINT & 0.283 & 0.25\\
PKS 1402+044 & 20408 & 9.6 & VFAINT & 0.03 & 0.26\\
B3 1428+422 & 7874 & 7 & FAINT & 0.201 & 0.25\\
3C 334 & 2097 & 17.6 & FAINT & 0.237 & 0.25\\
3C 327.1 & 13887 & 10.3 & VFAINT & 0.128 & 0.28\\
3C 418 & 21401 & 17.8 & VFAINT & 0.087 & 0.25\\
2123-463 & 4890 & 6.47 & FAINT & 0.010 & 0.28\\
         \enddata
         \tablenotetext{a}{After background flare removal, if any.}
         \tablenotetext{b}{See section \ref{subsec:lira_data_prep} for details.}
        \end{deluxetable*}

\subsection{Radio/X-ray Image Alignment}
An accurate offset inference requires the X-ray and radio cores to be precisely aligned. The high signal-to-noise (SNR) of the cores in our sample permitted us to estimate the X-ray centroid of the core in a straightforward manner using \texttt{dmstat} module in CIAO. We varied the size and position of the region used to compute the centroid to ensure that our region selection does not bias astrometric corrections. Furthermore, as a consistency check, we also applied wavelet decomposition (\texttt{wavdetect} module in CIAO) to independently estimate the centroid of the core. In all cases, centroids from both the methods were always within 0.05\as. For the radio, we fit an elliptical Gaussian using \texttt{imfit} task in CASA to obtain the location of the core. In all the cases the integrated flux was equal to the peak flux (indicating the core as an unresolved point source, well-fit by a 2D Gaussian model) and the error in the location was less than 0.1 mas. We note that while the observed location of the observed radio core relative to the central black hole (or true base of the jet) is frequency-dependent due to the ``core-shift'' effect \citep{blandford1979relativistic,konigl1981relativistic},
the associated uncertainty is at most a few milli-arseconds \citep[e.g.,][]{2008A&A...483..759K,2012A&A...545A.113P,plavin2019significant,2019Galax...7...20B}, and hence cannot produce the observed offsets we find in this study, which are 2 orders of magnitude larger. Because the radio core positions are extremely well-measured, the error in the radio-X-ray core alignment is mainly contributed by the X-ray observation, which we take to be 0.05\as.

\subsection{Measuring X-ray/Radio Offsets with LIRA}

Practically all the offsets reported in the literature are serendipitous discoveries that emerged from deep/high-count X-ray observations. Traditionally, smoothed longitudinal brightness profiles are used  to detect X-ray-to-radio morphological differences \citep[e.g.,][]{2007ApJ...657..145S,marshall2011x,Harris_2017}. Alternatively, centroid-based position differences can also be used to detect an offset in isolated features. Detecting an offset requires precise localization of the X-ray features, which is more easily achieved in high-count images.
 However, these methods begin to lose their accuracy when approaching the low-count regime: the emission from the core and other background fluctuations introduce significant uncertainties which these methods cannot quantify. LIRA\footnote{https://github.com/astrostat/LIRA} is a robust tool designed to address these problems, and we briefly summarize it below. A complete description of the algorithm is given in \citet{stein2015detecting}; see also \citet{McKeough_2016}.

\subsubsection{LIRA\label{subsubsec:lira_overview}}

The basic approach of LIRA is to model an image as a combination of two independent `components', each of which contributes to each individual pixel. The first model component is the user-specified `baseline component' which includes all features which we know or assume are present (for example, the point-source core and a constant background). Defined separately (on a per-pixel basis) is the `added component' which is not specified in advance, but which is inferred based on a comparison of the real data to the baseline model. 
LIRA adopts a multi-scale representation of an observation, which was initially used to estimate uncertainties in the deconvolution of X-ray images \citep[known as EMC2, ][]{esch2004image}. Using this representation, which allows flexible deviations from the baseline component, and Bayesian methodology, LIRA derives the added component by fitting  the full model to the observation. Specifically, it computes the posterior of the image parameters (i.e., pixel values in both the components) given the observation. 
Together, the superposition of the two model components (baseline and added) gives the underlying brightness distribution in each pixel, and the observed image can be reconstructed by convolving it with the PSF. For all our jet images, we presume a baseline model of a point source and a constant background. The added model would contain emission from any jet-related and other background features.

LIRA samples the resulting posterior using Markov chain Monte Carlo (MCMC) algorithm to generate images of the added component, which contains any emission that is unexplained by the baseline model.
In \citet{stein2015detecting}, a hypothesis test is defined to estimate the statistical significance of emission from images of the added  component within a chosen ROI \citep[see also][]{McKeough_2016}.  Here the null hypothesis states that the baseline model completely describes the observed emission and the alternative is that an additional component over the baseline is necessary.  The test uses the tail probability of the posterior of $\xi$ \citep[Eq. 11 in][]{stein2015detecting} for the observed image, where $\xi$ is fraction of the total emission (baseline+added) from the added component within a specified ROI. The tail's location is measured by fixing a small number for the upper tail probability ($\gamma$) of a combined posterior of $\xi$ that is obtained from several simulations of the baseline component.  This is essentially a test of whether the distributions of $\xi$ differ between the actual observation and simulated images of the baseline component.
A large discrepancy between the two distributions would therefore indicate a significant additional component (e.g., emission from the jet) in the observation. An example is shown in  Fig.~\ref{fig:lira_offet_demo}, which is further described in section \ref{subsubsec:lira_demo}.

\subsubsection{Setting the baseline model \label{subsec:lira_data_prep}}
For each source, we used the \texttt{specextract} module in CIAO to extract the spectrum of the core. 
A powerlaw model with two absorption components (host galaxy + our galaxy) was fit to this spectrum using SHERPA \citep{2001SPIE.4477...76F,2007ASPC..376..543D}, which is available as a python module in CIAO.  For sources where the hydrogen absorption column density (nH) of the host galaxy could not be constrained, we only used the galactic nH value\footnote{Retrieved using the WebPIMMS service} for the absorption and re-performed the fitting. We utilized the Levenberg–Marquardt optimization algorithm and $\chi^2$-statistic with Gehrels variance function for our fitting. We passed the fitted model to \texttt{simulate\_psf} module in CIAO and used MARX v5.4.0 \citep{2012SPIE.8443E..1AD} as the backend to simulate PSF of the core. We performed 100 iterations for each source and averaged them to generate the final PSF. We also matched the pixel boundaries of simulated PSF and its corresponding input image to ensure that we sampled the same part of the PSF as the core. 

Simulating a PSF using MARX includes reconstructing the aspect by convolving the PSF with a Gaussian. The variance of this Gaussian (\texttt{blur} parameter in \texttt{simulate\_psf}) is set to  0.07\as~by default. Historically, this value produced a narrower PSF at smaller radii and broader PSF wings than an actual ACIS-S observation \citep{2011ApJS..194...37P}, and this mismatch remains unexplained. A preliminary analysis conducted on some observations by the MARX team indicated that a value of $\approx0.25$\as~produced a better matching PSF at smaller radii while a slightly larger value of $\approx0.28$\as produced a better matching PSF wings for the ACIS-S instrument\footnote{\url{https://cxc.cfa.harvard.edu/ciao/why/aspectblur.html}}. However, the exact value depends on the source spectrum, and no analytical methods exist to compute it. Hence, to estimate the \texttt{blur} that best represented the observation at both smaller and larger radii, we 
simulated PSFs using \texttt{blur} values between 0.0 to 0.35 with a step size of 0.01.
We also generated their enclosed counts fraction (ECF) profiles using the \texttt{ecf\_calc} task in CIAO. The best-matching \texttt{blur} then corresponded to the simulated PSF with the smallest squared error between ECF profiles of itself and the observation. The resulting values ranged between 0.2-0.3 (see Table \ref{table:xray_obs}).

We adopted the approach outlined in \citet{McKeough_2016} to generate a baseline model for each source. Using SHERPA, we fit a 2D gaussian plus a constant background convolved with the simulated PSF of the core to the sub-pixel image. We then simulated 50 images of the fitted model using SHERPA's \texttt{fake} function. Changing the \texttt{blur} value also affected the baseline model. Using the default \texttt{blur} resulted in a Gaussian with its full width at half-maximum (FWHM) exceeding a single pixel. However, the best-matching \texttt{blur} resulted in a delta-like function, which is also consistent with the expectation that the core is a point source.

\subsubsection{Detecting and Locating X-ray Components with LIRA\label{sec:lira_method}}
Although we can use LIRA to estimate the statistical significance of any potential emission within an ROI (beyond the baseline model), we have to choose an ROI \textit{prior} to any analysis \citep{McKeough_2016}. 
Virtually all the jet knots and hotspots detected in deep \emph{Chandra} imaging are unresolved, with only a few exceptions \citep[e.g.,][]{jester2006new,kataoka2008chandra} where the resolved knots are very nearly point-like. Therefore, we treat jet features detected using LIRA as unresolved %
and use a centroid to locate the feature, just as is done for the well-sampled X-ray core of the source. To reduce bias, we construct an ROI that only encloses the radio source, something which may not always enclose the entire associated X-ray emission (e.g., due to an offset). Nevertheless, we can still test whether this centroid spatially coincides with its corresponding radio peak to detect any offset and thereby estimate its lower limit.

Each MCMC image realization created by LIRA will have slightly different pixel values, and the exact centroid value will vary with each image. To accurately capture the centroid position for the X-ray emission of the added model within our radio-based ROI, we average a large number of MCMC images and then compute the centroid on that average to determine the X-ray feature position. 
Although all the jets in our sample are previously reported as detected, we re-estimate the significance of the emission within each ROI using LIRA. This is important to ensure that our chosen ROI contains statistically significant emission from the jet.

As all the features in our sample have less than 20 counts, following \citet[][Table 1]{stein2015detecting}, we set $\alpha=0.01$ and $\gamma=0.005$. Put another way, we marked the emission inside a specified ROI as statistically significant only if \textit{p}-value$\leq0.01$ (of the hypothesis test using LIRA; see section \ref{subsubsec:lira_overview} for details). Finally, for each source, after discarding the first 1000 draws, we generated 2000 MCMC images of the added model for each of the observed image and all the 50 simulated baseline images.

\subsection{X-ray flux measurement}
For each source, a flux image (in counts/sec) was generated by dividing the averaged LIRA image with the exposure map. This image was used to derive the reported X-ray fluxes for detected X-ray knots. 
We first created a weighted instrument map using the \texttt{mkinstmap} task in CIAO. The required spectral weights were generated by passing an absorbed powerlaw model ( with $\Gamma=2$) to the \texttt{make\_instmap\_weights} task; the galactic absorption column density was used for this purpose. We then passed this instrument map along with the aspect histogram to the \texttt{mkexpmap} task to generate the final exposure map. The pixel values inside each ROI were summed to obtain the photon flux for each component. Finally, 
the photon flux was converted to the energy flux density at 1 keV by assuming $\Gamma=2$ and was absorption-corrected using the WebPIMMS service. We note that the exact choice of $\Gamma$ (for reasonable values 1.5$-$2.5) has a minor impact on the reported flux (at most $\pm$6\%) and thus has no significant impact on the observed trends reported in section \ref{subsec:flux_ratio}.

\section{Results\label{sec:results}}
\subsection{An example for detecting offsets using LIRA \label{subsubsec:lira_demo}}
We present here as an example of our method the results of our analysis of PKS 0605$-$08, a very bright quasar. It has been observed by \textit{Chandra} for a total of 62.8 ks (in three splits). The merged observation has \siml900 counts in its jet region, which does not qualify it as a valid low-count source for our main analysis.  However, it has a known radio/X-ray offset of \siml0.9\as in the terminal knot \citep{sambruna2004survey}, and thus this is an ideal observation to evaluate our method, by breaking the observations into smaller units mimicking low-counts observations.
 \begin{figure*}[h]
    \gridline{
        \fig{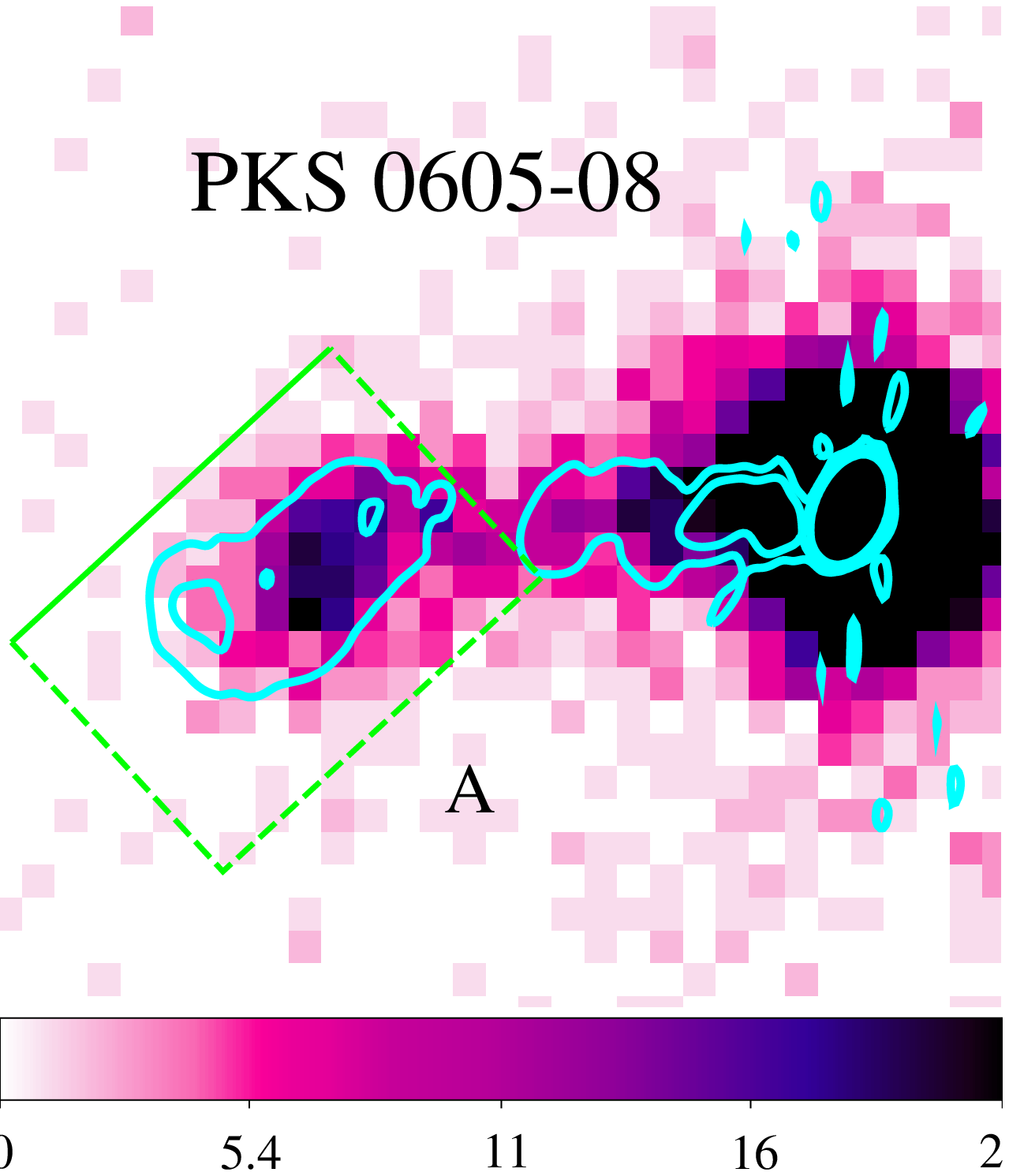}{0.33\textwidth}{(a)}
        \fig{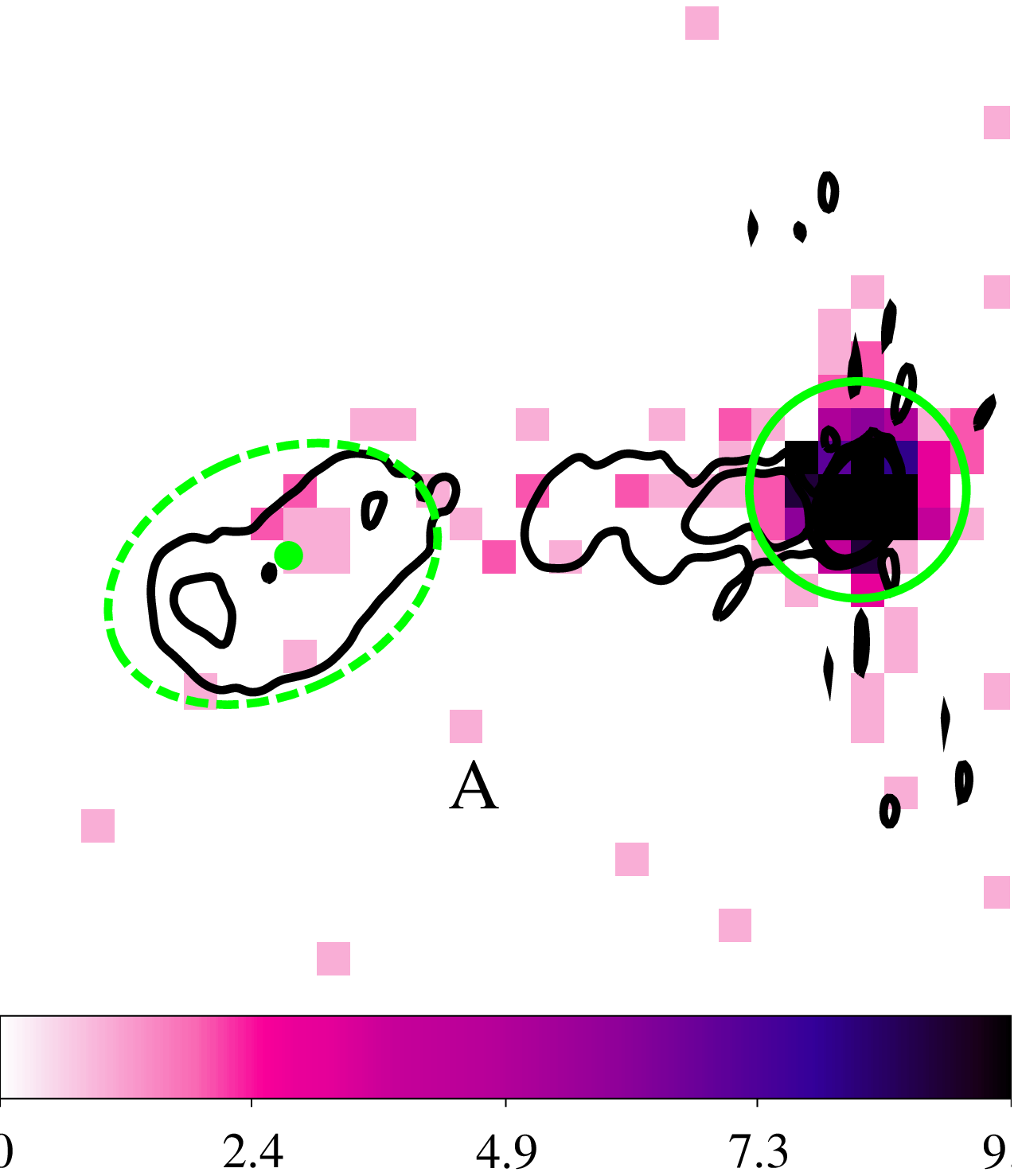}{0.33\textwidth}{(b)}
        \fig{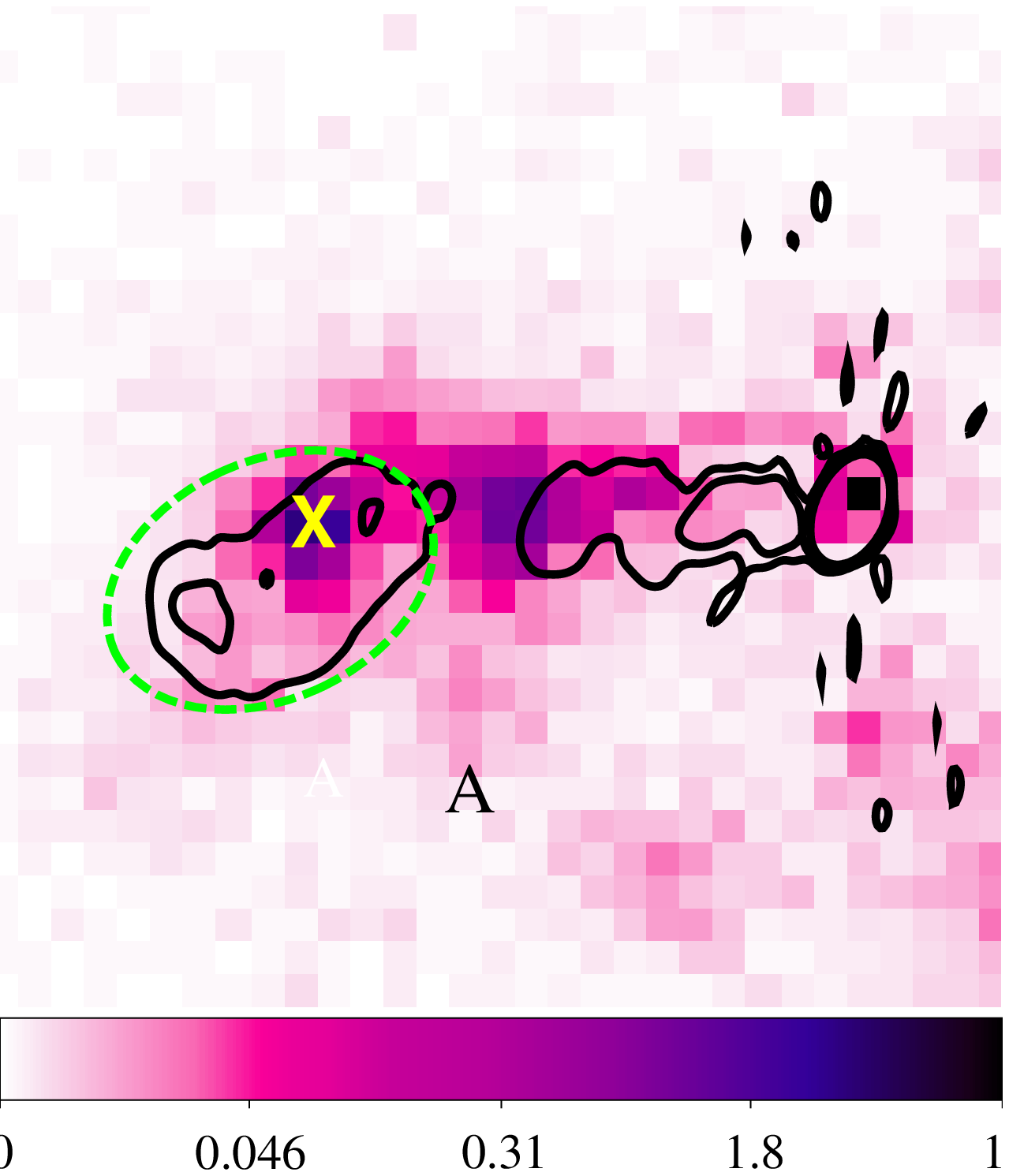}{0.33\textwidth}{(c)}
    }
    \gridline{
        \fig{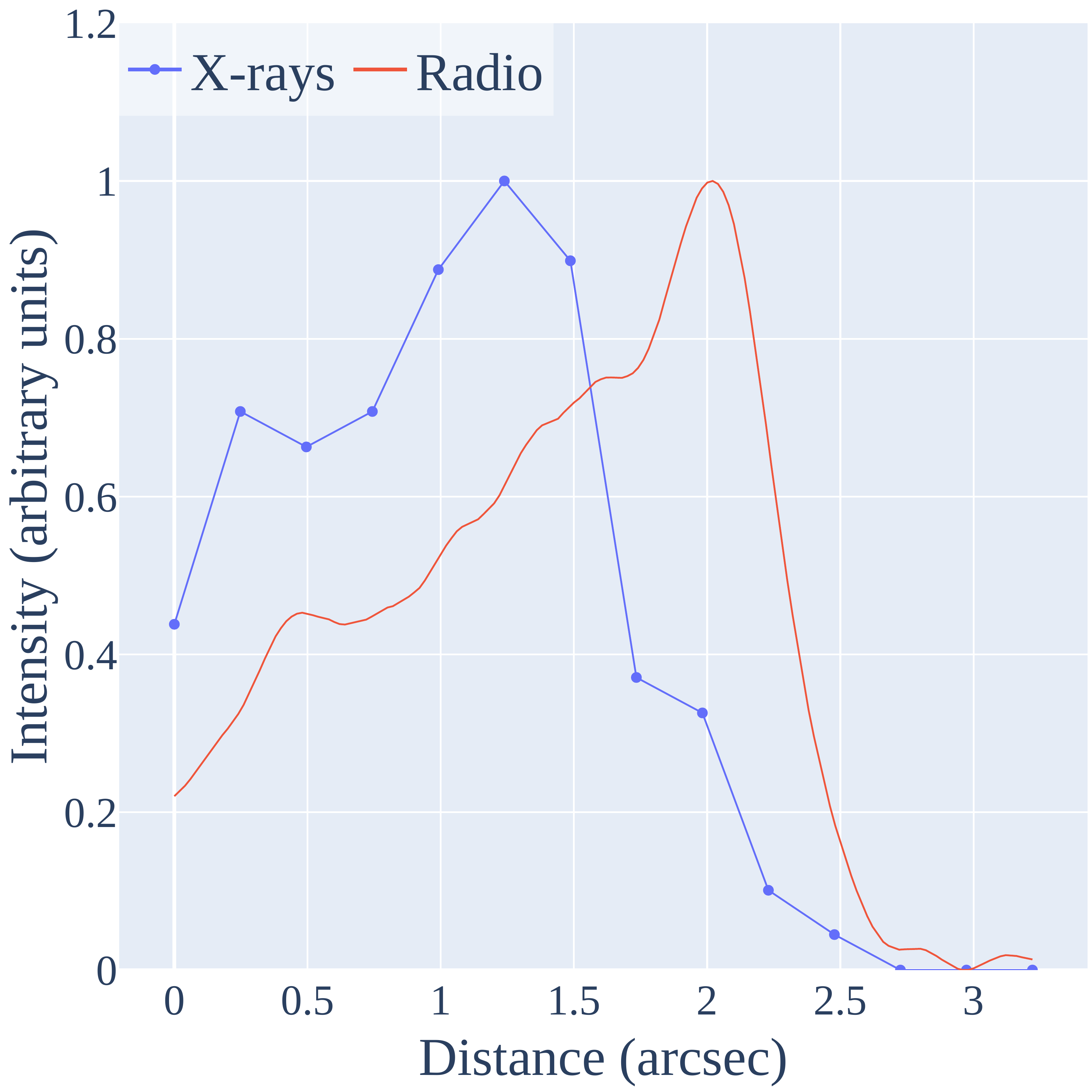}{0.4245614035\textwidth}{(d)}
        \fig{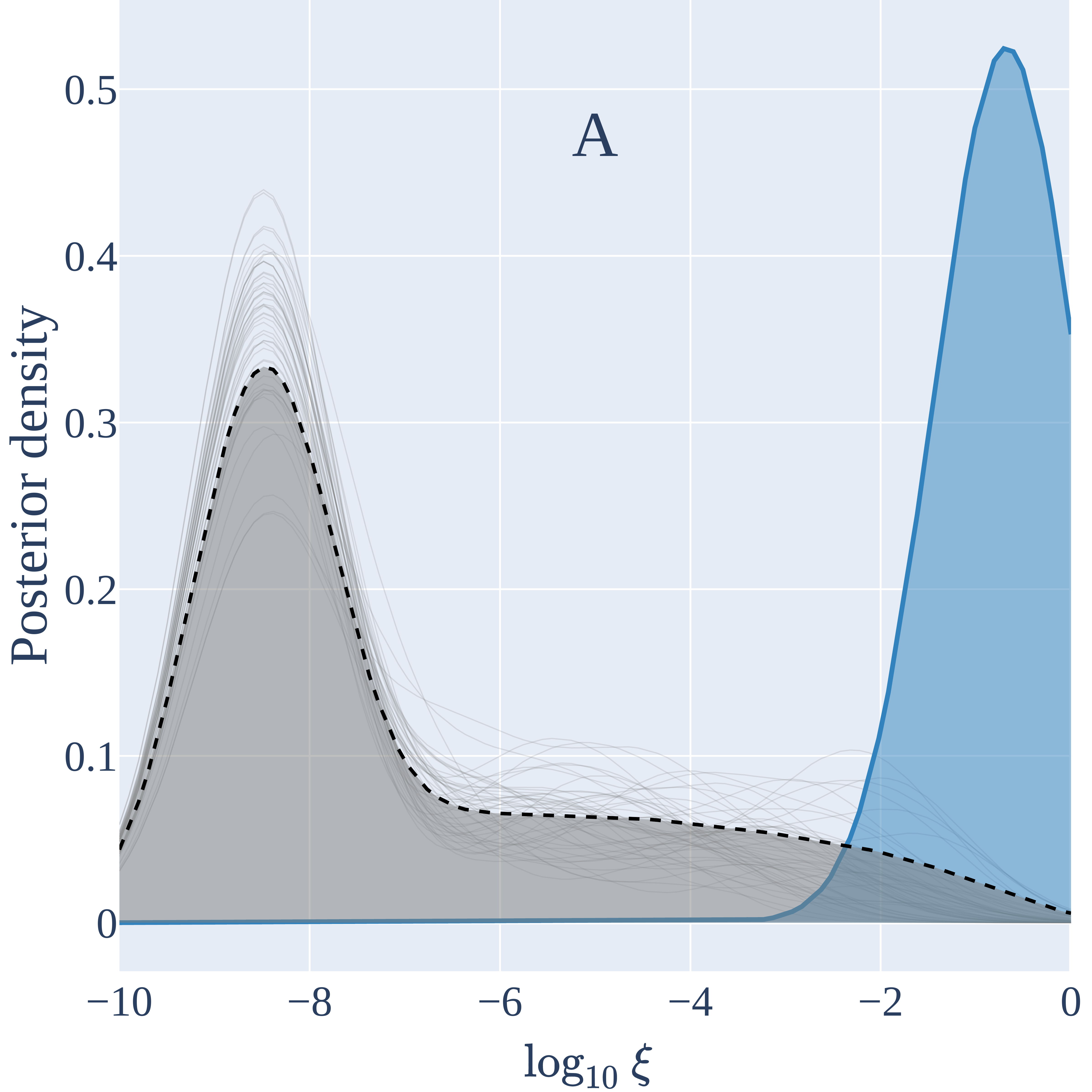}{0.4245614035\textwidth}{(e)}
    }
\caption{Using LIRA to infer an offset in a low count \textit{Chandra} image of PKS 0605-08 -- With the VLA 5 GHz radio contours overlaid, (a) A merged 62.8 ks observation binned by a factor of 1/2. (b) A 2 ks observation binned by a factor of 1/2. The solid circle indicates the region used to extract the spectrum of the core and the Green dot indicates the centroid of the dashed-green ellipse. (c) The average added component of the 2 ks observation. The Yellow \texttt{X} marks the centroid of the dashed-green ellipse, which differs from the radio peak by \siml0.9\as. (d) The longitudinal intensity profile of knot B computed using the deep observation [dashed-green rectangle in (a)]. The X-rays peak upstream of the radio \citep{sambruna2004survey}, thereby confirming the offset as inferred using LIRA. (e) The posterior distributions of $\xi$ for the dashed-green ellipse in the observed (teal) and the baseline replicates (gray; black dashed line indicates the average); a discrepancy between the black and the teal curves indicates a statistically significant excess emission. 
\label{fig:lira_offet_demo}}
\end{figure*} 

We here focus on knot A, which has a previously reported X-ray-to-radio peak-to-peak offset  \citep{sambruna2004survey}, based on a moderately deep exposure (8.7 ks). The offset measured using the merged observation using \texttt{dmstat} is \siml0.9\as. We extracted events from the first 2 ks of observation ID 2132 to create a ``low-count" observation and followed the procedure described above to generate an image for the averaged additional component. In Fig.~\ref{fig:lira_offet_demo} we show the merged 62.8 ks \emph{Chandra} image in panel (a) with the box showing the knot A region. Panel (b) shows an example image from only 2 ks of the total observation and (c) shows the averaged added component.  We constructed an ROI (dashed-Green ellipse in panels b, c) based on the radio that only enclosed knot A, and first verified that the emission inside it is statistically significant ($p\leq0.005$). We then averaged all the additional component images of the observation and used it to compute the centroid of the ROI. The resulting centroid (yellow \texttt{X} in Fig.~\ref{fig:lira_offet_demo}c) lies \siml0.9\as~upstream of the radio peak, in agreement with what is measured from the merged observation. This is not a lower limit because the chosen ROI contains all the X-ray emission associated with knot A, as shown in panel (a). Panel (d) shows longitudinal brightness profiles of radio (red) and X-rays (blue) [taken along the dashed rectangle shown in panel (a)], which confirm the offset. Panel (e) shows knot A's distributions of posterior density of $\xi$ derived by running LIRA on the observed image (teal) and on each of the 50 simulated baseline models (gray; their average is shown in dashed-black). A large discrepancy (quantified using an upper limit on the \textit{p}-value) between the black and teal curves indicates a statistically significant emission.

While aligning the radio and X-ray images, we found that the centroid estimates from \texttt{wavdetect} and \texttt{dmstat} modules were within 0.05\as~of each other. As an independent check of this error, we sub-divided all the observations of PKS 0605-05 (2132, 11431 and 12056 totaling 62.8 ks) into units of 1 ks each and re-estimated the centroid of the core with the \texttt{dmstat} module. The resulting centroids were always within 0.05\as~of the centroid measured on the merged observation.

\begin{figure}
    \centering
    \includegraphics[width=0.5\textwidth]{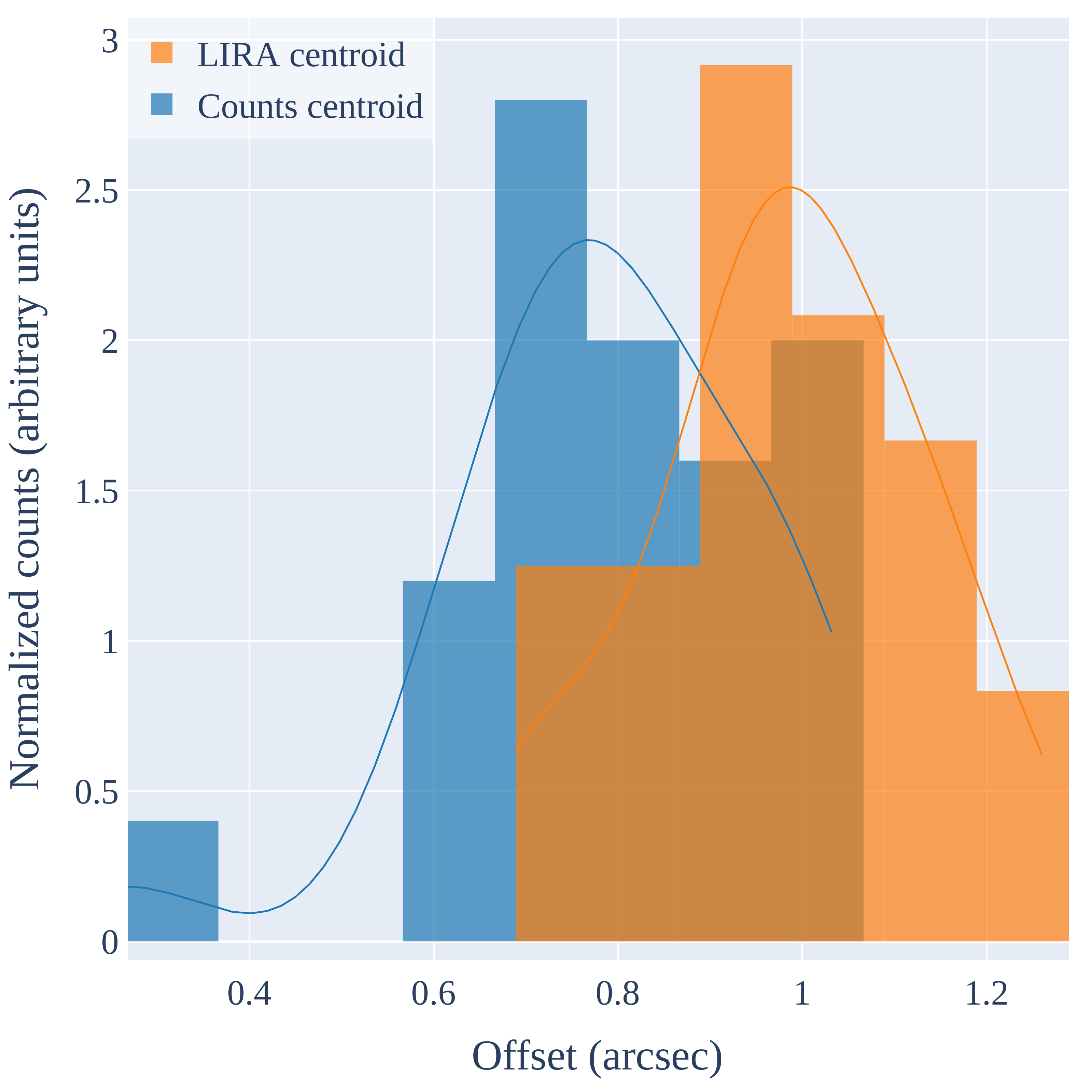}
    \caption{Histograms and corresponding kernel density curves of offsets measured using our method (orange) and using \texttt{dmstat} directly on the counts-image (blue).}
    \label{figure:error_hist}
\end{figure}

To  compare the accuracy of our method against computing an offset directly on the counts-image, we again sub-divided the two observations into units of 1.2 ks each. This time unit was chosen to match the average counts (\siml8; see table \ref{table:results}) in the features of our main analysis with that in knot A (\siml7) of PKS 0605-08. We used both our method and \texttt{dmstat} on a set of 25 randomly chosen sub-units to measure the offsets. Figure \ref{figure:error_hist} shows histograms and corresponding kernel density estimate curves of the measured offsets. The blue color indicates offsets measured directly on the counts image. It is apparent that the blue histogram is biased towards lower values (peaking at \siml0.77) and also extends to values as small as 0.26. In comparison, the offsets measured using our method (orange) peaks at 0.97, which is consistent with the offset measured on the merged observation, and the distribution is clearly peaked at this value with no major asymmetry or bias apparent. In cases with more significant core and background counts in the knot region, these differences would likely be more pronounced.

\subsection{A New X-ray Jet and Two Non-detections \label{subsec:new_detections}}

For all cases, we verified that the chosen ROI contained statistically significant emission from the jet before measuring an offset. All the ROIs were consistent with previously reported detections of X-ray emission except in 3C 327.1 (\textit{p}-value$\leq$0.08) and PKS 1402+044 (\textit{p}-value$\leq$0.04). We could not confirm the presence of any significant X-ray emission from those two sources. The contours of the \textit{Chandra} X-ray image of 3C 327.1,  initially presented in Figure 18 of \citet{massaro2013chandra}, suggested a jet-like structure that aligned with the radio jet. However, we observed a similar jet-like structure in a few of its baseline replicates, and our analysis with LIRA also  did not indicate any significant X-ray emission from the jet.

\begin{figure*}[h]
    \gridline{
        \fig{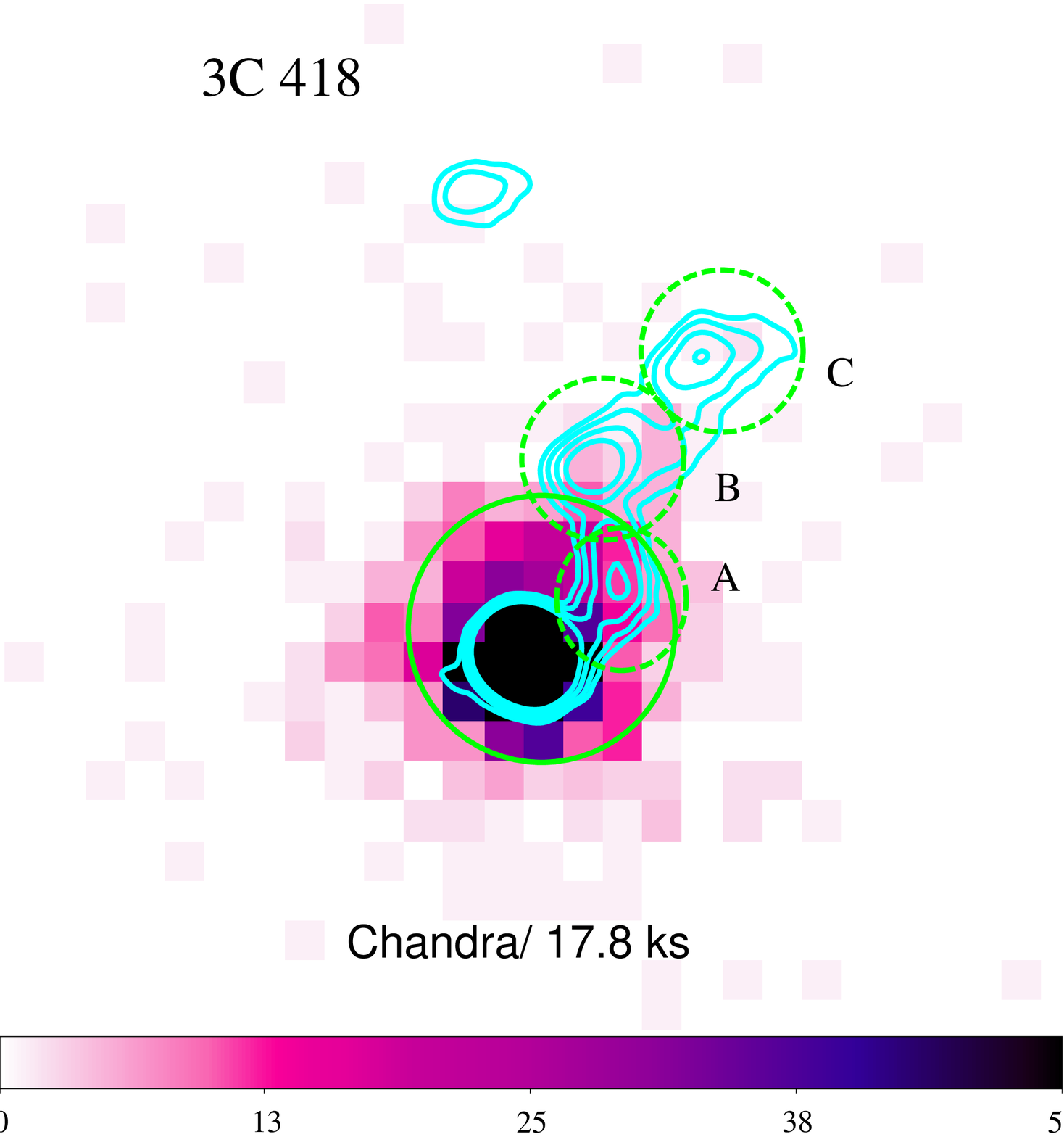}{0.5\textwidth}{(a)}
        \fig{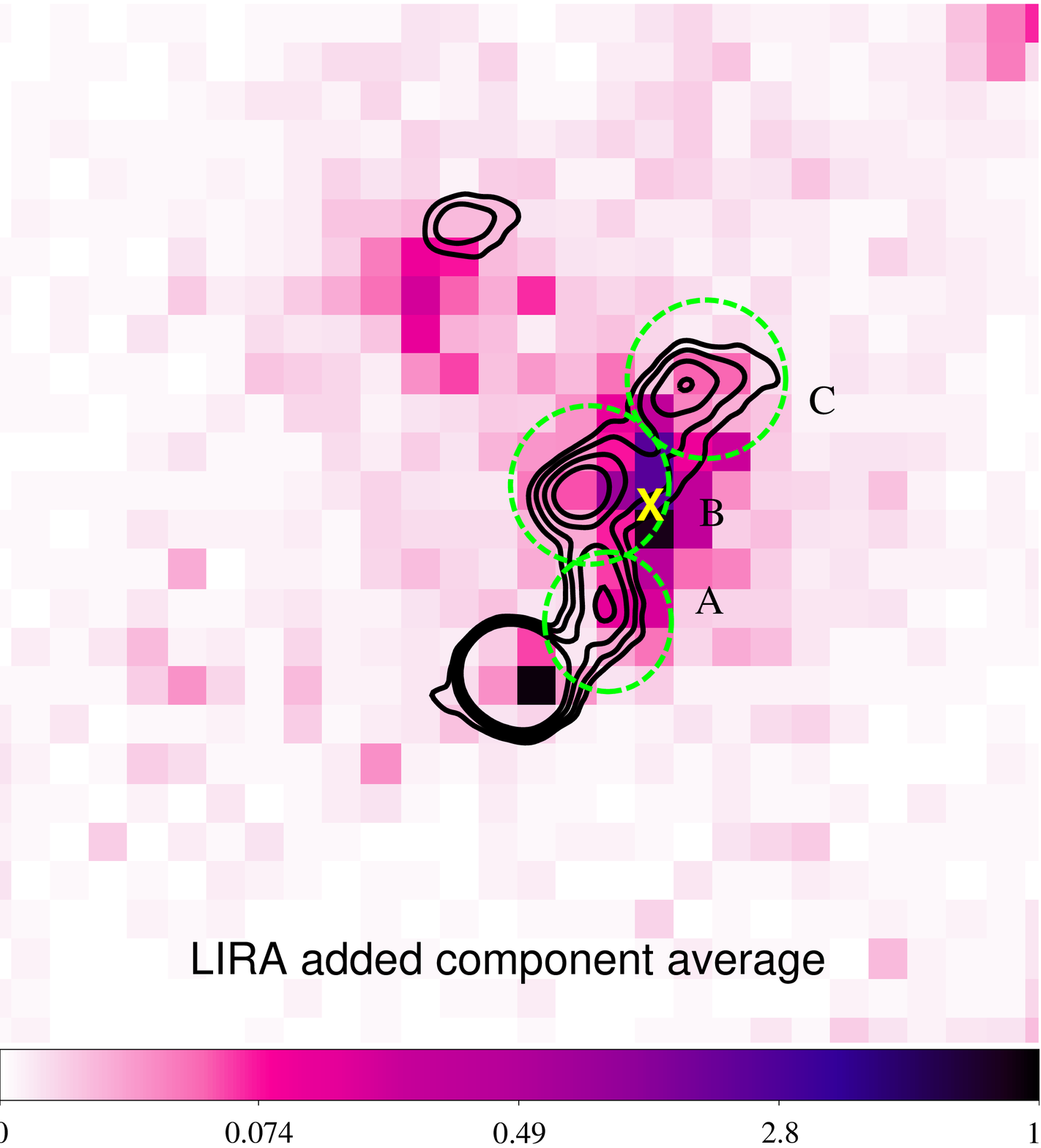}{0.5\textwidth}{(b)}
    }
    \gridline{
    \fig{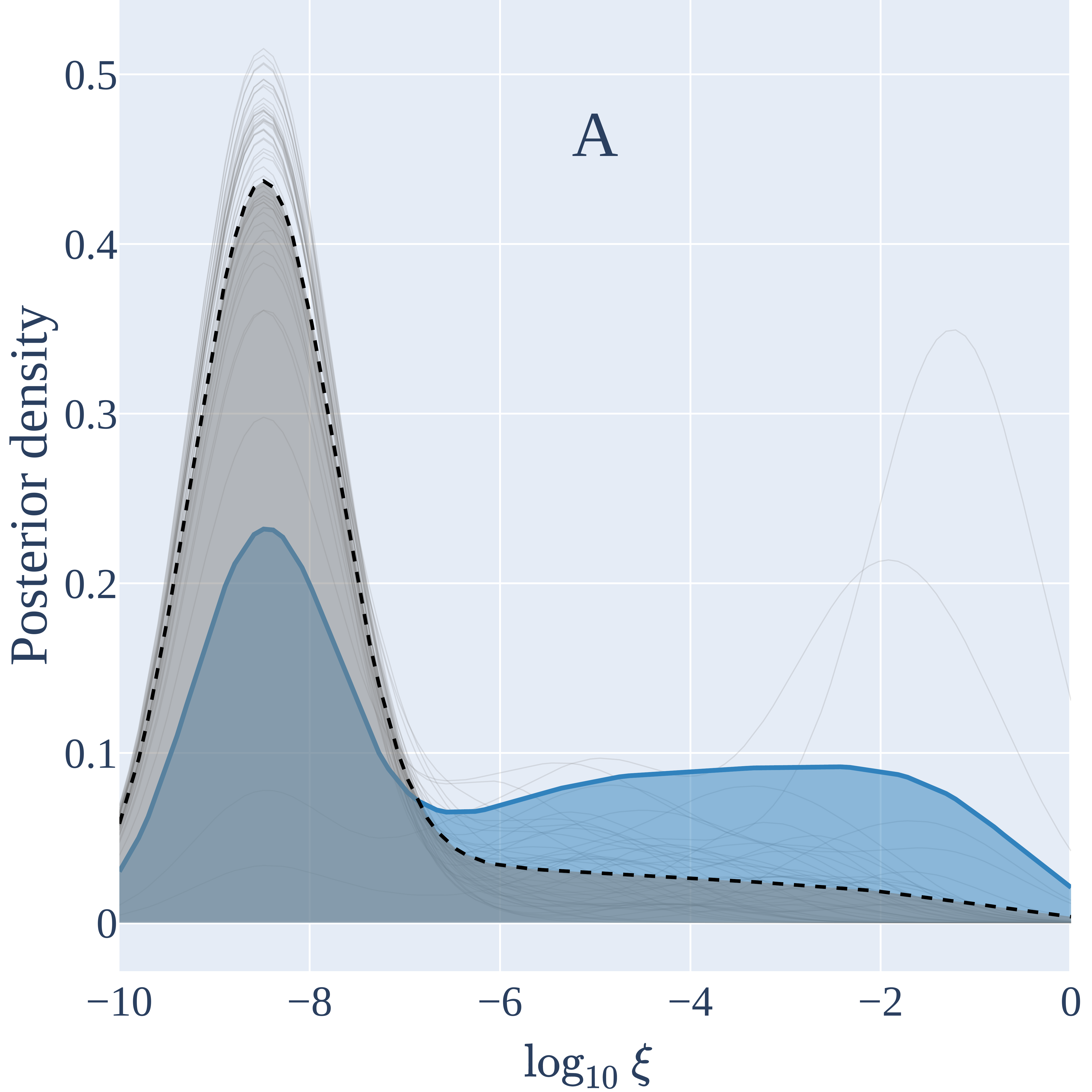}{0.33\textwidth}{(c)}
     \fig{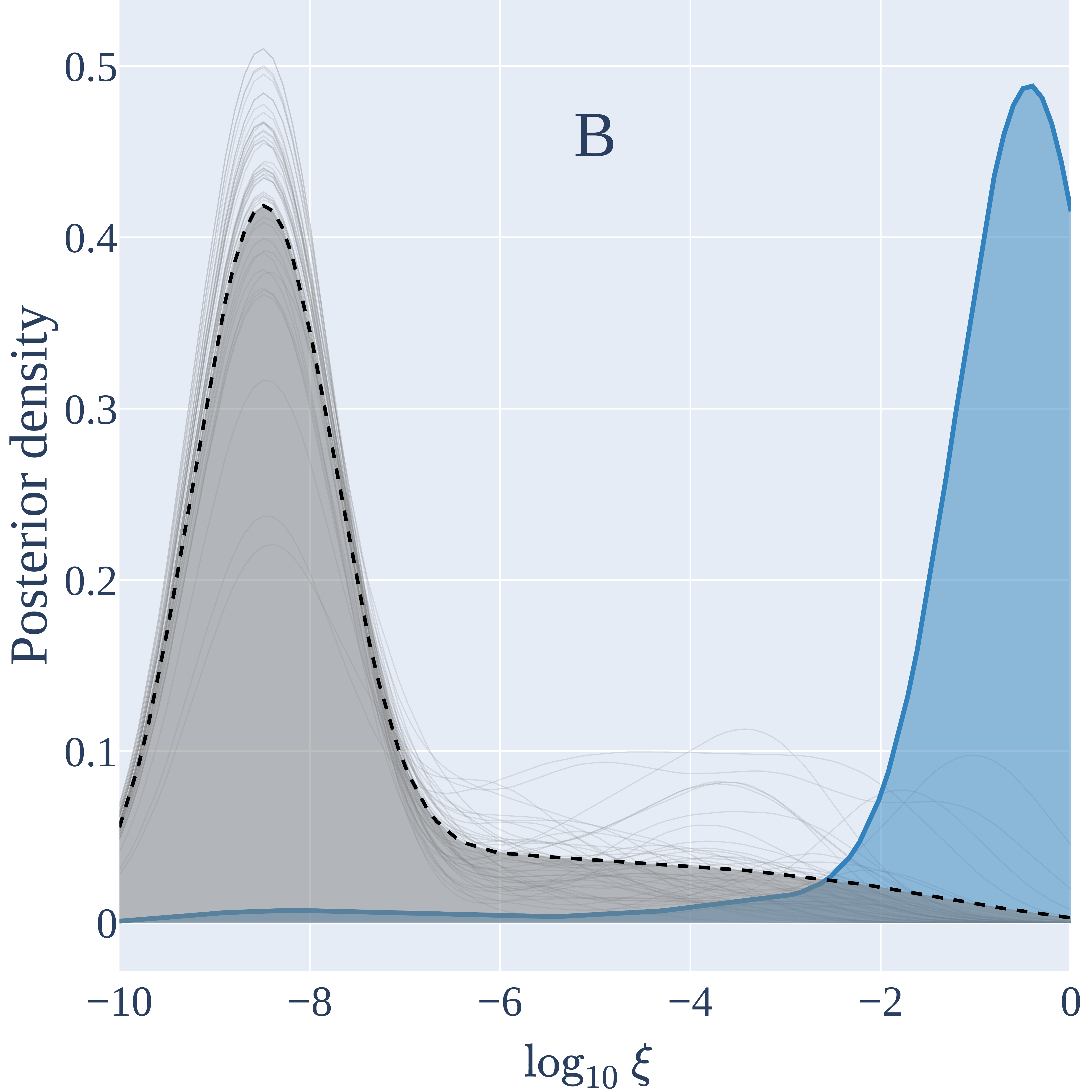}{0.33\textwidth}{(d)}
      \fig{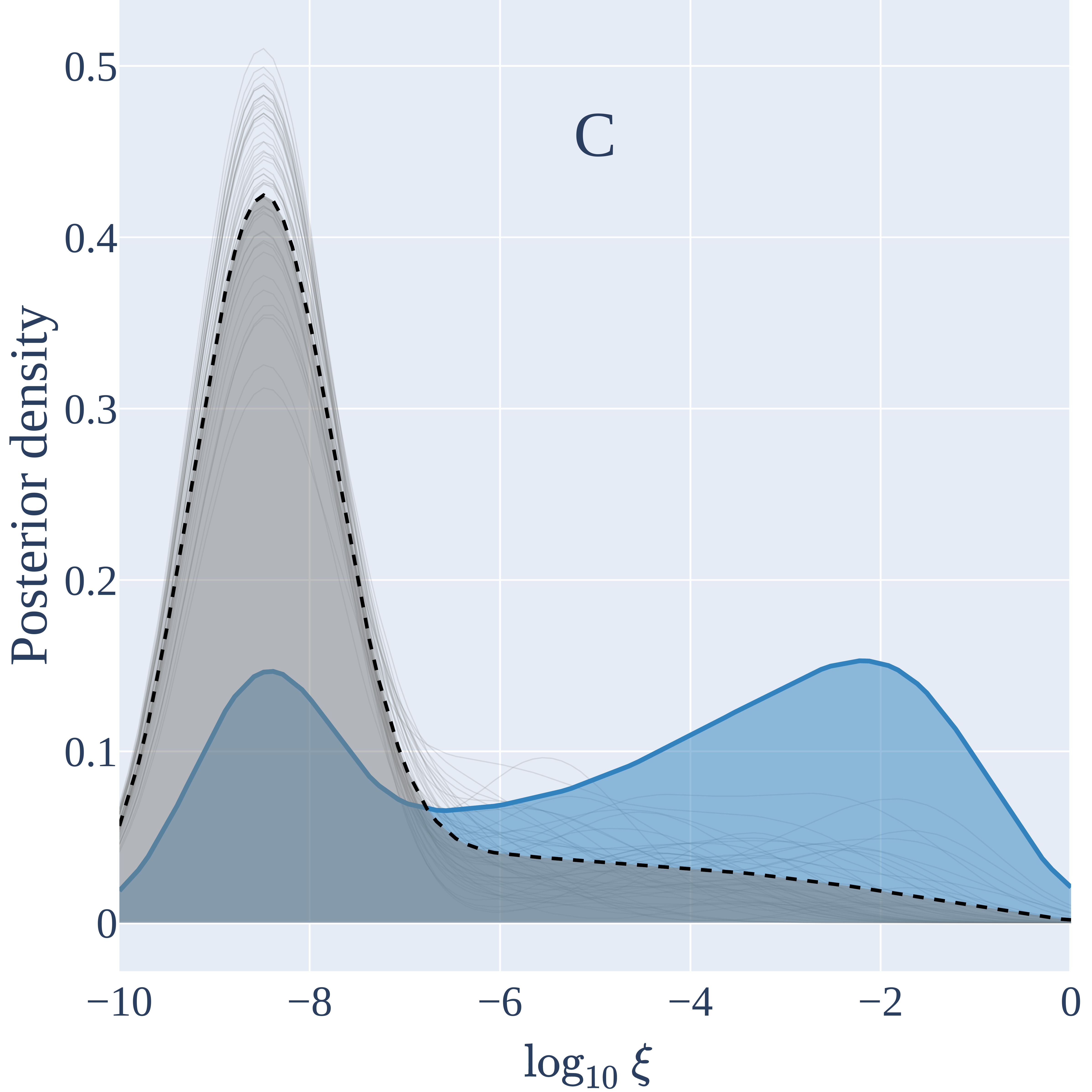}{0.33\textwidth}{(e)}
    }
    
\caption{Results for the analysis for 3C 418, a quasar. With the VLA 8.4 GHz radio contours overlaid in the top panel--(a) The \textit{Chandra} observation, binned by a factor of 0.5. (b) The average added component where the Red \texttt{X} marks the centroid of the added component within each ROI (dashed-green).  (c)-(e) The posterior distributions of $\xi$ (fraction of the total emission in the added component) in each ROI  for the observed image (shaded blue), 50 simulated images of the baseline model (gray) and their average (gray shaded area with a dashed-black curve). \Radiocontours{ 2, 4, 8, 15} \label{fig:results_3C418}}
\end{figure*}

We have detected new jet-linked emission from 3C 418 (see Fig.~\ref{fig:results_3C418}). This source shows a 3\as~knotty jet in the radio with multiple bends. The jet makes a 70\degree~projected bend to the northwest at knot A and then makes another 60\degree~projected bend to the northeast at knot B. Although its \textit{Chandra} observation was first published in \citet{jimenezgallardo2020textitchandra}, as the jet was within  the PSF of the core, no jet detection was indicated. Our analysis using LIRA revealed a significant X-ray emission in an ROI constructed around knot B. We did not find evidence for X-ray emission from knots A and C. Interestingly, the centroid of the X-ray emission in knot B lies on the outer edge of the jet. This morphology suggests the jet itself may not be producing the observed X-ray emission. It may instead be originating from a jet-deflecting stationary obstacle (e.g., a dense gas cloud) that gets shock-heated by the jet \citep[e.g.,][]{worrall2016x}. Finally, in any of our sources, we found no significant emission from the core region that exceeded emission from a single point source (e.g., from an unresolved jet).

\startlongtable
\tablecolumns{10}
\begin{deluxetable*}{llL|clc|lccc}
  \tablecaption{Results for the offset analysis\label{table:results}}
  \tablehead{
    \colhead{Name} &
    \colhead{Component}&
    \colhead{Region\textsuperscript{a}}&
    \multicolumn{3}{c}{Offset}&
    \colhead{Counts}&
    \colhead{Optical/IR\textsuperscript{d}}&
    \colhead{Spectral type\textsuperscript{e}}&
    \colhead{\textit{p}-value}\\[-5pt]
    \colhead{}&
    \colhead{}&
    \colhead{arcsecond$^2$}&
    \colhead{\arcsec}&
    \colhead{Type\textsuperscript{b}}&
    \colhead{kpc\textsuperscript{c}}&
    \colhead{}&
    \colhead{}
  }
  \startdata
3C 17  &  S10.8  &  e(0.65\times0.70)  &  0.18  &  Xf  &  0.66  &  2  &  N  &  -  &  0.009\\
PKS 1311-270  &  B  &  c(0.6\times0.6)  &  0.25  &  Xf  &  2.11  &  8  &  N  &  MSC (3)  &  0.005\\
3C 275.1  &  A  &  e(0.45\times0.7)  &  0.65  &  Xf  &  4.16  &  19  &  N  &  MSC (3)  &  0.005\\
3C334  &  B  &  e(0.7\times0.5)  &  0.22  &  Xf  &  1.46  &  14  &  N  &  MSC (3)  &  0.005\\
3C 17  &  S11.3  &  c(0.7\times0.7)  &  0.15  &  Xf+Bnd  &  0.55  &  5  &  Y (a)  &  Amb  (2)  &  0.005\\
0529+075  &  B  &  e(0.5\times0.7)  &  0.56  &  Xf+Bnd  &  4.7  &  8  &  -  &  -  &  0.005\\
3C 179  &  D  &  c(0.5\times0.5)  &  0.16  &  Xf+Bnd  &  1.23  &  4  &  N  &  MSC (3)  &  0.007\\
TXS 0833+585  &  B  &  c(0.4\times0.6)  &  0.18  &  Xf+Bnd  &  1.54  &  8  &  -  &  -  &  0.005\\
PKS 1046-409  &  C  &  e(0.68\times1.0)  &  0.54  &  Xf+Bnd  &  3.66  &  6  &  -  &  -  &  0.009\\
3C 280.1  &  D  &  e(0.85\times0.5)  &  0.54  &  Xf+Bnd  &  4.65  &  5  &  N  &  MSC (3)  &  0.005\\
3C 6.1  &  NHS  &  c(1.25\times1.25)  &  0.19  &  Xf  &  1.49  &  14  &  N  &  MSC (1)  &  0.005\\
3C 179  &  HS  &  c(0.6\times0.6)  &  0.3  &  Xf  &  2.31  &  11  &  N  &  MSC (4)  &  0.005\\
4C+55.17  &  HS  &  e(0.4\times0.6)  &  0.23  &  Xf  &  1.84  &  9  &  N  &  MSC (5)  &  0.006\\
PKS 1311-270  &  HS  &  e(0.57\times0.90)  &  0.24  &  Xf  &  2.02  &  13  &  N  &  MSC (3)  &  0.005\\
PKS 2123-463  &  HS  &  c(0.97\times0.97)  &  0.89  &  Xf  &  7.62  &  7  &  -  &  -  &  0.006\\
3C 6.1  &  SHS  &  c(0.8\times0.8)  &  0.25  &  Xf+Bnd  &  1.96  &  4  &  N  &  MSC (1)  &  0.007\\
Centaurus B  &  A  &  c(1.0\times1.0)  &  0.57  &  Rf  &  0.15  &  8  &  -  &  -  &  0.005\\
PKS 2123-463  &  A  &  c(0.6\times0.6)  &  0.35  &  Rf  &  3  &  10  &  -  &  -  &  0.009\\
4C+11.45  &  C  &  e(0.44\times0.55)  &  0.19  &  Rf+Bnd  &  1.61  &  8  &  N  &  MSC (3)  &  0.005\\
3C 17  &  S3.7  &  c(0.5\times0.5)  &  0.12  &  (T) Xf  &  0.44  &  13  &  Y (a)  &  Amb (2)  &  0.005\\
3C133  &  B  &  c(0.7\times0.7)  &  0.09  &  (T) Xf  &  0.03  &  8  &  Y (b)  &  MSC (3)  &  0.005\\
B3 1428+422  &  A  &  c(1.0\times1.0)  &  0.09  &  (T) Xf  &  0.6  &  7  &  N  &  MSC (6)  &  0.005\\
4C+55.17  &  A  &  c(0.4\times0.4)  &  0.07  &  (T) Xf  &  0.56  &  19  &  -  &  -  &  0.005\\
3C334  &  H  &  e(1.0\times0.7)  &  0.12  &  (T) Xf+Bnd  &  0.8  &  8  &  N  &  MSC (3)  &  0.005\\
TXS 0833+585  &  A  &  c(0.6\times0.6)  &  0.07  &  (T) Rf  &  0.6  &  11  &  -  &  -  &  0.006\\
3C213.1  &  NHS  &  c(0.65\times0.56)  &  0.04  &  Co-s  &  0.133  &  3  &  Y (d)  &  MSC (3)  &  0.006\\
4C+25.21  &  HS  &  e(0.75\times0.36)  &  0.02  &  Co-s  &  0.65  &  4  &  Y  &  MSC(3)  &  0.005\\
3C 220.2  &  SHS  &  e(0.42\times0.35)  &  0.08  &  Unclear  &  0.68  &  3  &  Y (c)  &  MSC (2)  &  0.008\\
\cutinhead{Potential high-redshift IC/CMB jets}
4C+25.21  &  A  &  e (0.5\times0.6)  &  -  &  -  &  -  &  7  &  -  &  -  &  0.007 \\
PKS 2123-463  &  Jet  &  e(1.1\times0.5)  &  -  &  -  &  -  &  5  &  -  &  -  &  0.009 \\
\cutinhead{Non-detections}
PKS 1402+041  &  A  &  e(1.2\times0.65)  &  -  &  -  &  -  &  14  &  -  &  -  &  0.04\\
PKS 1402+041  &  HS  &  c(0.6\times0.6)  &  -  &  -  &  -  &  0  &  -  &  -  &  0.5\\
3C 327.1  &  A  &  r(3.3\times1.6)  &  -  &  -  &  -  &  14  &  -  &  -  &  0.08\\
3C 327.1  &  B  &  e(0.75\times0.5)  &  -  &  -  &  -  &  2  &  -  &  -  &  0.13\\
\cutinhead{New detection}
3C 418 & A & c(0.4\times0.4) &  -  &  -  &  -  &  141  &  -  &  -  &  0.25\\
3C 418 & B & c(0.5\times0.5) &  -  &  -  &  -  &  48  &  -  &  -  &  0.06\\
3C 418 & C & c(0.5\times0.5) &  -  &  -  &  -  &  5  &  -  &  -  &  0.4\\
  \enddata
  \tablenotetext{a}{The size and shape of  the chosen region following the convention used in \citet{2011ApJS..197...24M}. Here ``c" denotes a circle, ``e" an ellipse and ``r" a rectangle.}
  \tablenotetext{b}{The relative position of the X-ray centroid to its radio peak. ``Xf" and ``Rf" denote that the X-ray centroid lies upstream and downstream the radio peak, respectively with offsets $>0.15$\as. Tentative offsets (0.05\as-0.15\as) are indicated with (T). ``Co-s" indicates that they are approximately co-spatial or are within 0.05\arcsec~of each other. The ``+Bnd" modifier indicates that the knot lies at an apparent bend in the jet. }
  \tablenotetext{c}{The lower bound on the offset. See text for more details.}
  \tablenotetext{d}{``Y" indicates an optical or IR detection and ``N" a non-detection, ``-" indicates lack of any HST observations.}
  \tablenotetext{e}{``MSC" indicates that a second spectral component is required to explain the observed X-ray emission. ``Amb" indicates that, within the error limits, the existing data is consistent with both one and two spectral component interpretations. ``-" indicates lack of optical data.}
  \tablerefs{Optical/IR detections: (a) \citet{Massaro_2009}. (b) \citet{2006ApJ...643..660F}. (c) This paper. (d) \citet{2006ApJ...643..660F}. Component spectral type: (1) \citet{2004ApJ...612..729H}. (2) \citet{Massaro_2009}. (3) This paper. (4) \citet{sambruna2002survey}. (5) \citet{tavecchio2007chandra}. (6) \citet{2012ApJ...756L..20C}.}
  \end{deluxetable*}

\subsection{Summary of all 22 sources}
The main motivation for this study was to search for offsets in low-count \textit{Chandra} observations that had not been analyzed previously. We have used the method outlined above to search for offsets in 22 low-count \emph{Chandra} observations of jets. Although our main focus is on offsets in knots of MSC jets, we have also included hotspots and an FR-I source (Centaurus-B) for completeness, and in total 37 individual features were evaluated. The total number of counts in each feature ranged between 2 and 19 with 8.3$\pm$4.7 counts on average. Two knots in the inner jet in 3C 418 were excluded from this average as they were within the psf of the core with a large number of counts.

To claim an offset, we require the displacement between radio and X-ray position to be $>$0.15$''$ given our (conservative) estimate of the registration error between radio and X-ray images and the X-ray positions which are both $\approx0.05\arcsec$.  With this requirement we have found clear offsets in 14 of the 22 jets analyzed, or 19 out of 28 features. Six sources each with one feature have a tentative or possible offset (0.05-0.15\as), and there is no evidence of offsets in one source ($<$0.05\as~separation). Of the clearly detected offsets, 16 are ``X-ray first'' type and 3 are ``radio-first''. Of the tentative cases, 5 out of 6 are ``X-ray first" type and one is ``radio-first''.

The radio and X-ray observations in this study are not simultaneous (time baselines range from 0 to 24 years). Nevertheless we assume a negligible change in jet structure between radio and X-ray observations. Large-scale structural changes in the jet may be caused, for example, by kpc scale proper motions \cite[e.g.,][]{meyer2017proper,2019ApJ...871..248S} or changes in the jet orientation. However for proper motions to produce the observed offsets would require superluminal apparent motions on kpc scales with speeds of hundreds to thousands of times the speed of light, which is clearly unreasonable.
Changes in the jet orientation, for example, by precession, will not affect our results as they happen on much larger time scales \citep[e.g.,][]{2019MNRAS.482..240K}.

\subsection{Optical/IR data and Spectral Analysis}
To understand the spectral nature of the analyzed components, we also analyzed their corresponding HST data. Figures \ref{fig:hst_3C6.1}-\ref{fig:hst_3c334} display HST images overlaid with their respective radio contours. Components of four sources have detections at optical/IR wavelengths. Three are previously known, and we have newly identified IR emission from the southern hotspot of 3C 220.2. Furthermore, except in 3C 17, the rest of the HST detections are hotspots.

 We fit a double-powerlaw phenomenological model  \citep{uchiyama2006shedding} to generate spectral energy distributions (SED). Where necessary, we used upper limits on the optical/IR fluxes for undetected components. For sources where HST data was available, except in the case of 3C 17, all the components required a second spectral component to explain the X-ray emission (see Table \ref{table:results}). For 3C 17, the existing signal to noise ratio (SNR) of the UV data does not permit differentiating between one or two spectral components \citep{Massaro_2009}. The right panels of figures \ref{fig:hst_3C6.1}-\ref{fig:hst_3c334} show the SEDs whenever a fit was made.

\subsection{Notes on Individual Sources}
Here we give details on each of the 22 jets analyzed. Each source has a corresponding figure showing our analysis results, analogous to Figure \ref{fig:results_3C418} for 3C 418. In particular, except for 3C 6.1, we show in panel (a) the sub-pixel Chandra image of the jet, with radio contours overlaid in cyan. The dashed-green regions indicate the regions used to test the significance of emission and measure an offset unless otherwise stated. Solid-green circles indicate the region used to extract the spectrum of the core. In panel (b), we show the average of 2000 MCMC images from LIRA, with radio contours overlaid in black. A yellow \texttt{X} indicates the centroid for each region that is measured on the average added component. The rest of the images show the posterior density of $\xi$ for each analyzed region. Because of the large size of 3C 6.1, we show a zoomed-in version of its hotspots in (c) followed by the plots of posterior densities. In what follows, unless otherwise stated, the offsets are values that are projected onto the plane of the sky.

 Table~\ref{table:results} summarizes the results of our analysis, with the name of the source in column 1, name of the analyzed feature in column 2, size of the region in column 3, offset (arcseconds) in column 4, type of the offset in column 5, sky-projected offset (kpc) in column 6, optical/IR detection status in column 7, the spectral type of the component in column 8 and the \textit{p}-value in column 9. In column 6, ``Xf" and ``Rf" denote that the X-ray peaks closer and farther away from the core by more than 0.15\as, respectively, compared to the radio. Tentative offsets (0.05\as-0.15\as) are indicated with (T). ``Co-s" indicates a feature where the X-ray centroid and radio peak are within 0.05\as~of each other; ``Bnd" indicates that the feature lies at an apparent bend in the jet.

\FloatBarrier
\begin{figure*}[h]
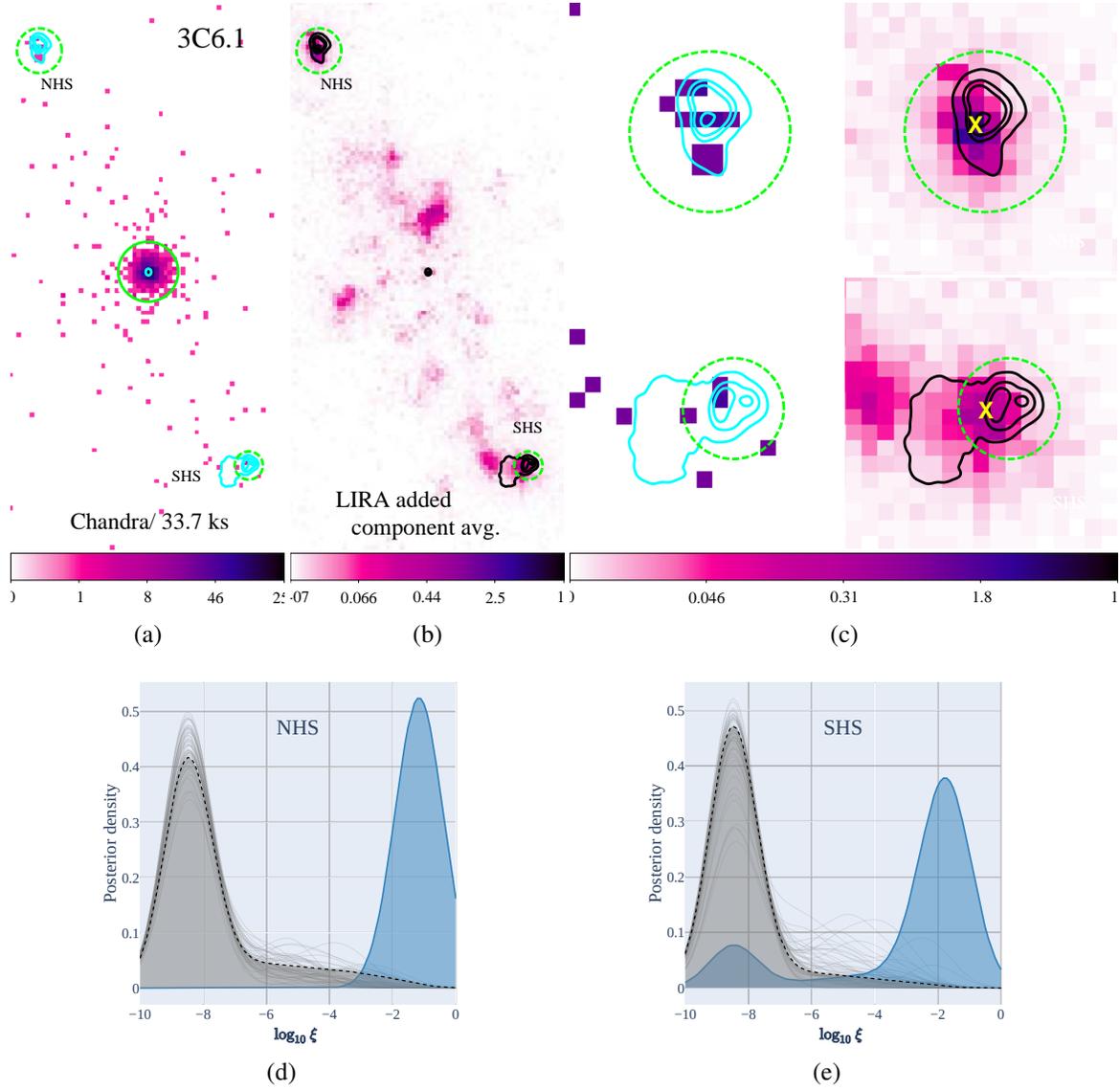

    \gridline{
        \fig{{3C6.1_lira_paper_count}.eps}{0.25\textwidth}{(a)}
        \fig{{3C6.1_lira_paper_add}.eps}{0.25\textwidth}{(b)}
        \fig{{3C6.1_lira_paper_offset}.eps}{0.5\textwidth}{(c)}
    }
    \gridline{
    \fig{{3C6.1_NHS_lira_post}.pdf}{0.33\textwidth}{(d)}
     \fig{{3C6.1_SHS_lira_post}.pdf}{0.33\textwidth}{(e)}
    }
\caption{Results for the analysis of 3C 6.1, an FR-II NLRG--with the VLA 8.4 GHz radio contours overlaid in the top panel--(a) The \textit{Chandra} observation, binned by a factor of 0.5. (b) The average added component. (c) Same as (a) and (b) but zoomed in to NHS (top) and SHS (bottom) respectively. The Red \texttt{X} marks the centroid of the added component within each ROI (dashed-green). The X-rays peak closer to the core than the radio in NHS. The jet appears to make a 90 deg turn to the west before reaching SHS where the X-rays peak upstream of the radio
. (d)-(e) The distributions of $\xi$ in each ROI  for the observed image (shaded blue), 50 simulated images of the baseline model (gray) and their average (gray shaded area with a dashed-black curve). \Radiocontours{1.5, 8, 15, 40} 
\label{fig:results_3C6.1}}
\end{figure*} 
\textbf{3C 6.1} (Fig. \ref{fig:results_3C6.1}) -- This is an FR-II NLRG source with two X-ray hotspots. We find offsets of 0.19\as~(1.46 kpc) in the northern hotspot (NHS) and 0.25\as~(1.66 kpc) in the southern hotspot (SHS) with X-rays lying upstream of the radio.

\begin{figure*}[h]
    \gridline{
        \fig{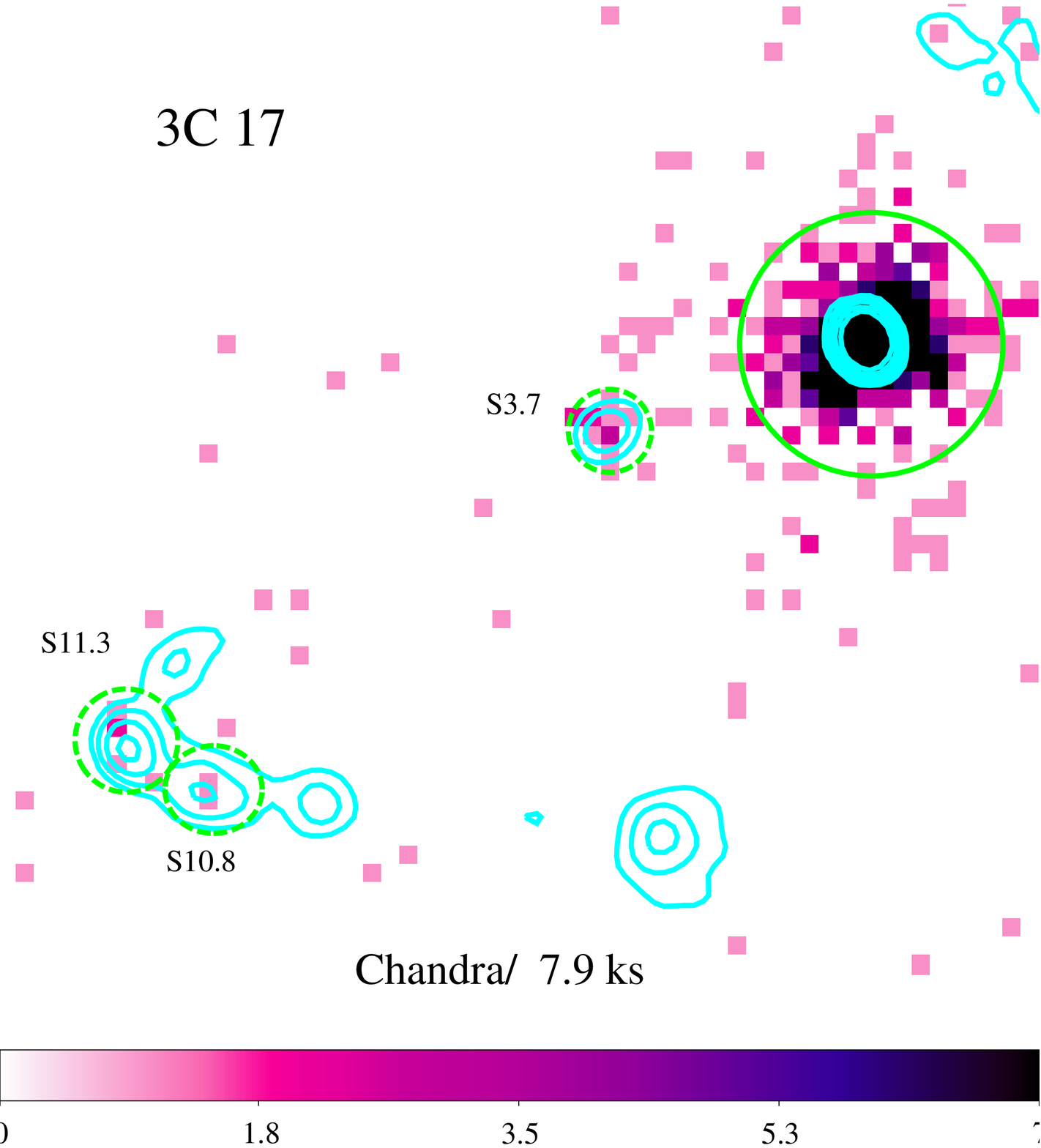}{0.5\textwidth}{(a)}
        \fig{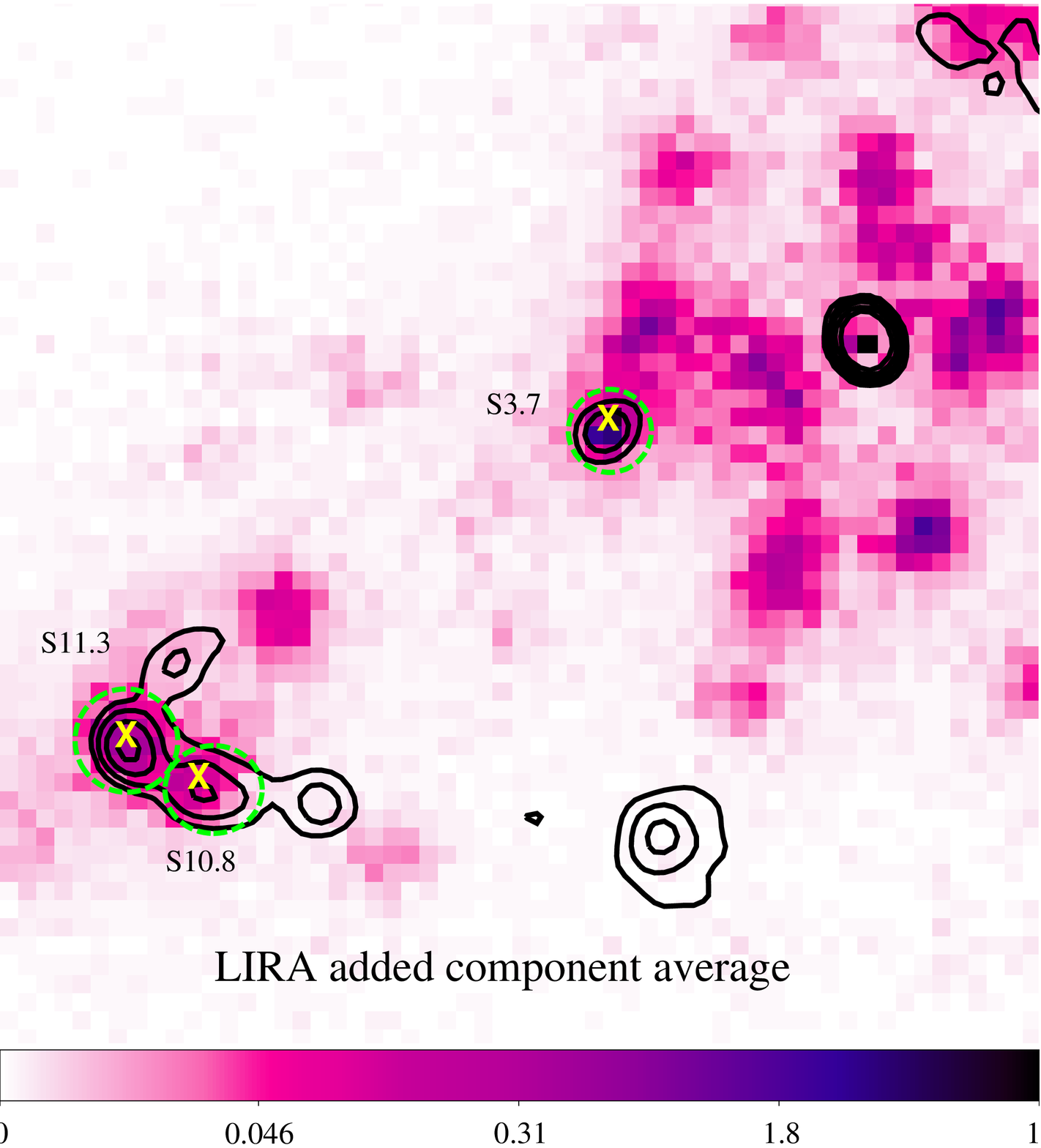}{0.5\textwidth}{(b)}
    }
    \gridline{
    \fig{{3C17_S3.7_lira_post}.pdf}{0.33\textwidth}{(c)}
     \fig{{3C17_S11.3_lira_post}.pdf}{0.33\textwidth}{(d)}
     \fig{{3C17_S10.8_lira_post}.pdf}{0.33\textwidth}{(e)}
    }
\caption{Same as in Fig. \ref{fig:results_3C418}, for 3C 17, an FR-II BLRG.}

\end{figure*} 
\textbf{3C 17} (Fig. \ref{fig:results_3C17}): This is an FR-II BLRG source. We adopt the knot names from \citet{Massaro_2009}. The X-ray centroid and radio peak lie within 0.15\as~of each other in knot S3.7. Knots S11.3 and S10.8 show Xf-type projected offsets of 0.15\as~(0.55 kpc) and 0.18\as~(0.6 kpc) respectively. The jet bending at S11.3 presumably produces a internal shock where a strong shock at the bend produces X-rays while a weaker shock downstream produces radio \citep[e.g.,][]{2005MNRAS.360..926W}. \citet{2018ApJS..238...31M} suggest that a dense intracluster medium might be responsible for bending the jet.

\begin{figure*}[h]
    \gridline{
        \fig{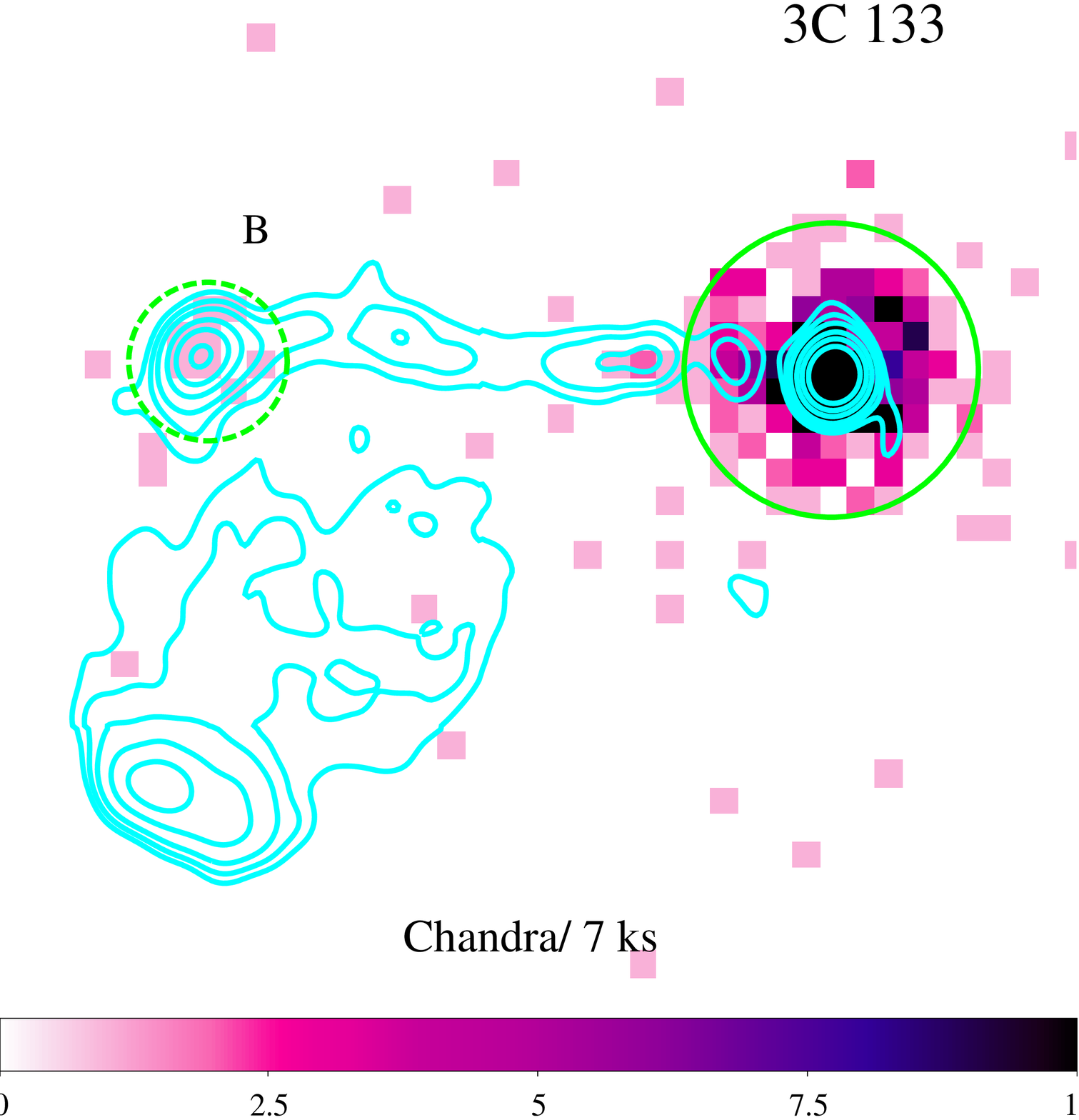}{0.5\textwidth}{(a)}
        \fig{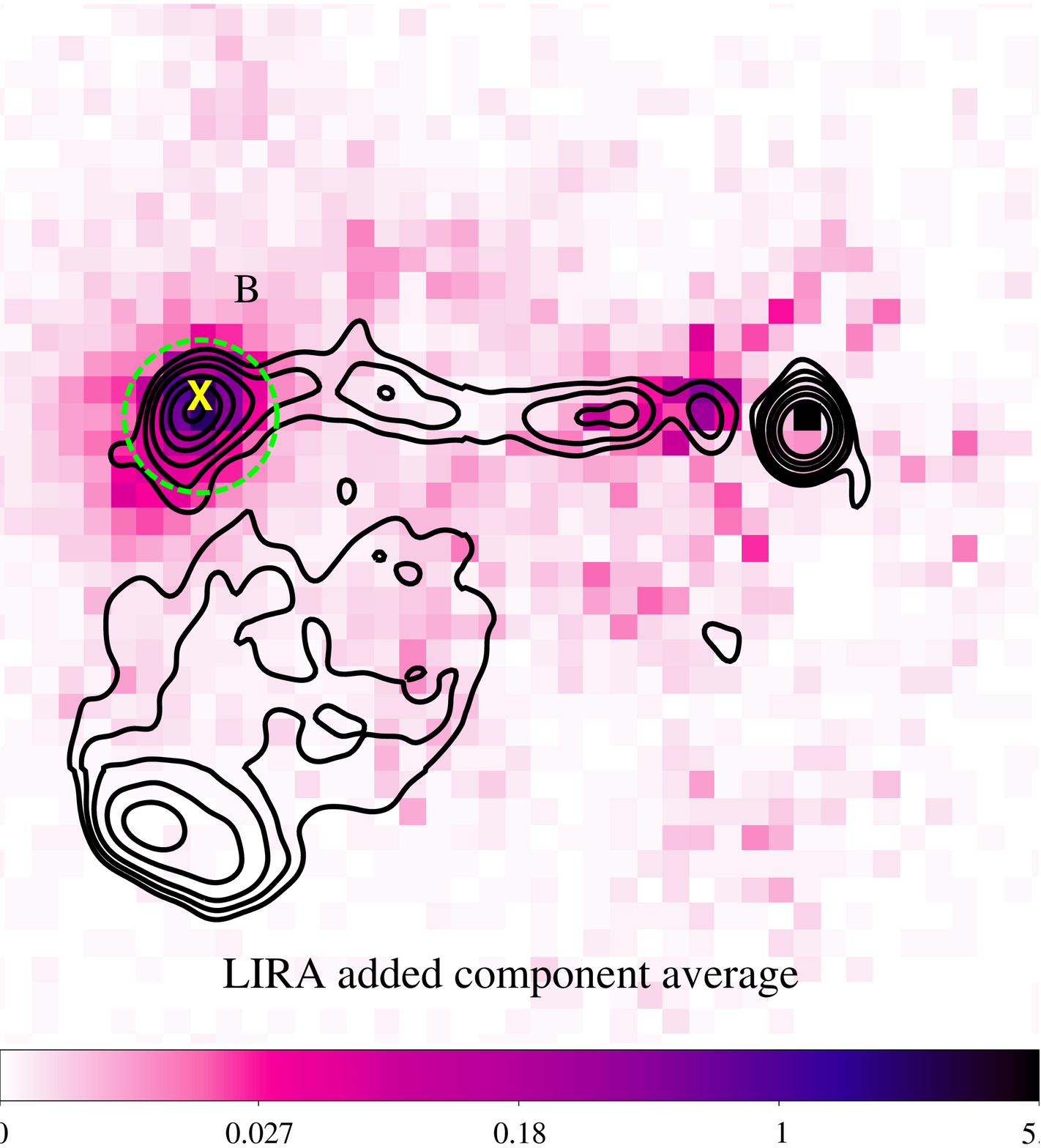}{0.5\textwidth}{(b)}
    }
    \gridline{
    \fig{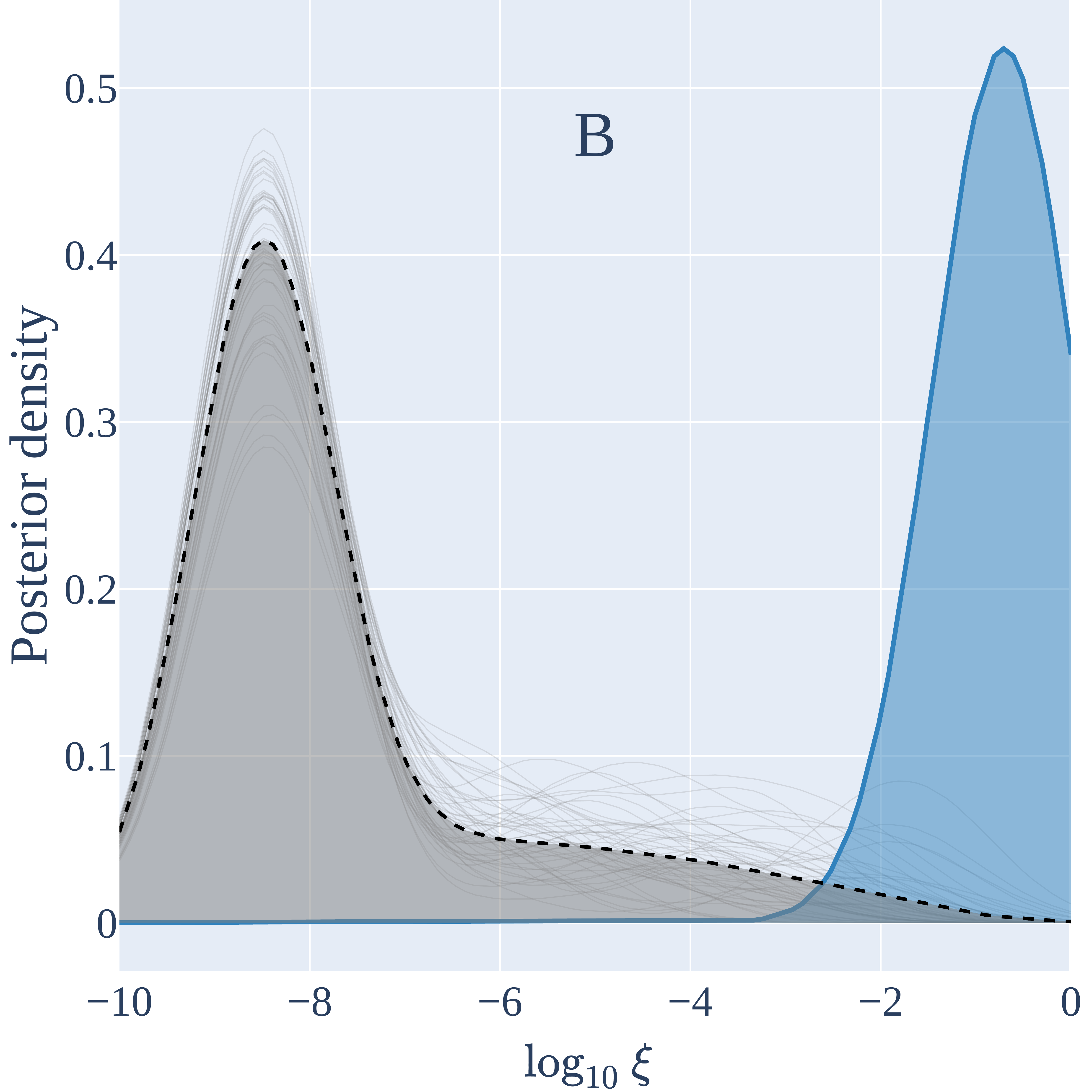}{0.33\textwidth}{(c)}
    }
\caption{Same as in Fig. \ref{fig:results_3C418}, with the VLA 4.8 GHz radio contours overlaid, for 3C 133, a HERG. \Radiocontours{ 2, 4, 8, 10, 20, 40, 80}\label{ref:results_3C133}}
\end{figure*} 
\textbf{3C 133} (Fig. \ref{ref:results_3C133}): This is an FR-II HERG source. The X-ray centroid in Knot A, which lies at a 90\degree~projected apparent bend, lies within in 0.15\as~of its radio peak. Despite the X-ray emission from the knots in the inner-jet, we excluded them from the offset analysis as their separation was less than two ASCI-S pixels.

\begin{figure*}[h]
    \gridline{
        \fig{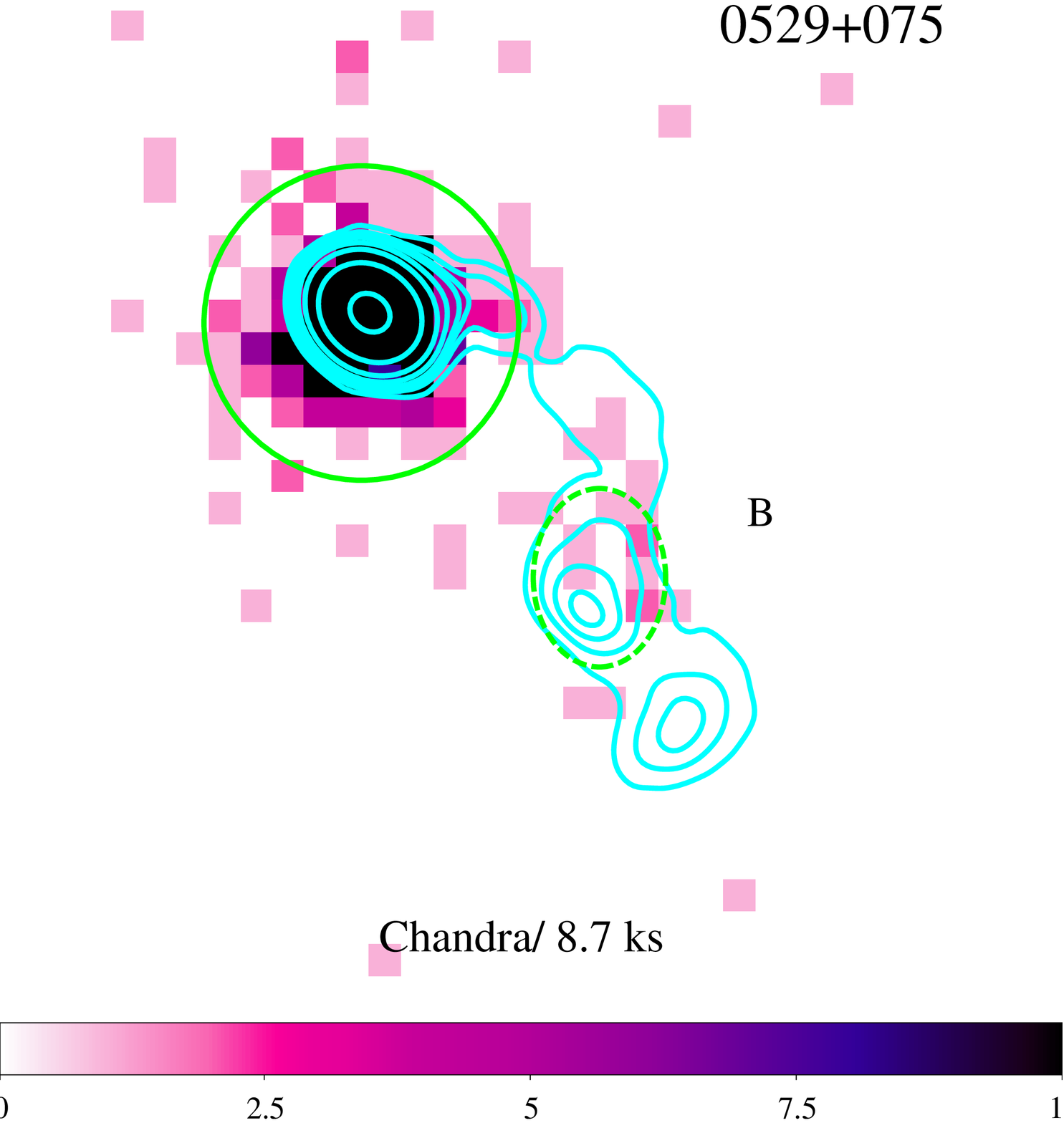}{0.5\textwidth}{(a)}
        \fig{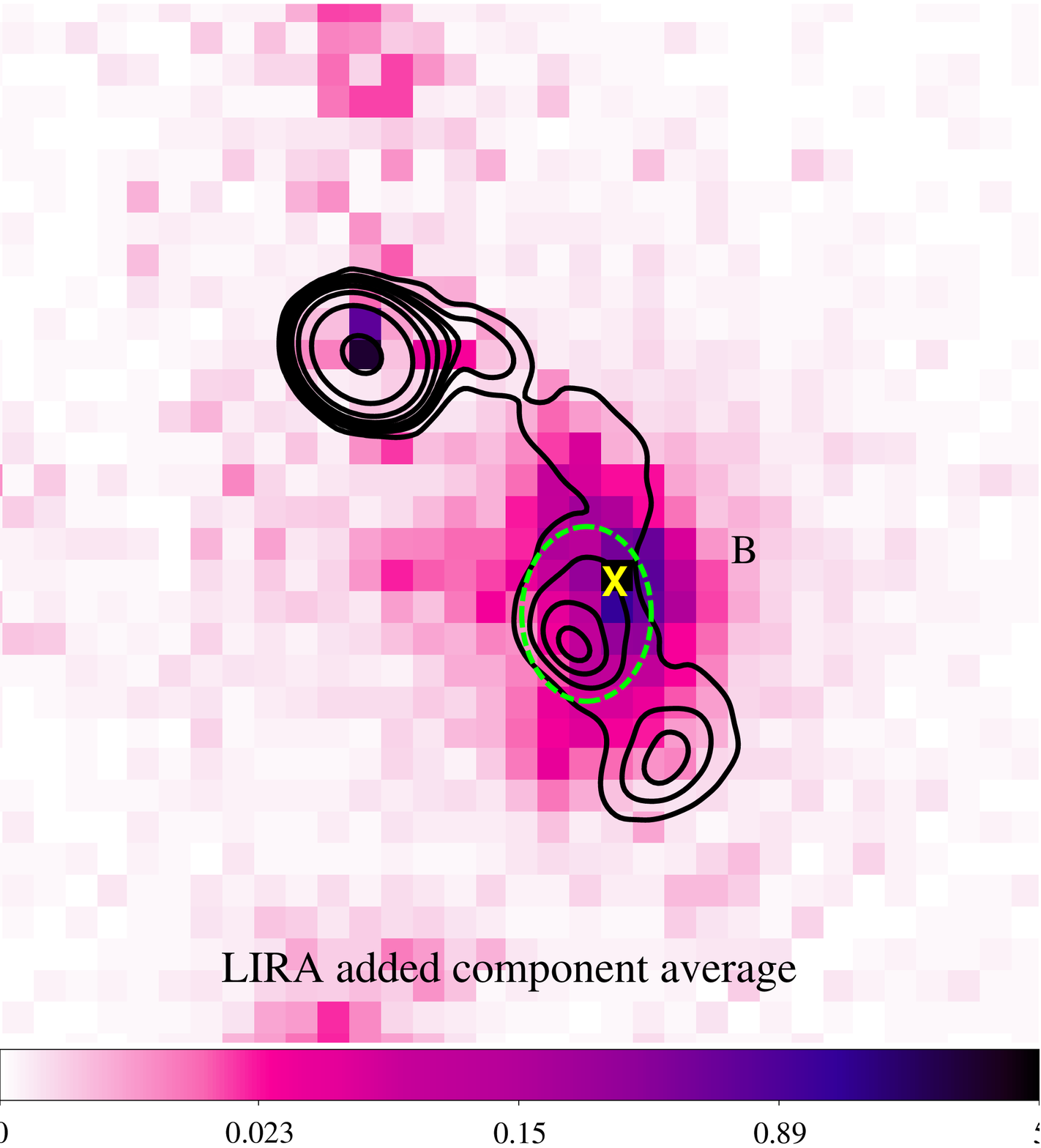}{0.5\textwidth}{(b)}
    }
    \gridline{
    \fig{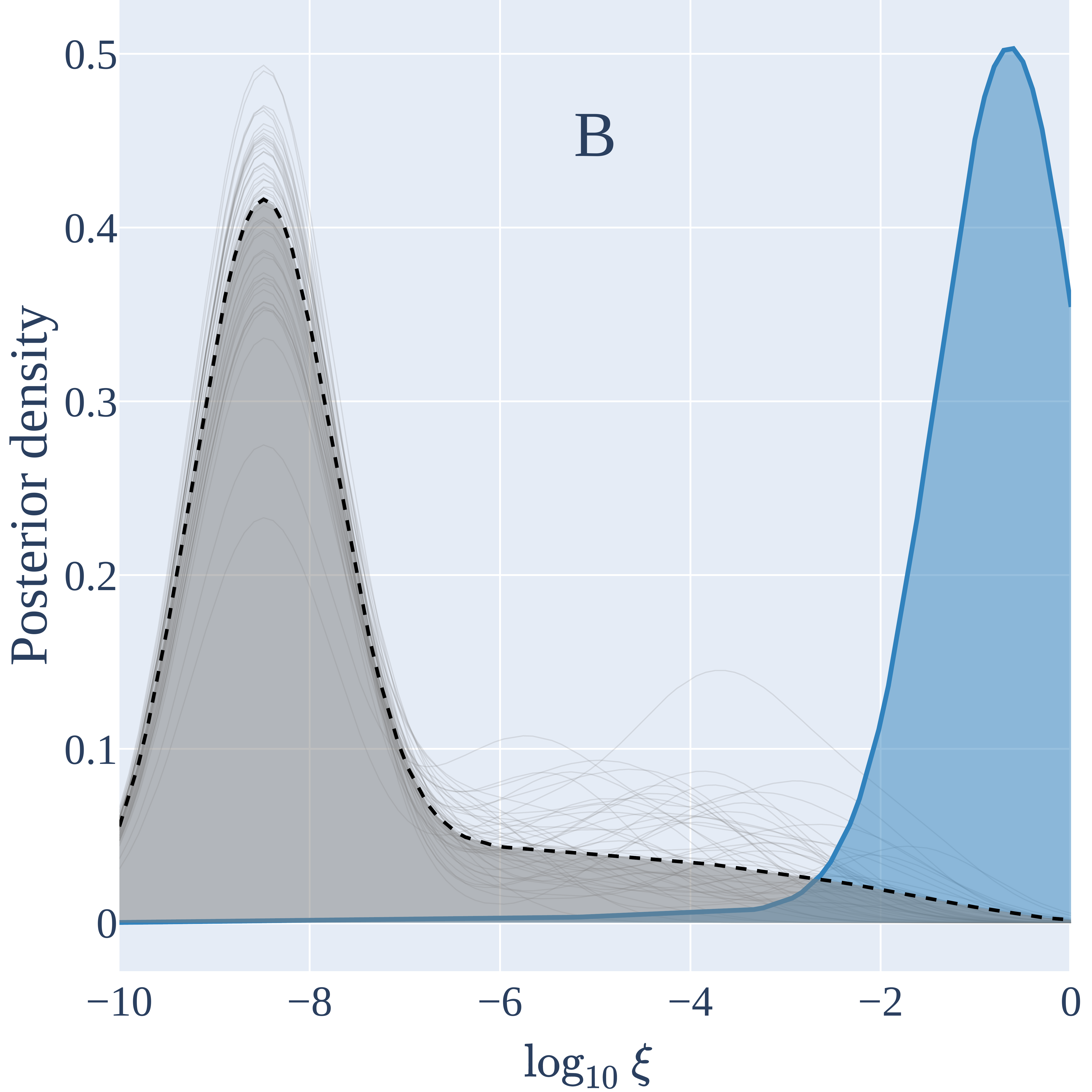}{0.33\textwidth}{(c)}
    }
\caption{Same as in Fig. \ref{fig:results_3C418}, with the VLA 4.8 GHz radio contours overlaid,, for 0529+075, a highly-superluminal CDQ.  \Radiocontours{1, 2, 3, 4, 8, 20, 100, 1000} \label{fig:results_0529+075}}
\end{figure*} 
\textbf{0529+075} (Fig. \ref{fig:results_0529+075}): This is a core-dominated quasar. The inner jet makes a sharp apparent bend to the south and later turns to the southeast at knot B where it shows an Xf-offset of 0.56\as~(4.7 kpc).

\begin{figure*}[h]
    \gridline{
        \fig{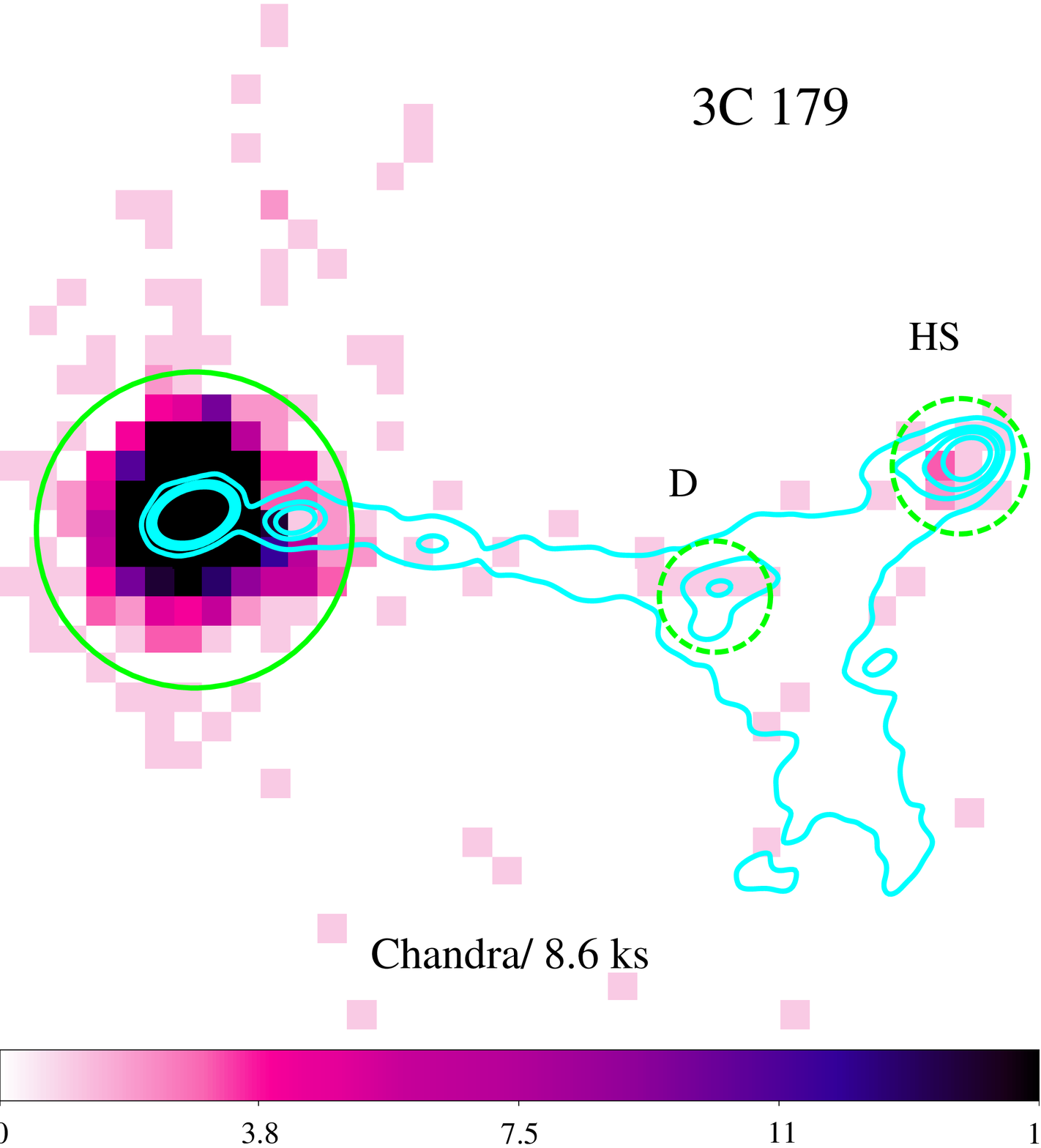}{0.5\textwidth}{(a)}
        \fig{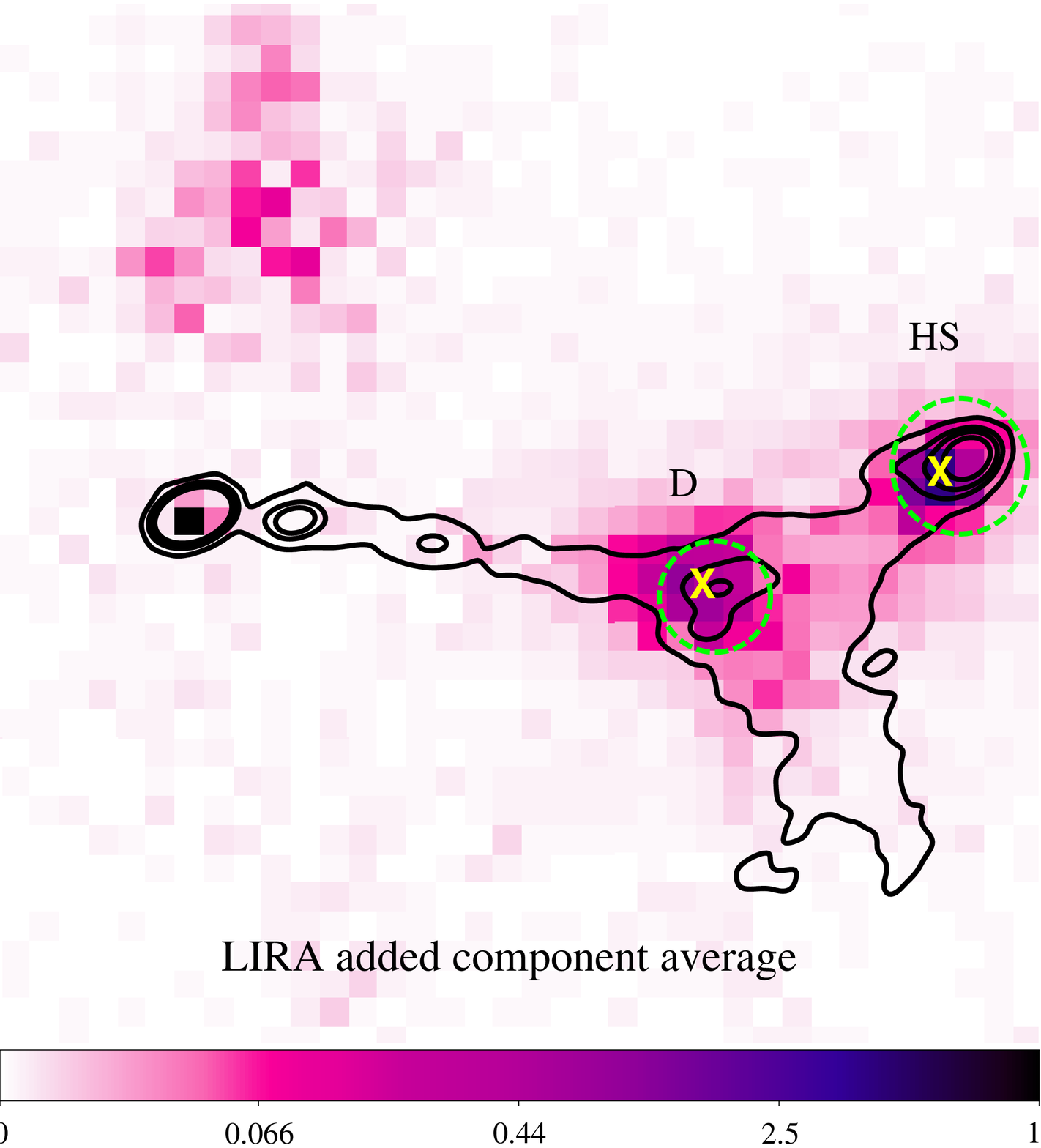}{0.5\textwidth}{(b)}
    }
    \gridline{
    \fig{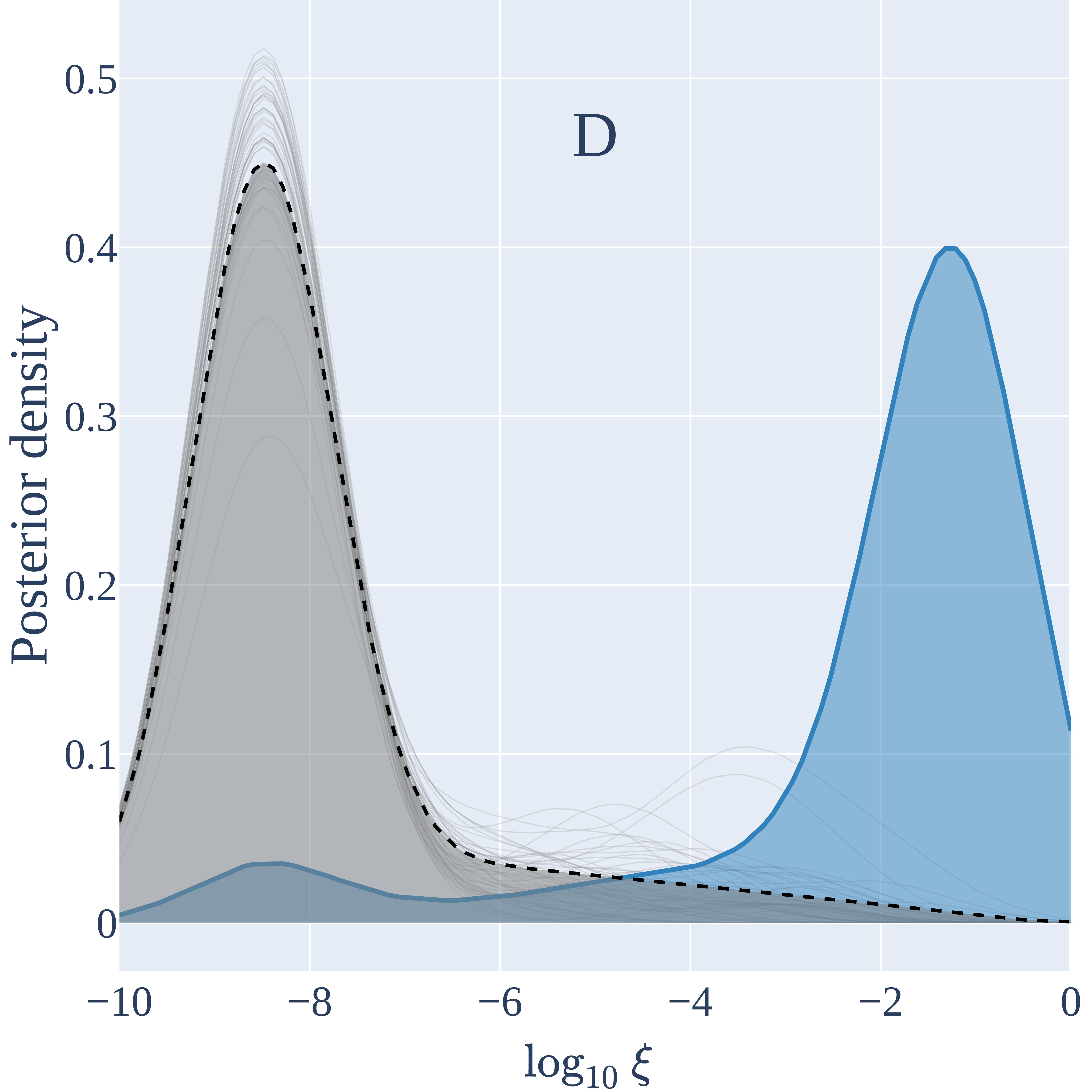}{0.33\textwidth}{(c)}
     \fig{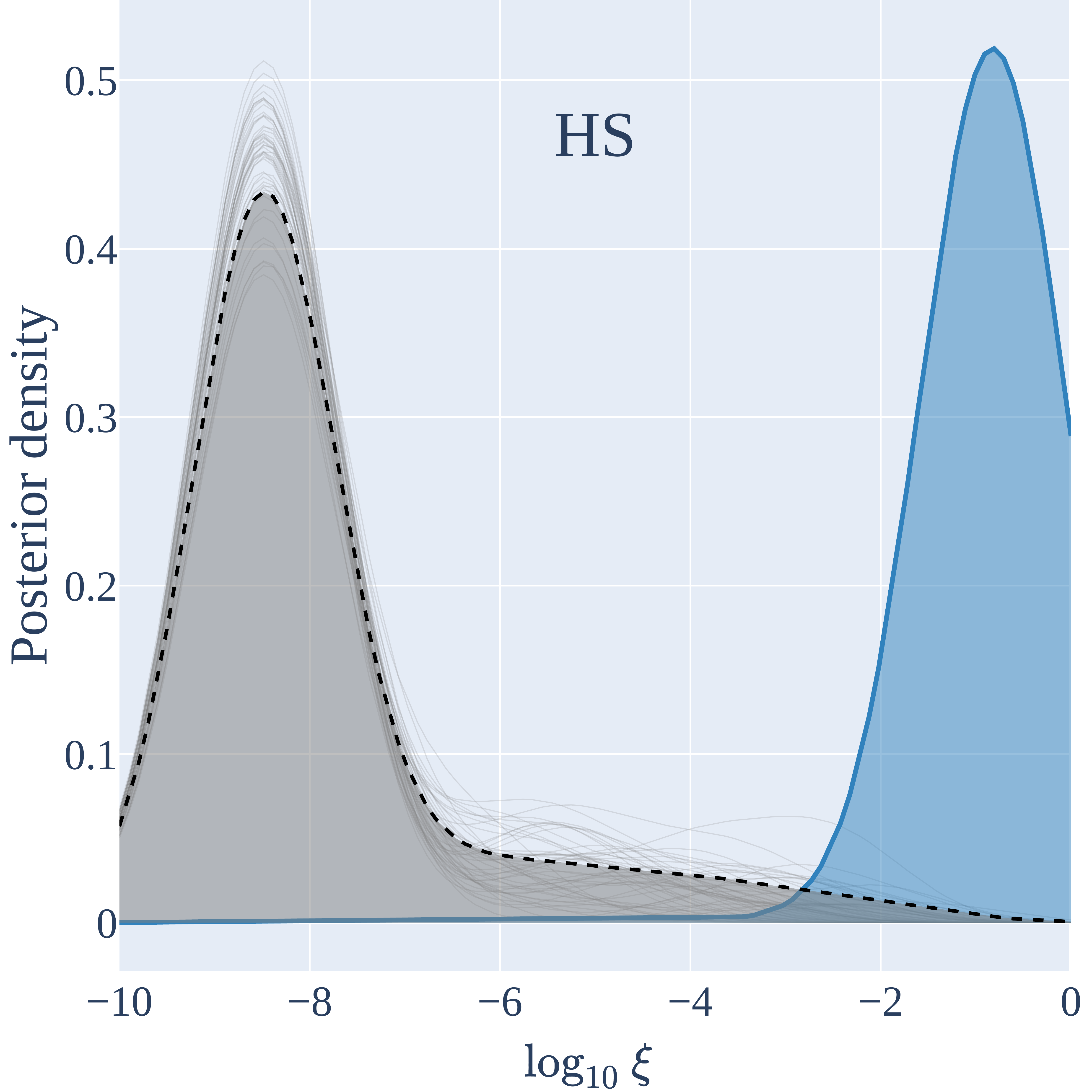}{0.33\textwidth}{(d)}
    }
    
\caption{Same as in Fig. \ref{fig:results_3C418}, with the VLA 4.8 GHz radio contours overlaid,, for 3C 179, an LDQ. \Radiocontours{ 0.3, 2, 4, 8}
\label{fig:results_3C179}}
\end{figure*} 
\textbf{3C 179} (Fig. \ref{fig:results_3C179}): This is a lobe-dominated quasar. The jet branches out into two directions--northwest and southwest--at knot D, where the X-ray centroid lies upstream of the radio peak by 0.16\as~(1.23 kpc). The northwestern branch terminates in a hotspot (HS) while the southwestern one diffuses away. \citet{1992A&A...259L..61A} suggest that a dense inter-galactic medium at D may be responsible for this splashing. The hotspot also shows an Xf-offset of 0.3\as~(2.31 kpc). 

\begin{figure*}[h]
    \gridline{
        \fig{{4C+25.21_lira_count}.eps}{0.5\textwidth}{(a)}
        \fig{{4C+25.21_lira_add}.eps}{0.5\textwidth}{(b)}
    }
    \gridline{
    \fig{{4C+25.21_B_post}.pdf}{0.33\textwidth}{(c)}
     \fig{{4C+25.21_C_post}.pdf}{0.33\textwidth}{(d)}
    }
    
\caption{Same as in Fig. \ref{fig:results_3C418}, for 4C+25.21, a high redshift quasar. \Radiocontours{0.5, 2, 4, 8, 20}\label{fig:results_4C+25.21}}
\end{figure*}
\textbf{4C+25.21} (Fig. \ref{fig:results_4C+25.21}): This is a high-redshift quasar (z=2.68). The X-ray centroid lies within 0.15\as~ of the radio peak in the hotspot. We also detect significant X-ray emission along a putative direction of the jet between the inner jet and the hotspot. This morphology suggests that IC/CMB presumably produces these X-rays (see section \ref{subsec:flux_ratio} for details).

\begin{figure*}[h]
    \gridline{
        \fig{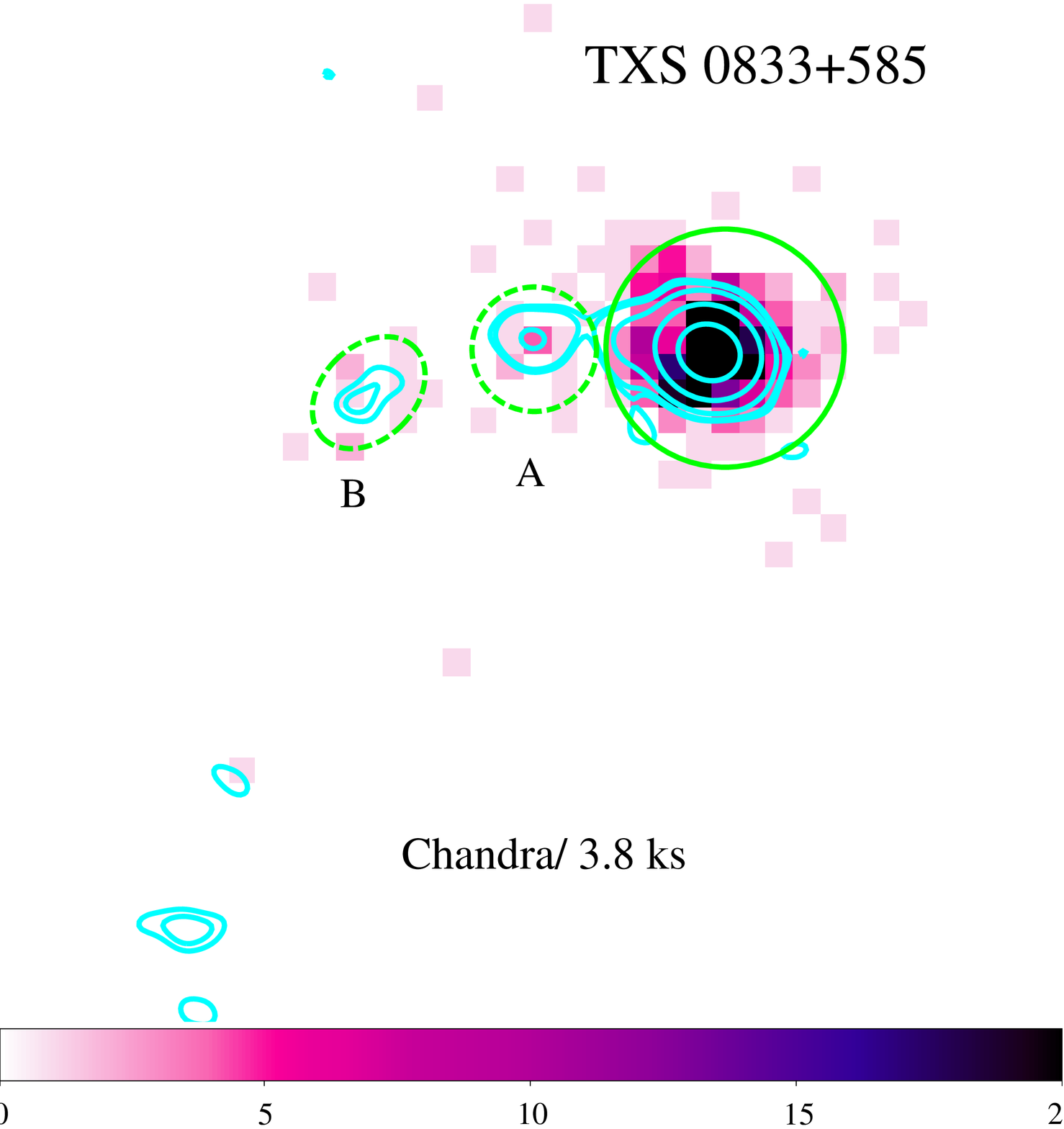}{0.5\textwidth}{(a)}
        \fig{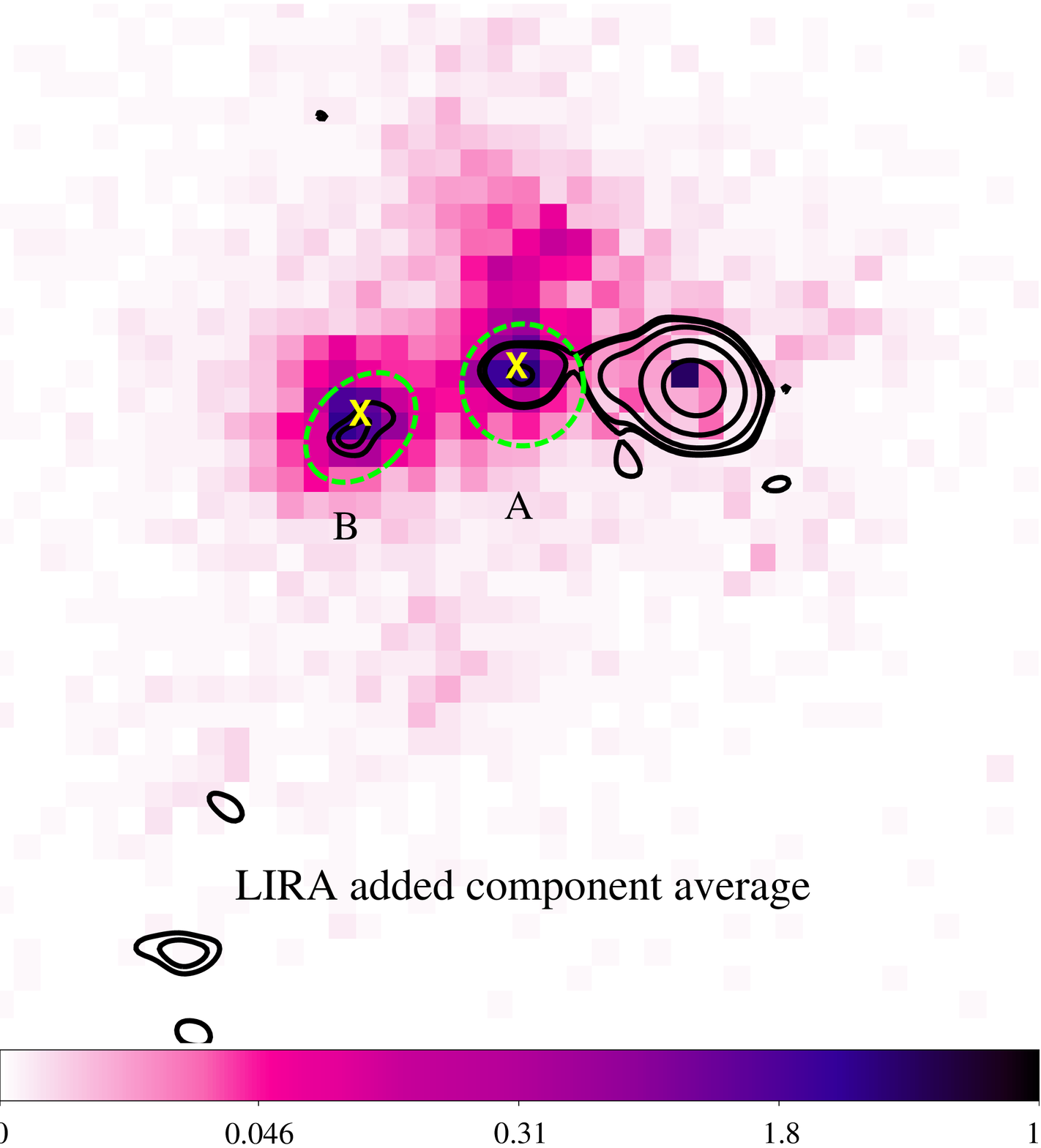}{0.5\textwidth}{(b)}
    }
    \gridline{
    \fig{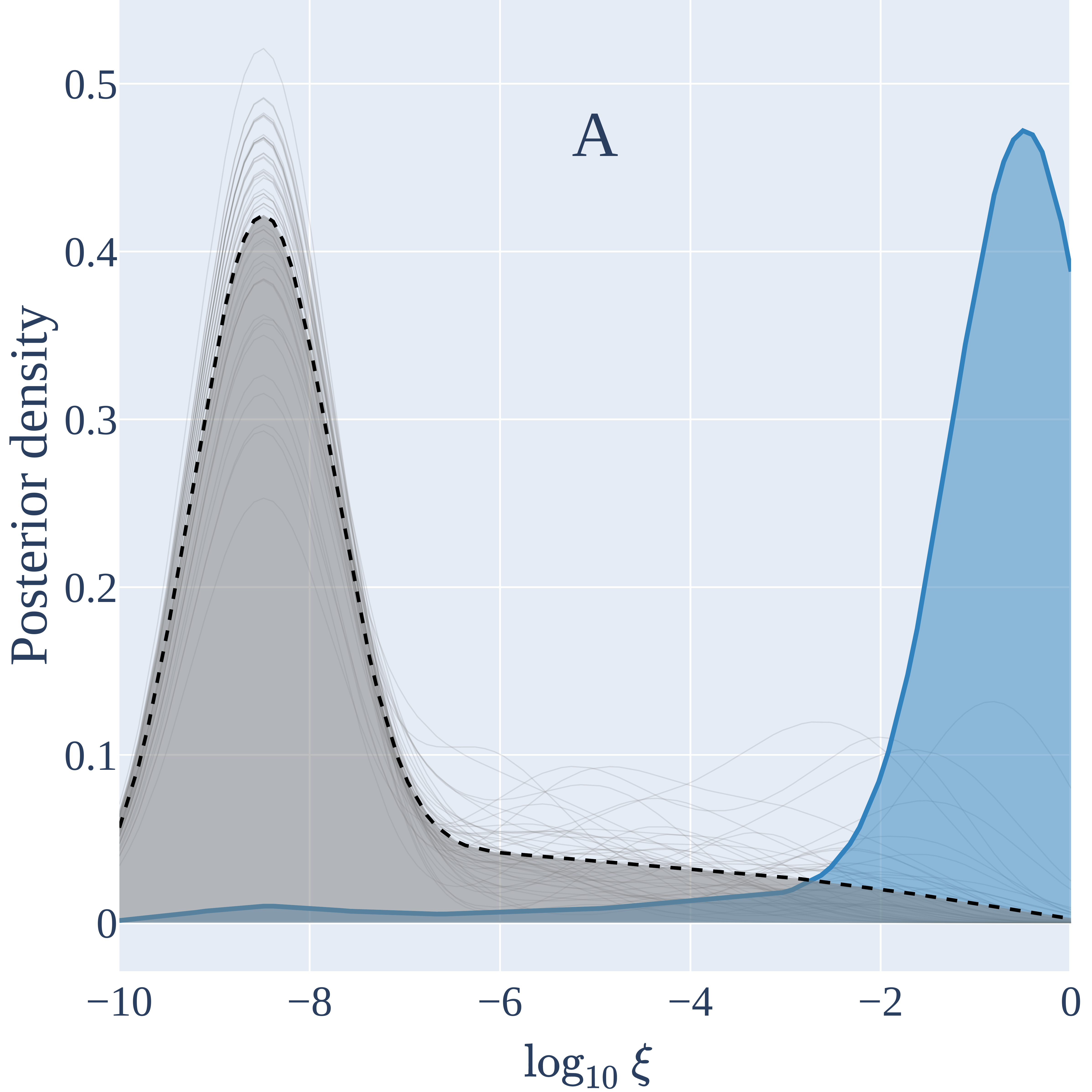}{0.33\textwidth}{(c)}
     \fig{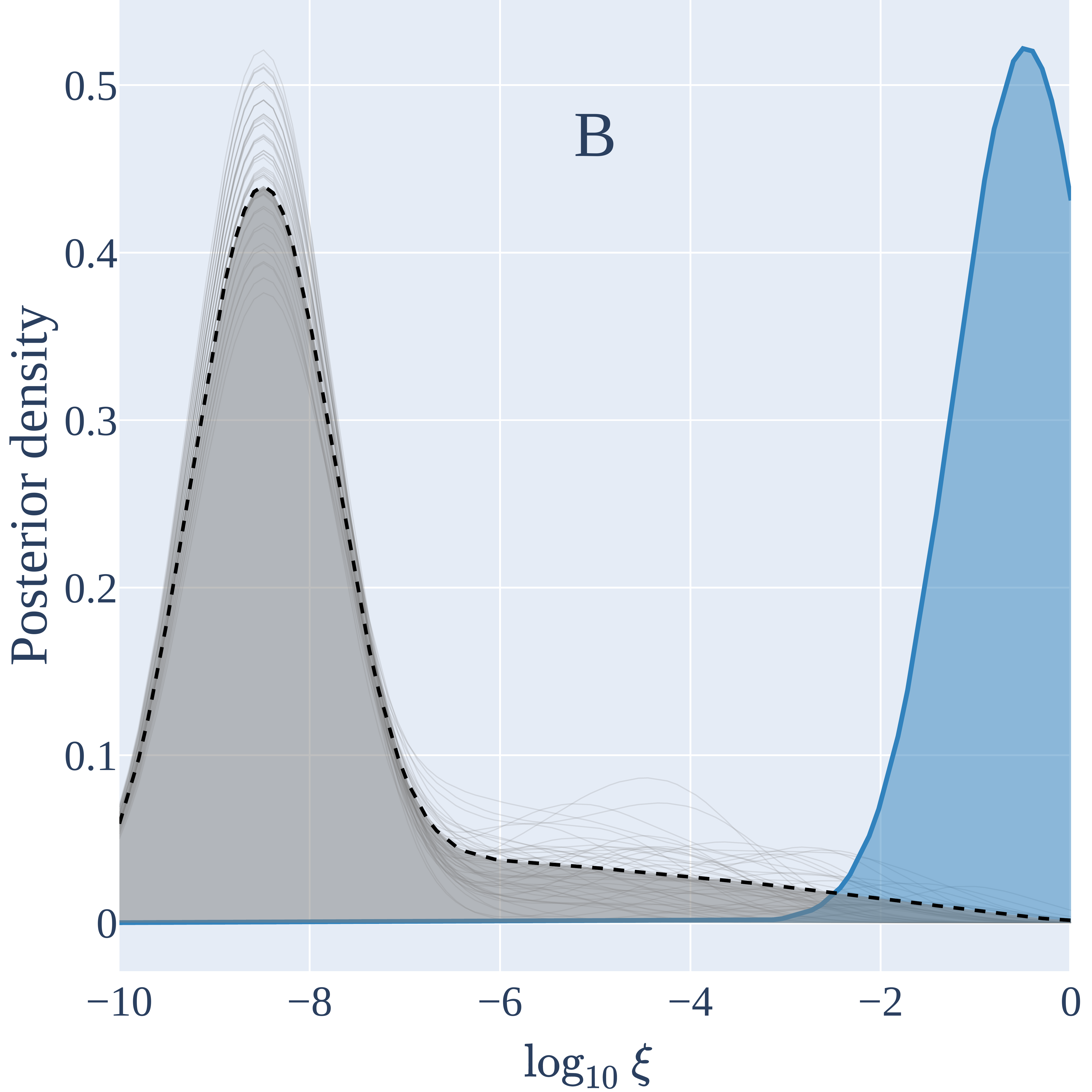}{0.33\textwidth}{(d)}
    }
    
\caption{Same as in Fig. \ref{fig:results_3C418}, with the VLA 4.8 GHz radio contours overlaid, for TXS 0833+585, a high-redshift quasar. \Radiocontours{0.34, 0.5, 2, 20, 200}\label{fig:results_TXS0833+575}}
\end{figure*}
\textbf{TXS 0833+585} (Fig. \ref{fig:results_TXS0833+575}): This is a hight redshift quasar (z=2.1). The X-ray centroid and the radio peak approximately coincide in knot A. In contrast, it lies upstream of the radio in knot B where the jet makes a 90\degree~projected turn to the south. The flux ratio increases with distance from the core and may indicate that IC/CMB dominates the X-ray emission mechanism (see section \ref{subsec:flux_ratio} for details).

\begin{figure*}[!]
    \gridline{
        \fig{{3C213.1_lira_paper_count}.eps}{0.5\textwidth}{(a)}
        \fig{{3C213.1_lira_paper_add}.eps}{0.5\textwidth}{(b)}
    }
    \gridline{
    \fig{{3C213.1_NHS_lira_post}.pdf}{0.33\textwidth}{(c)}
    }
    
\caption{Same as in Fig. \ref{fig:results_3C418}, for 3C 213.1, an LERG. \Radiocontours{ 0.5, 2, 8, 40}\label{fig:results_3C213.1}}
\end{figure*}
\textbf{3C 213.1} (Fig. \ref{fig:results_3C213.1}): This an FR-II source. NHS is detected in radio, optical and X-rays. The X-ray centroid in NHS is within 0.05\as~(0.17 kpc) of the radio peak.

\begin{figure*}[h]
    \gridline{
        \fig{{3C220.2_lira_paper_count}.eps}{0.5\textwidth}{(a)}
        \fig{{3C220.2_lira_paper_add}.eps}{0.5\textwidth}{(b)}
    }
    \gridline{
    \fig{{3220.2_shs_lira_post}.pdf}{0.33\textwidth}{(c)}
    }
    
\caption{Same as in Fig. \ref{fig:results_3C418}, for 3C 220.2, a quasar. \Radiocontours{0.5, 2, 5, 20, 80} \label{fig:results_3C220.2}}
\end{figure*} 
\textbf{3C 220.2} (Fig. \ref{fig:results_3C220.2}): This is a quasar. The X-ray centroid in SHS is within 0.15\as~(1.28 kpc) of its radio peak. However, unlike most of analyzed jet-features, the position angle of the tentative offset lies perpendicular to the putative direction of the jet. This morphology suggests that hot gases on kpc-scales may be emitting the X-rays rather than the hotspot itself.

\begin{figure*}[h]
    \gridline{
        \fig{{4C+55.17_lira_paper_count}.eps}{0.5\textwidth}{(a)}
        \fig{{4C+55.17_lira_paper_add}.eps}{0.5\textwidth}{(b)}
    }
    \gridline{
    \fig{{4C+55.17_A_lira_post}.pdf}{0.33\textwidth}{(c)}
     \fig{{4C+55.17_HS_lira_post}.pdf}{0.33\textwidth}{(d)}
    }
    
\caption{Same as in Fig. \ref{fig:results_3C418}, for 4C+55.17, a quasar/FSRQ. \Radiocontours{ 2, 4, 8, 40, 100, 1000} \label{fig:results_4C+55.17}
}
\end{figure*}
\textbf{4C+55.17} (Fig. \ref{fig:results_4C+55.17}): The classification for this source is ambiguous. It was initially classified as a quasar while \citet{McConville_2011} suggest that it is a CSO. The X-ray centroid in knot A lies within 0.15\as~(0.14 kpc) of its radio peak. On the other hand, the hotspot (HS) shows an Xf-offset of 0.23\as~(1.84 kpc).

\begin{figure*}[h]
    \gridline{
        \fig{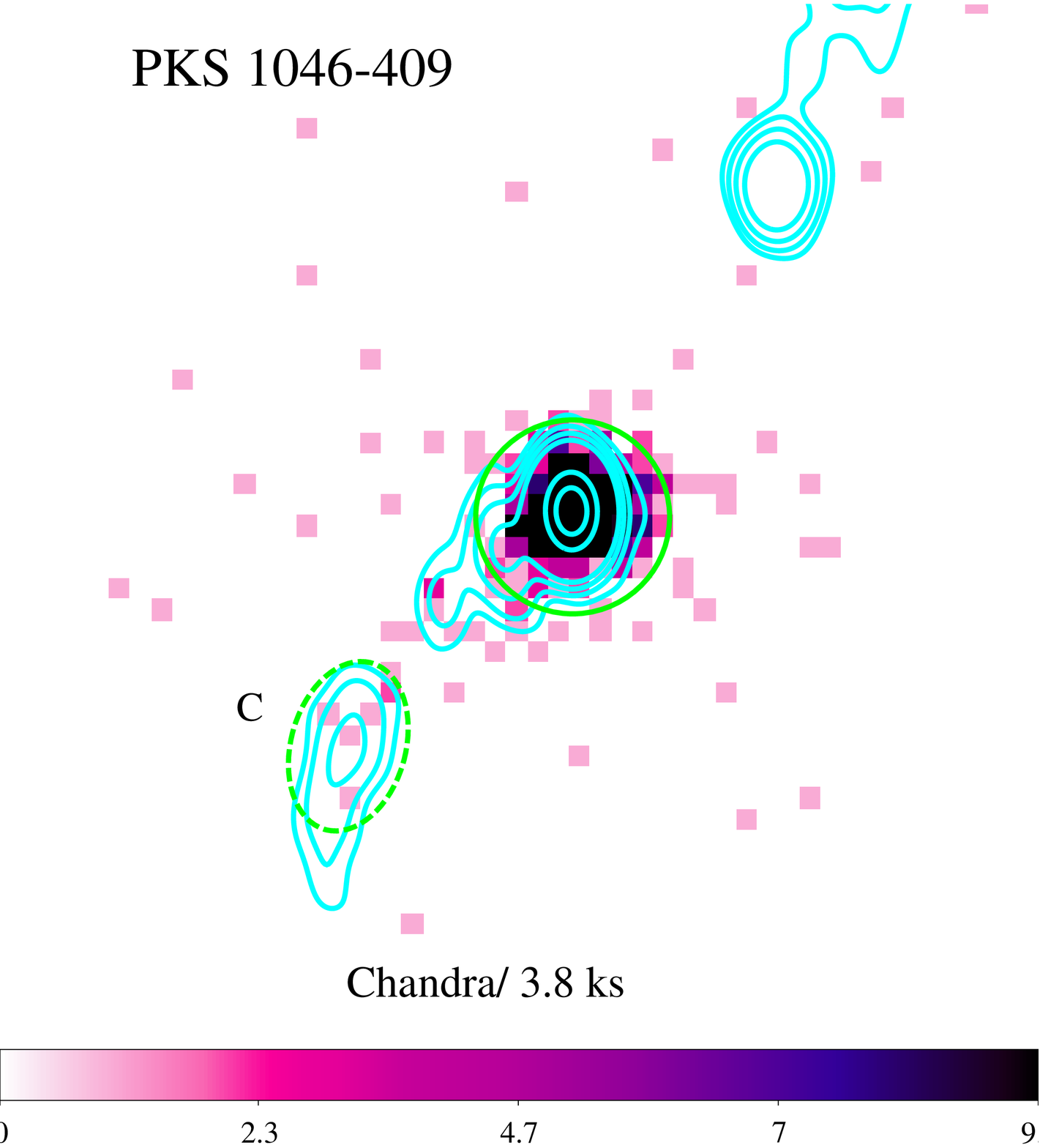}{0.5\textwidth}{(a)}
        \fig{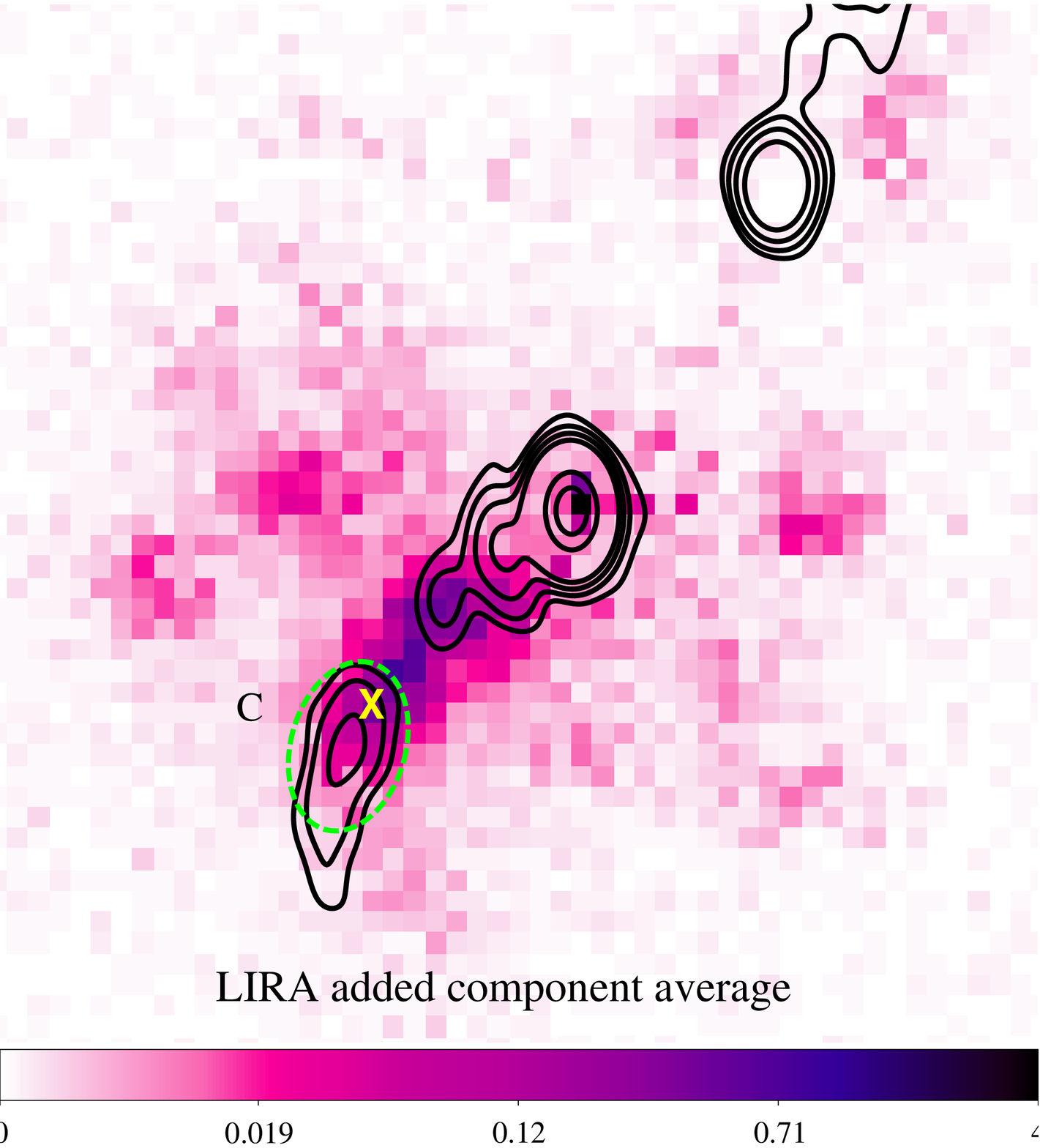}{0.5\textwidth}{(b)}
    }
    \gridline{
    \fig{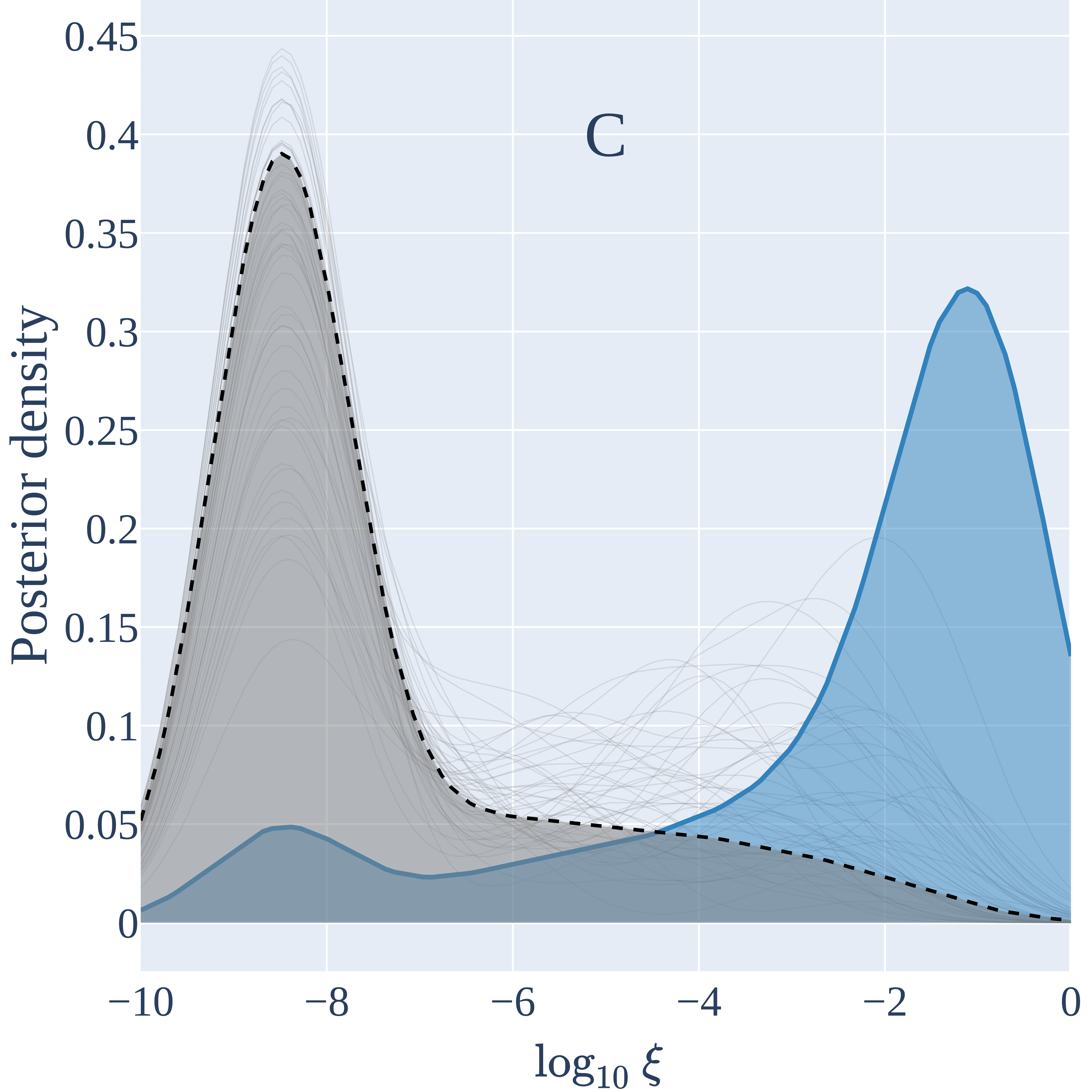}{0.33\textwidth}{(c)}
    }
    
\caption{Same as in Fig. \ref{fig:results_3C418}, with the ATCA 19 GHz radio contours overlaid, for PKS 1046-409, a CDQ. \Radiocontours{ 1, 2, 4, 8, 100, 200}\label{fig:results_PKS1046-409}} 
\end{figure*} 
\textbf{PKS 1046-409} (Fig. \ref{fig:results_PKS1046-409}): This is a core-dominated quasar. The jet makes a projected 20\degree~bend to the south at knot C, where the X-ray centroid lies upstream of the radio peak by 0.54\as~(4.65 kpc).

\begin{figure*}[h]
    \gridline{
        \fig{{3C275.1_lira_paper_count}.eps}{0.5\textwidth}{(a)}
        \fig{{3C275.1_lira_paper_add}.eps}{0.5\textwidth}{(b)}
    }
    \gridline{
    \fig{{3C275.1_A_lira_post}.pdf}{0.33\textwidth}{(c)}
    }
    
\caption{Same as in Fig. \ref{fig:results_3C418}, with the VLA 4.8 GHz radio contours overlaid, for 3C 275.1, an LDQ. \Radiocontours{ 0.5, 1, 2, 5, 70} 
\label{fig:results_3C275.1}}
\end{figure*}
\textbf{3C 275.1} (Fig. \ref{fig:results_3C275.1}): This is a lobe-dominated quasar. We find that knot A shows an Xf-offset offset of 0.65\as~(4.16 kpc).

\begin{figure*}[h]
    \gridline{
        \fig{{3C280.1_lira_paper_count}.eps}{0.5\textwidth}{(a)}
        \fig{{3C280.1_lira_paper_add}.eps}{0.5\textwidth}{(b)}
    }
    \gridline{
    \fig{{3C280.1_D_lira_post}.pdf}{0.33\textwidth}{(c)}
    }
    
\caption{Same as in Fig. \ref{fig:results_3C418}, with the VLA 4.8 GHz radio contours overlaid, for 3C 280.1, an LDQ. \Radiocontours{ 1, 4, 8, 20}\label{fig:results_3C280.1} 
}
\end{figure*}
\textbf{3C 280.1} (Fig. \ref{fig:results_3C280.1}): This is a lobe-dominated quasar. The jet makes a 30\degree~bend to the south at knot D. This knot shows an Xf-offset of 0.54\as~(4.65 kpc).  A bend in the jet at knot D may produce an internal shock and the onset of the bend could create a strong-shock which can produce X-rays while a weaker shock downstream can produce radio \citep[e.g.,][]{2005MNRAS.360..926W}.

\begin{figure*}[h]
    \gridline{
        \fig{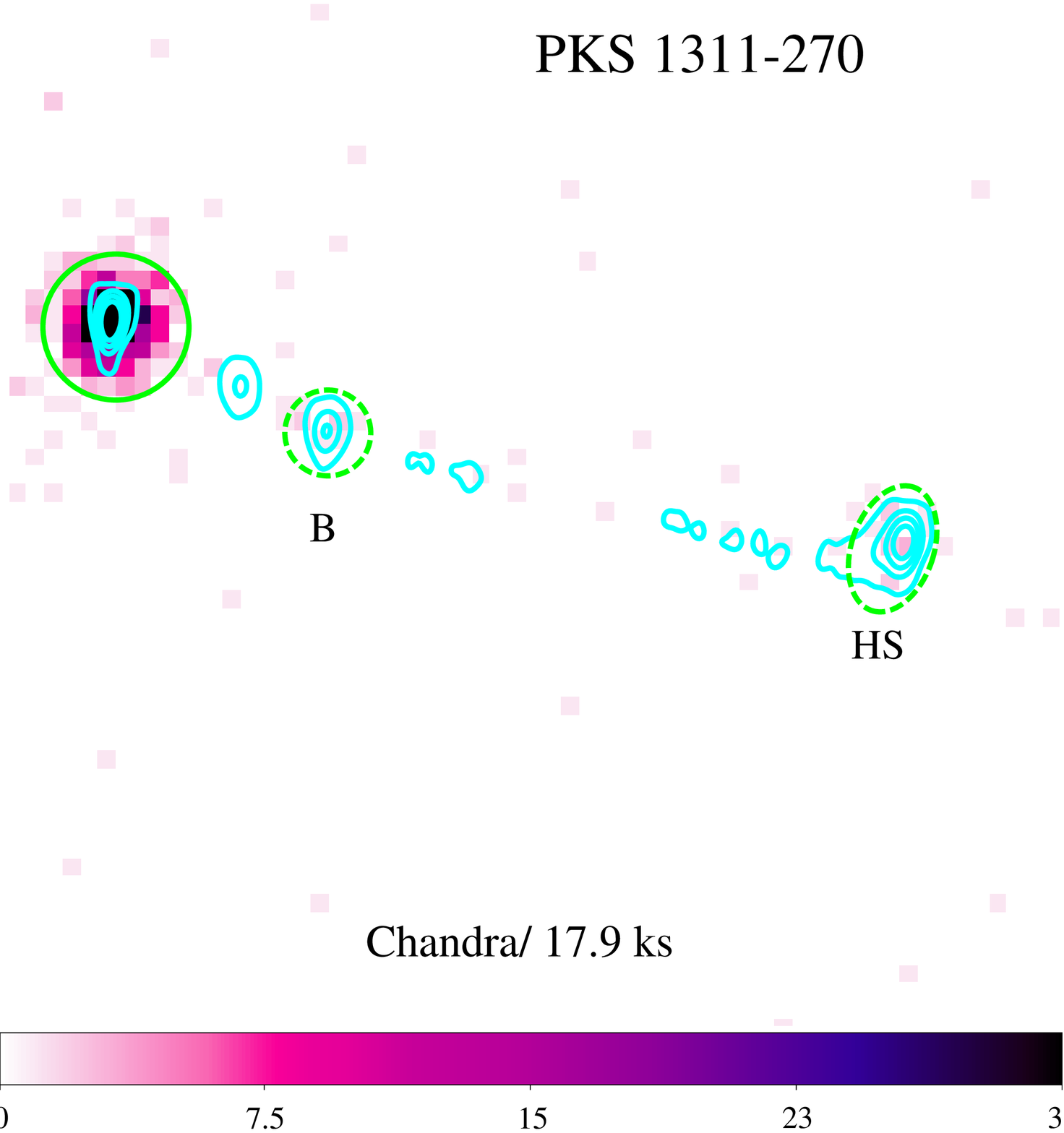}{0.5\textwidth}{(a)}
        \fig{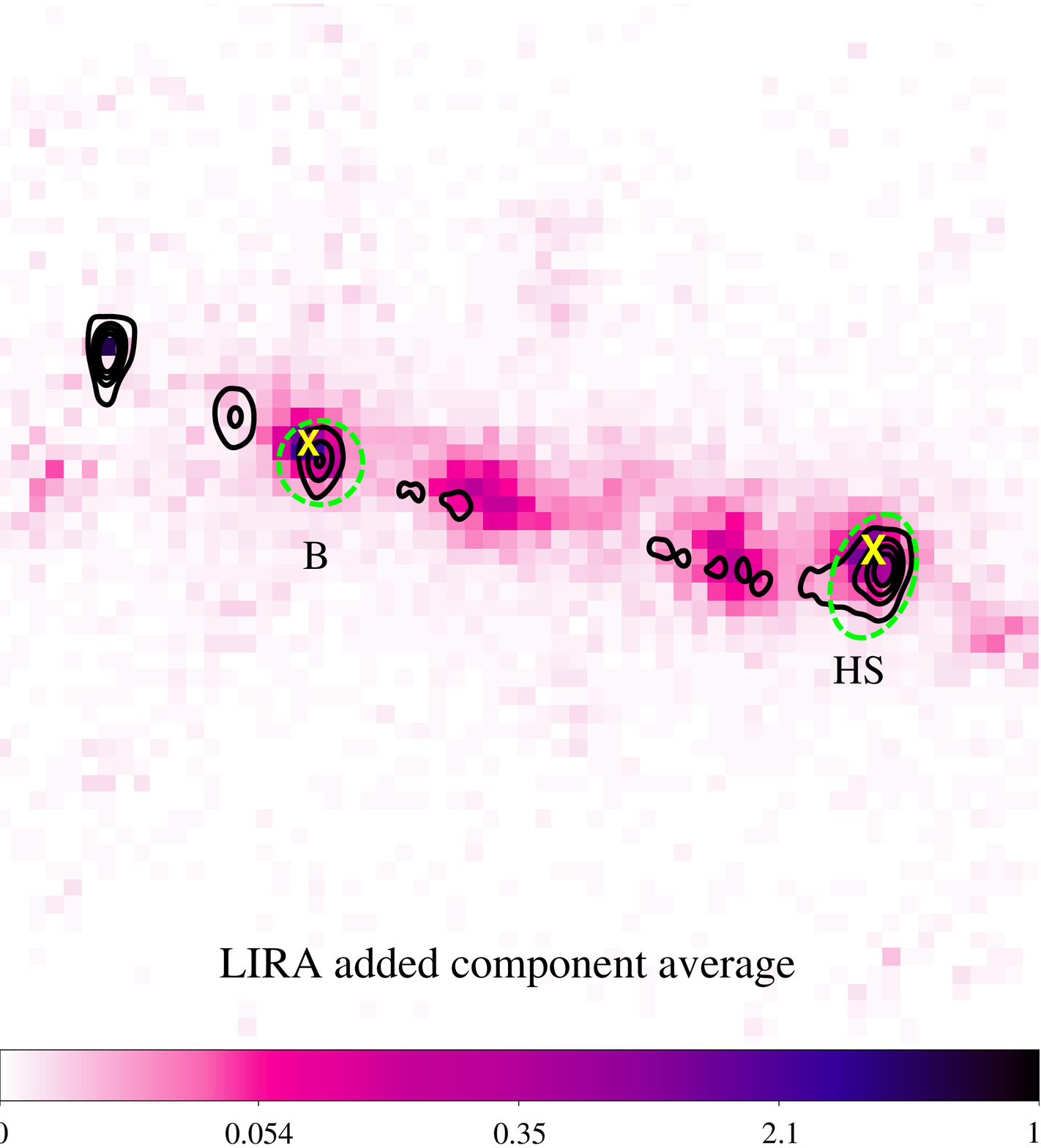}{0.5\textwidth}{(b)}
    }
    \gridline{
    \fig{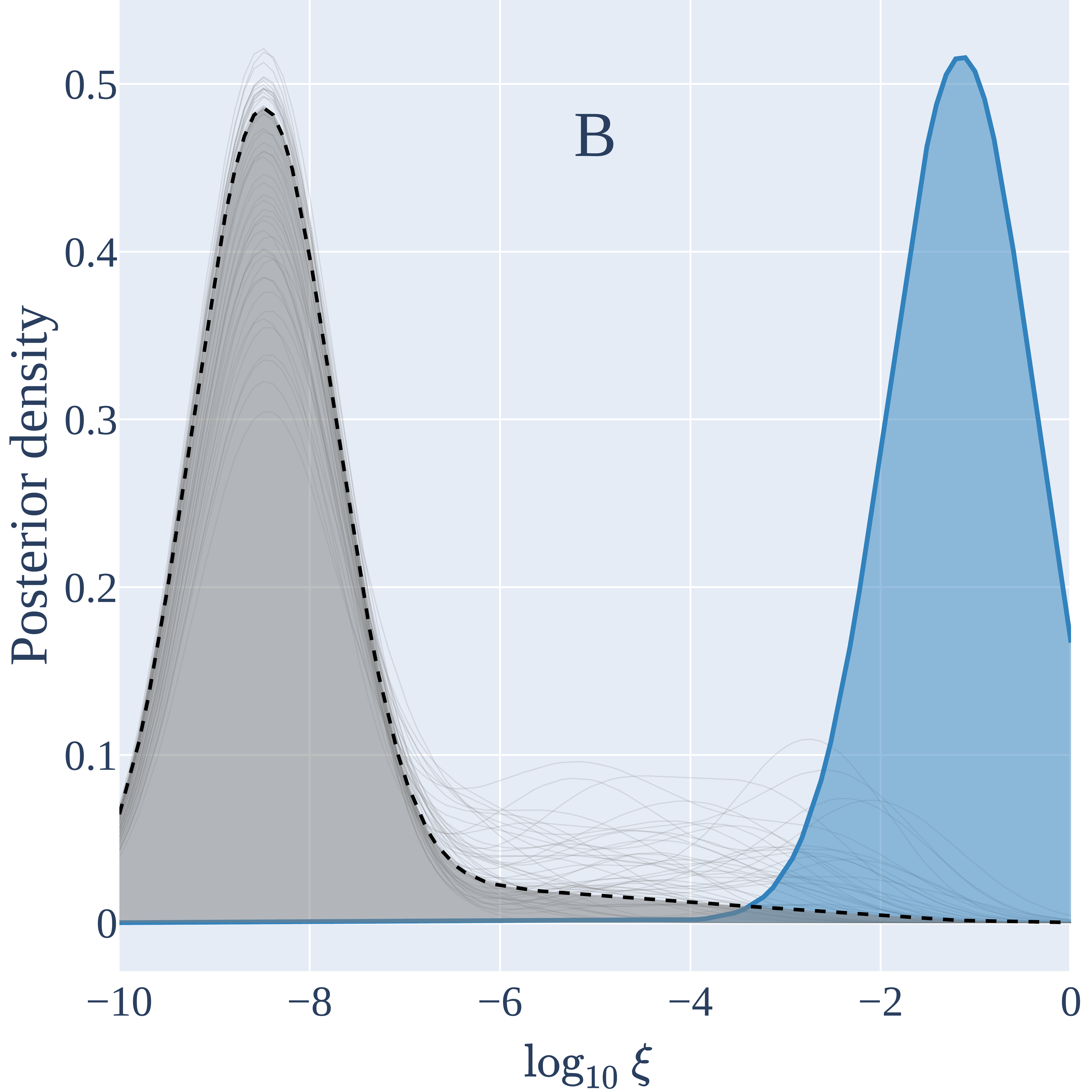}{0.33\textwidth}{(c)}
     \fig{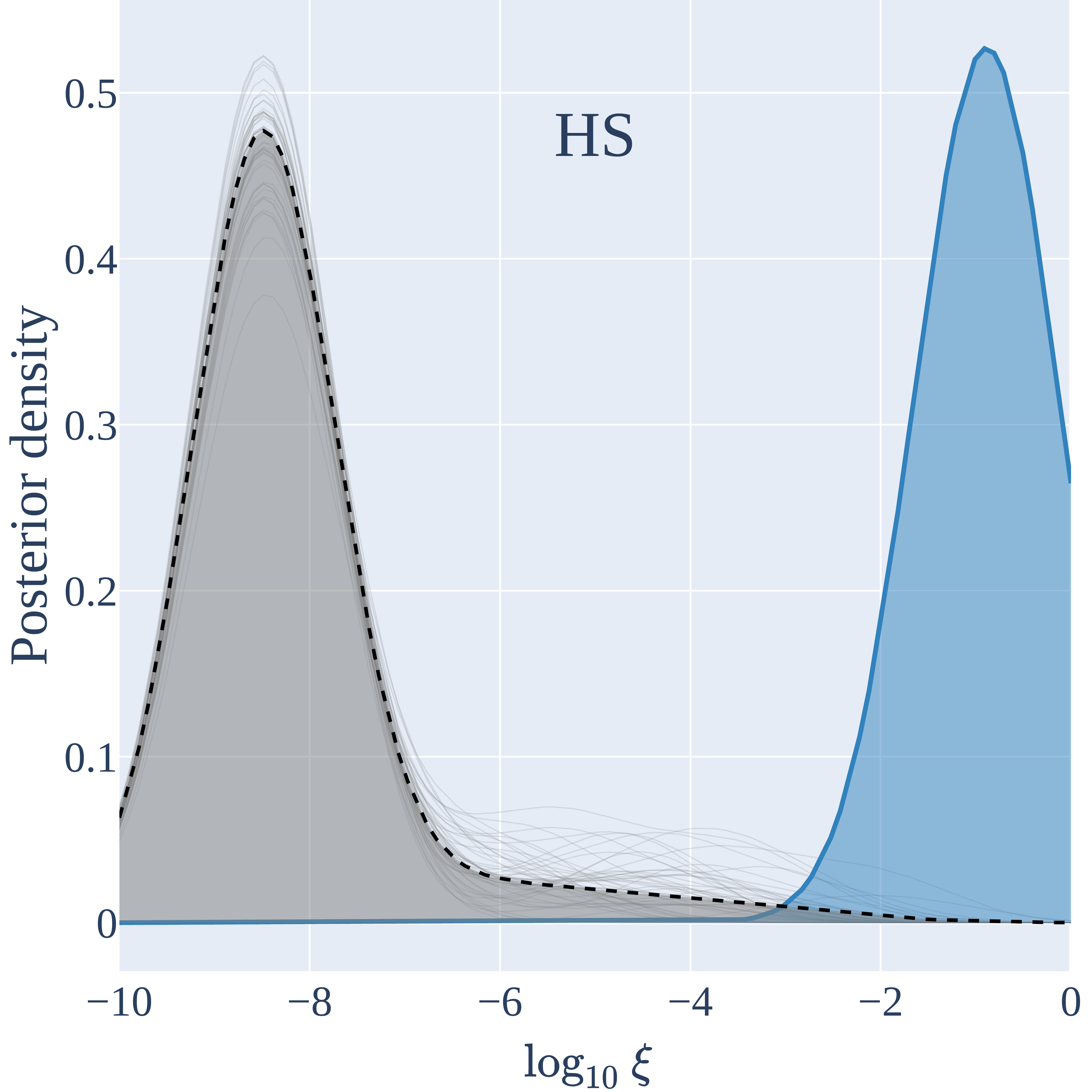}{0.33\textwidth}{(d)}
    }
    
\caption{Same as in Fig. \ref{fig:results_3C418}, for PKS 1311-270, a high-redshift quasar.  \Radiocontours{ 0.5, 4, 10, 20} \label{fig:results_PKS1311-270}}
\end{figure*}
\textbf{PKS 1311-270} (Fig. \ref{fig:results_PKS1311-270}): This is a high-redshift quasar (z=2.19). Knot B shows an Xf-offset of 0.25\as~(2.11 kpc). The scenario is slightly complicated in the case of HS. The radio jet makes a 35\degree~projected turn to the northeast before reaching HS.  Although the X-ray centroid lies closer to the core than its radio peak [by 0.24\as~(2.02 kpc)], the offset is inconsistent with the direction of the jet. It is possible that the jet first turns towards the X-ray centroid before reaching the radio hotspot and projection effects create such an appearance. Alternatively, a large-scale hot and diffuse gas may be emitting the X-rays near HS. 

\begin{figure*}[h]
    \gridline{
        \fig{{4C+11.45_lira_paper_count}.eps}{0.5\textwidth}{(a)}
        \fig{{4C+11.45_lira_paper_add}.eps}{0.5\textwidth}{(b)}
    }
    \gridline{
    \fig{{4C+11.45_C_lira_post}.pdf}{0.33\textwidth}{(c)}
    }
    
\caption{Same as in Fig. \ref{fig:results_3C418}, for 4C+11.45, a quasar.\Radiocontours{0.3, 4, 20, 50, 80} \label{fig:results_4C+11.45}}
\end{figure*}
\textbf{4C+11.45} (Fig. \ref{fig:results_4C+11.45}): This is a quasar. The jet makes a 50\degree~projected bend to the southeast at knot C, where the the X-ray centroid lies downstream of the radio peak by 0.19\as~(1.61 kpc).

\begin{figure*}[h]
    \gridline{
        \fig{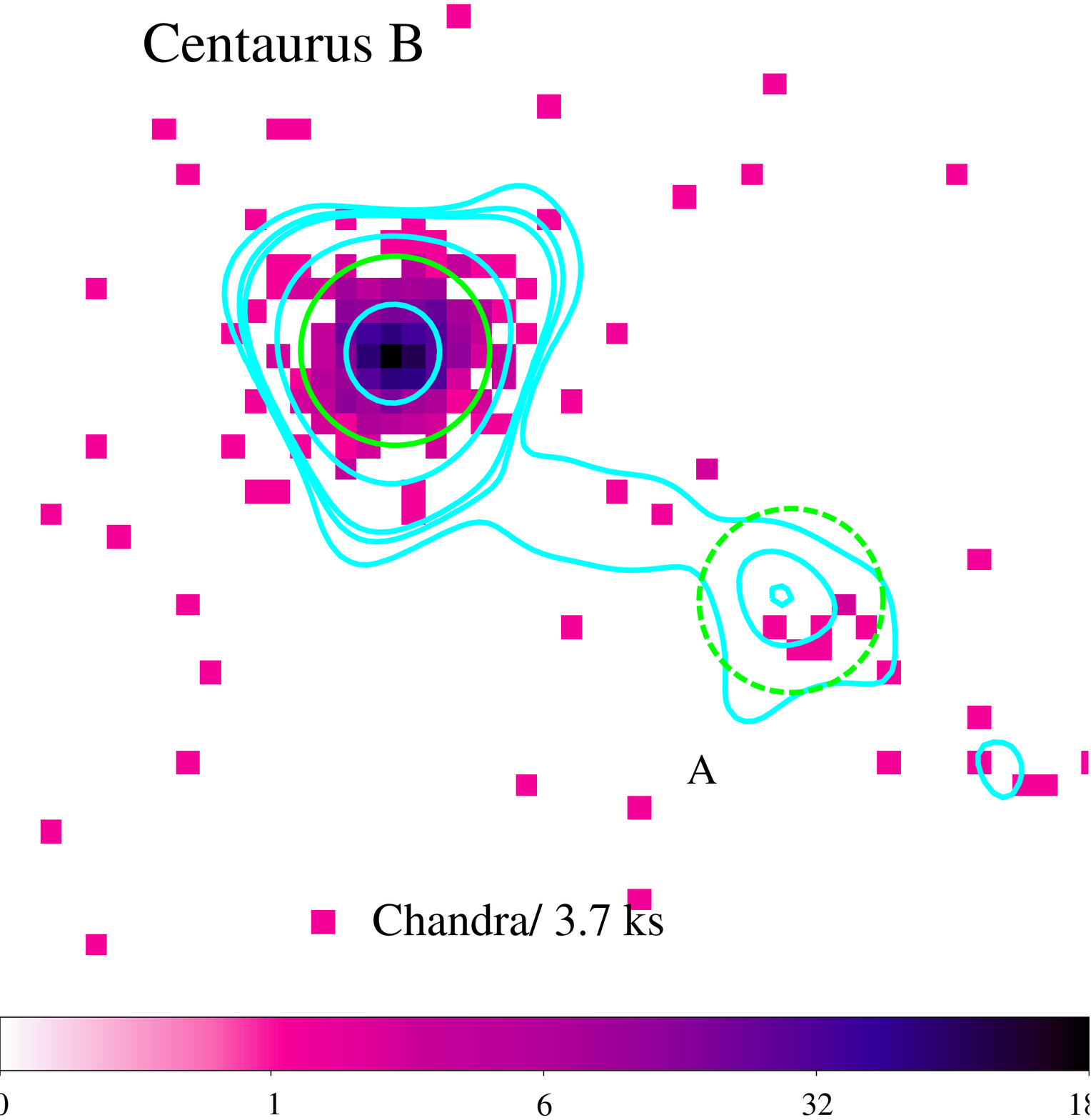}{0.5\textwidth}{(a)}
        \fig{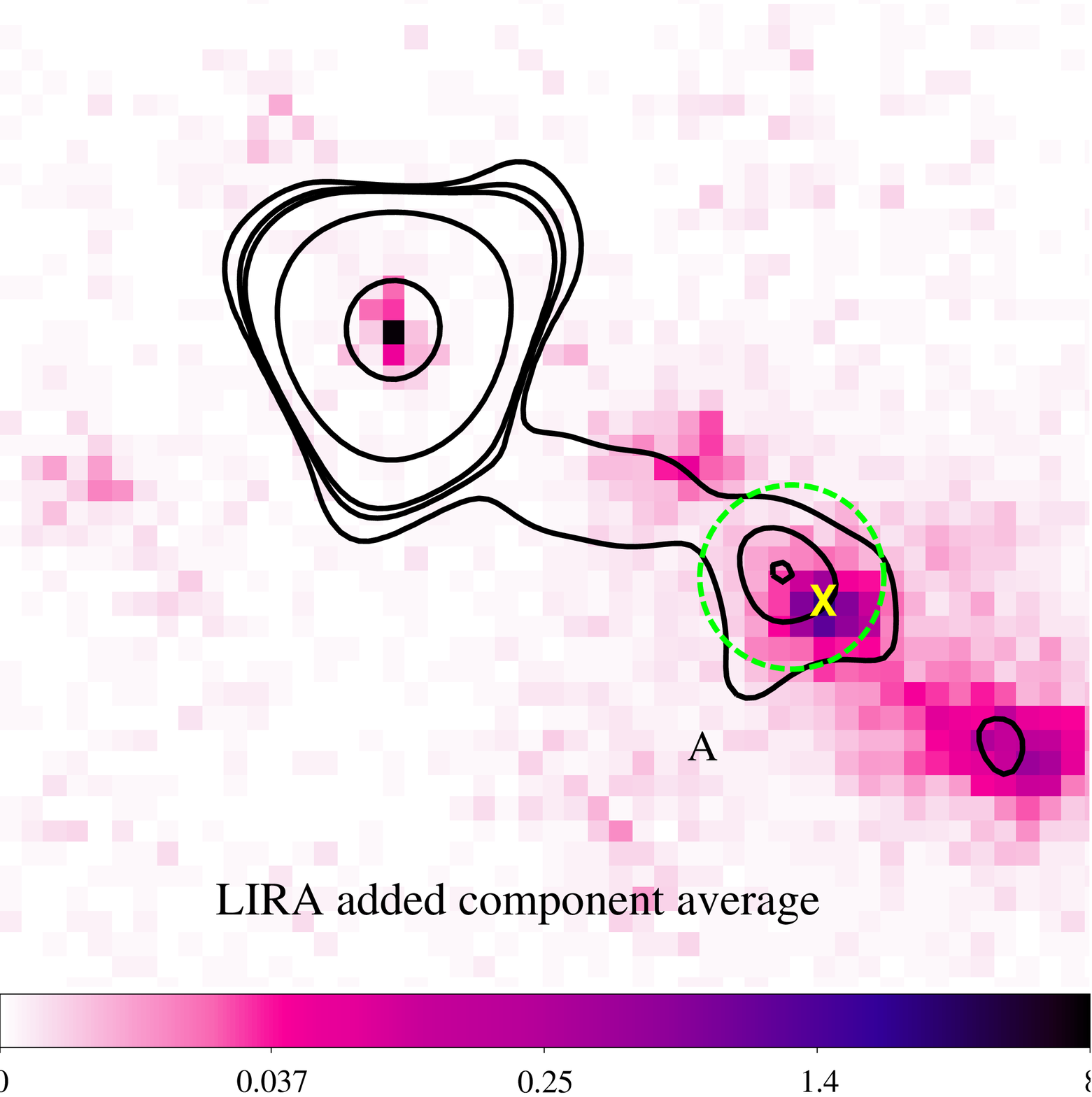}{0.5\textwidth}{(b)}
    }
    \gridline{
    \fig{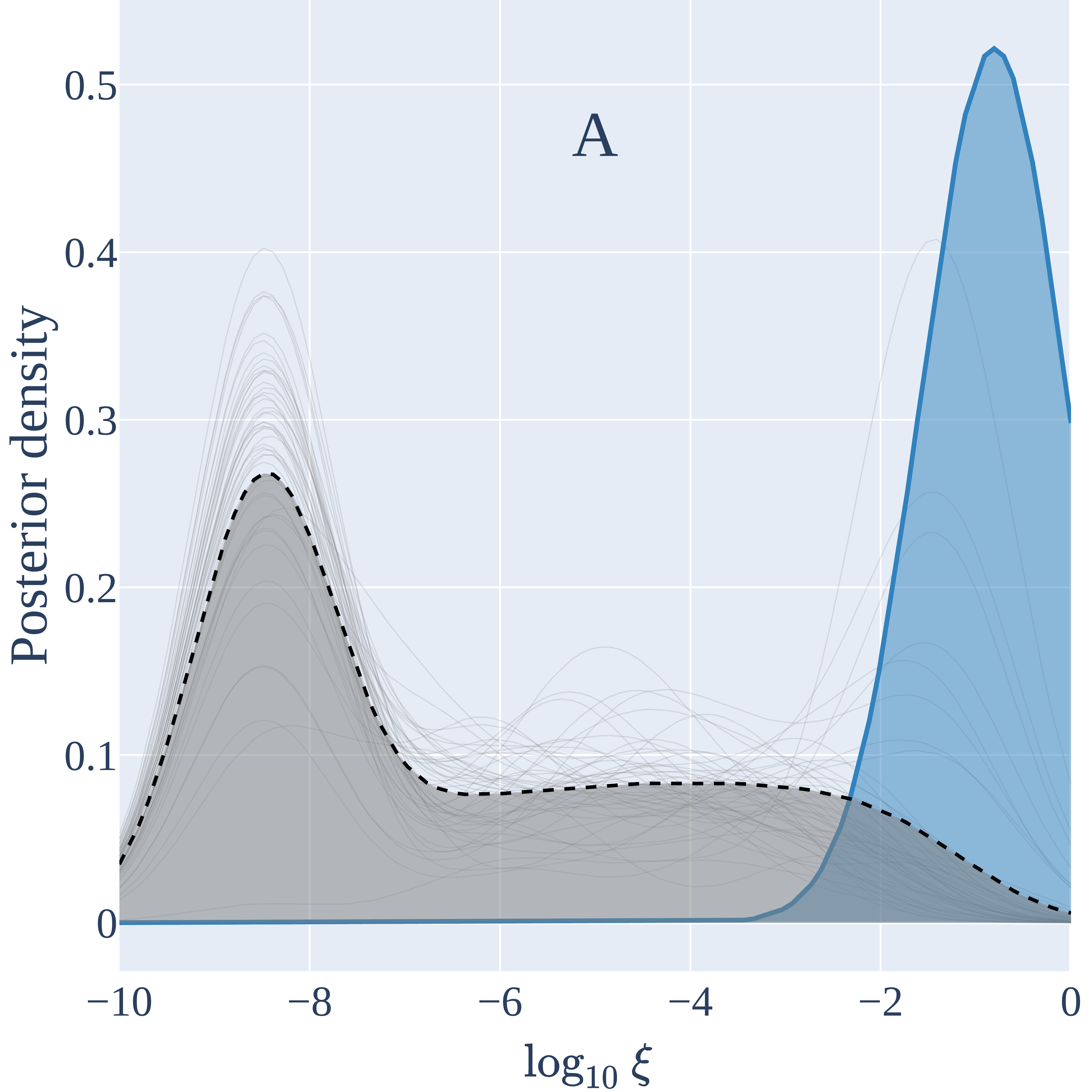}{0.33\textwidth}{(c)}
    }
    
\caption{Same as in Fig. \ref{fig:results_3C418}, with the ATCA 8.6 GHz radio contours overlaid, for Centaurus B, an FR-I RG. \Radiocontours{ 8, 15, 20, 100, 1000}\label{fig:results_CenB}}
\end{figure*}
\textbf{Centaurus B} (Fig. \ref{fig:results_CenB}): This is an FR-I source. Knot A shows an Rf-type offset of 0.57\as~(0.15 kpc). Such an offset is rarely observed in jets with an FR-I type morphology.

\begin{figure*}[h]
    \gridline{
        \fig{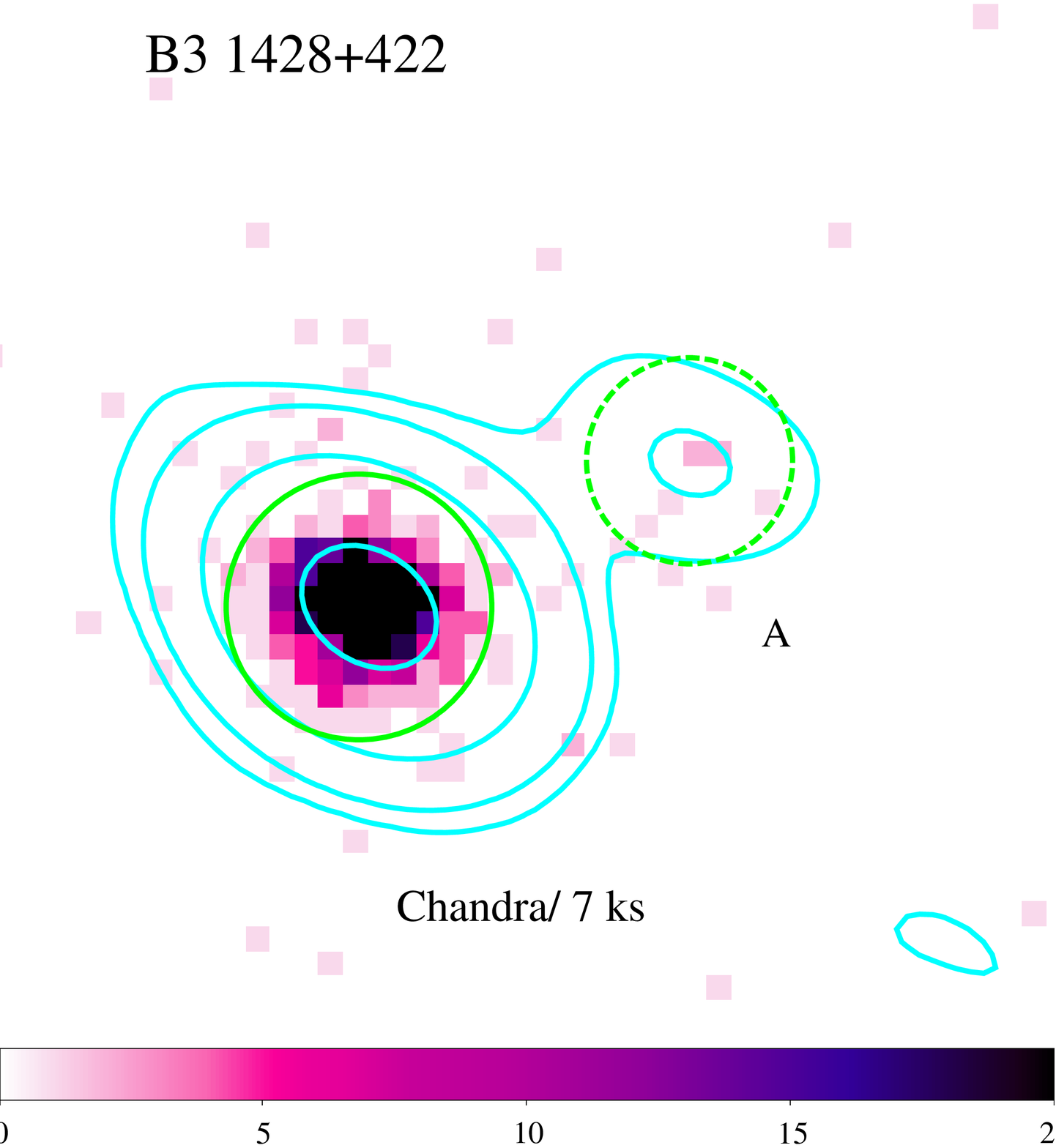}{0.5\textwidth}{(a)}
        \fig{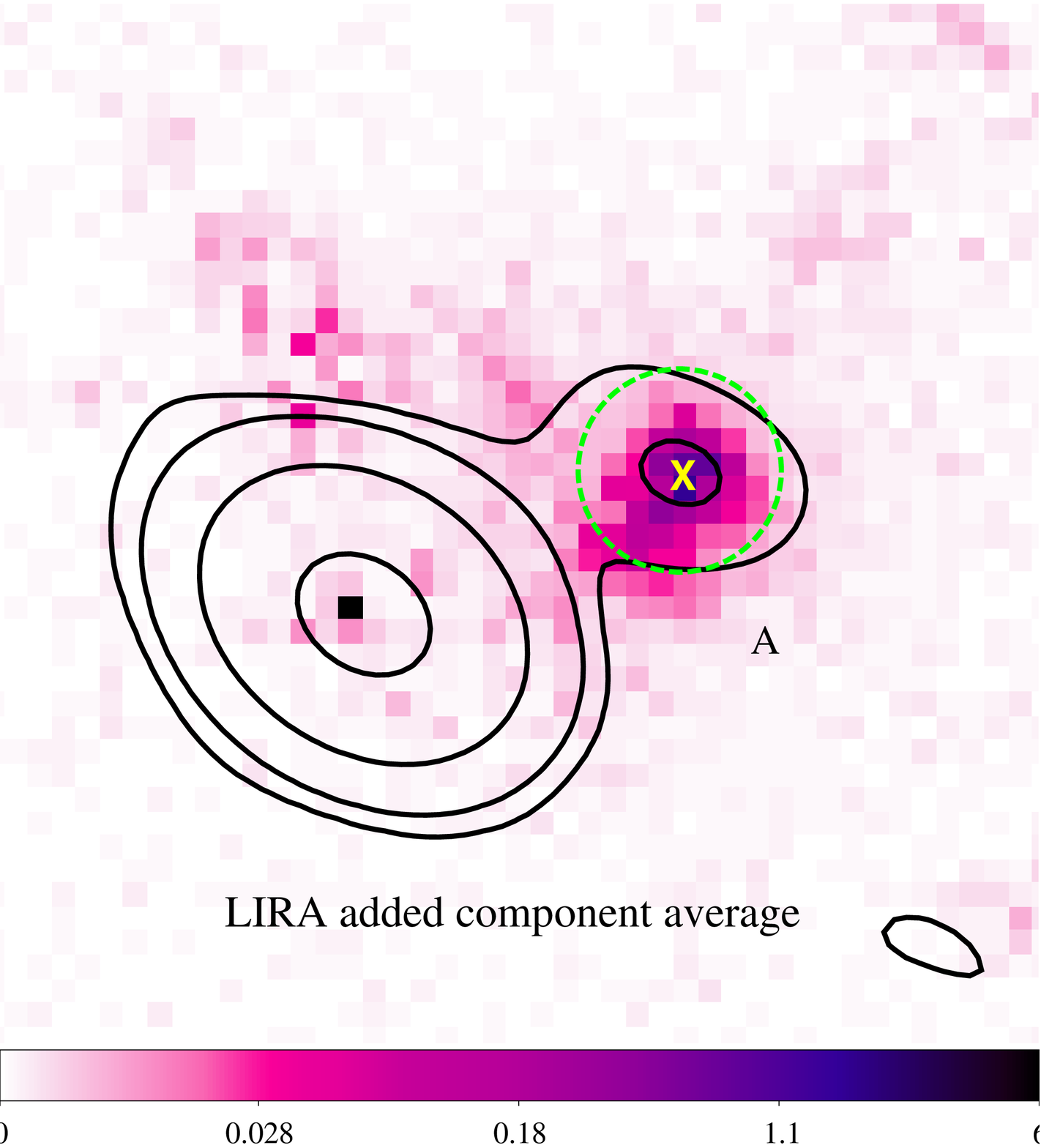}{0.5\textwidth}{(b)}
    }
    \gridline{
    \fig{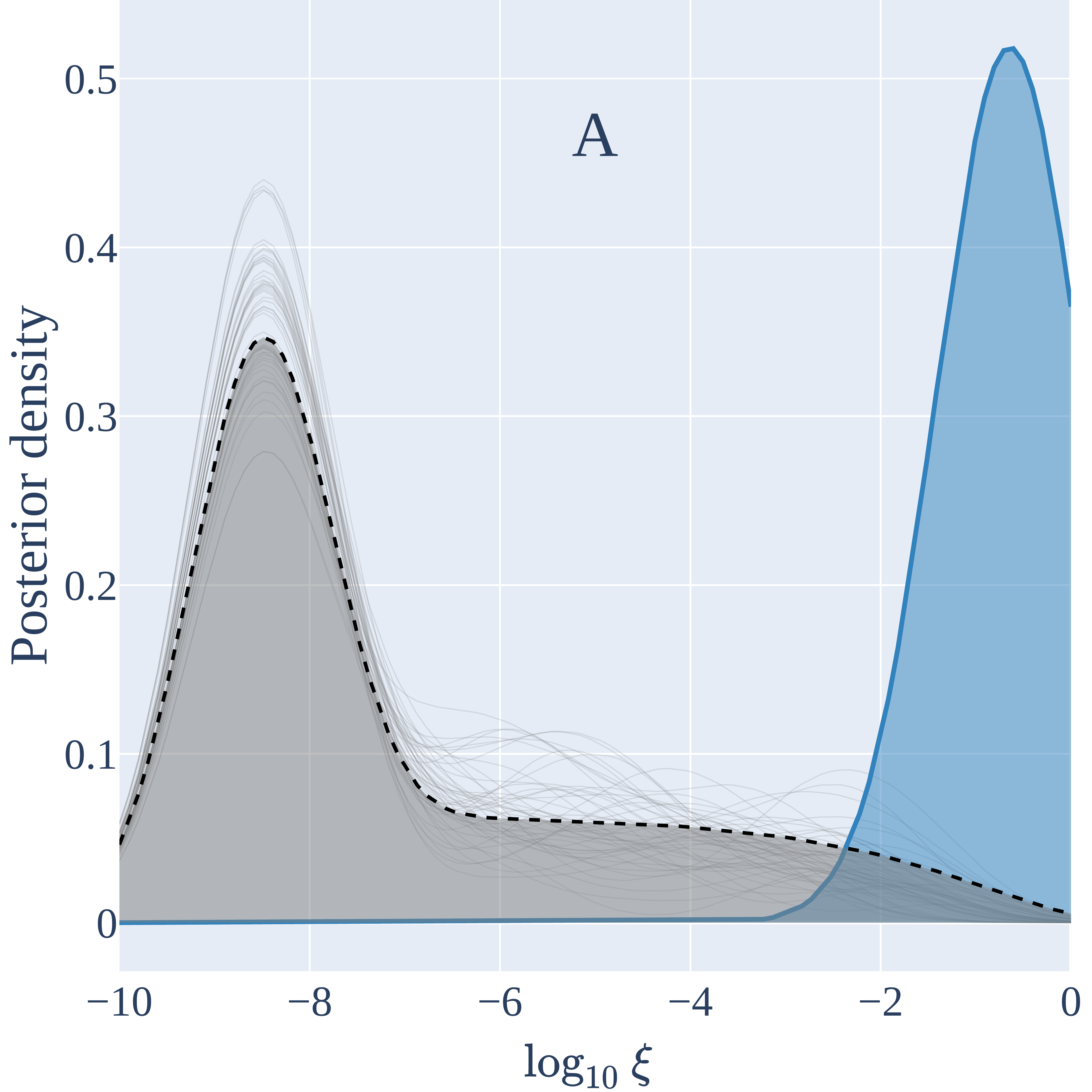}{0.33\textwidth}{(c)}
    }
\caption{Same as in Fig. \ref{fig:results_3C418}, with the VLA 1.4 GHz radio contours overlaid, for B3 1428+422, a high-redshift quasar. \Radiocontours{ 0.2, 1, 10, 100}\label{fig:results_B3_1428+422}}
\end{figure*}
\textbf{B3 1428+422} (Fig. \ref{fig:results_B3_1428+422}): This is one of the highest redshift (z=4.7) X-ray jet known to date. We find that the X-ray centroid in its jet-related feature, A, lies within 0.15\as~(1 kpc) of its radio peak.

\begin{figure*}[h]
    \gridline{
        \fig{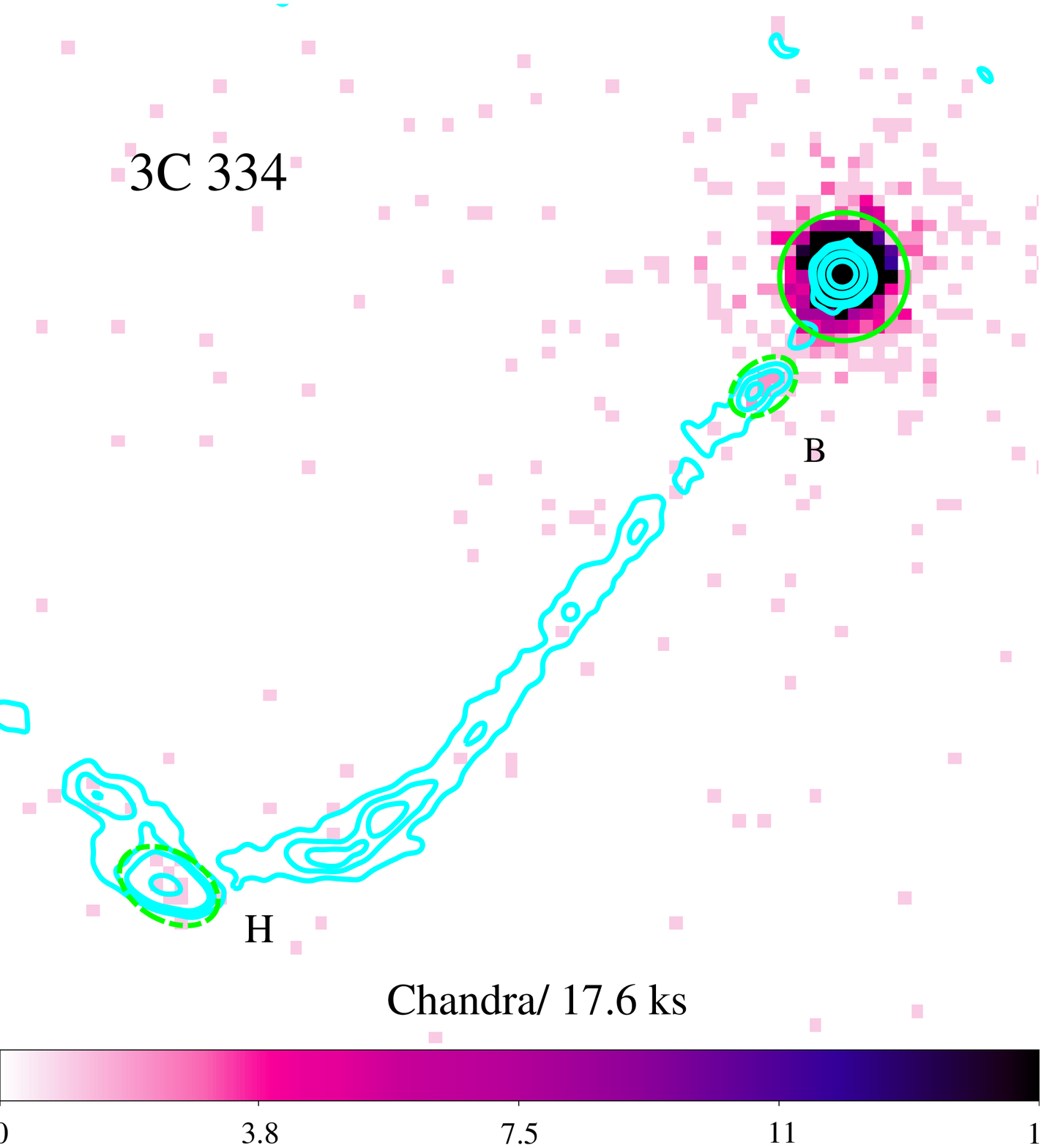}{0.5\textwidth}{(a)}
        \fig{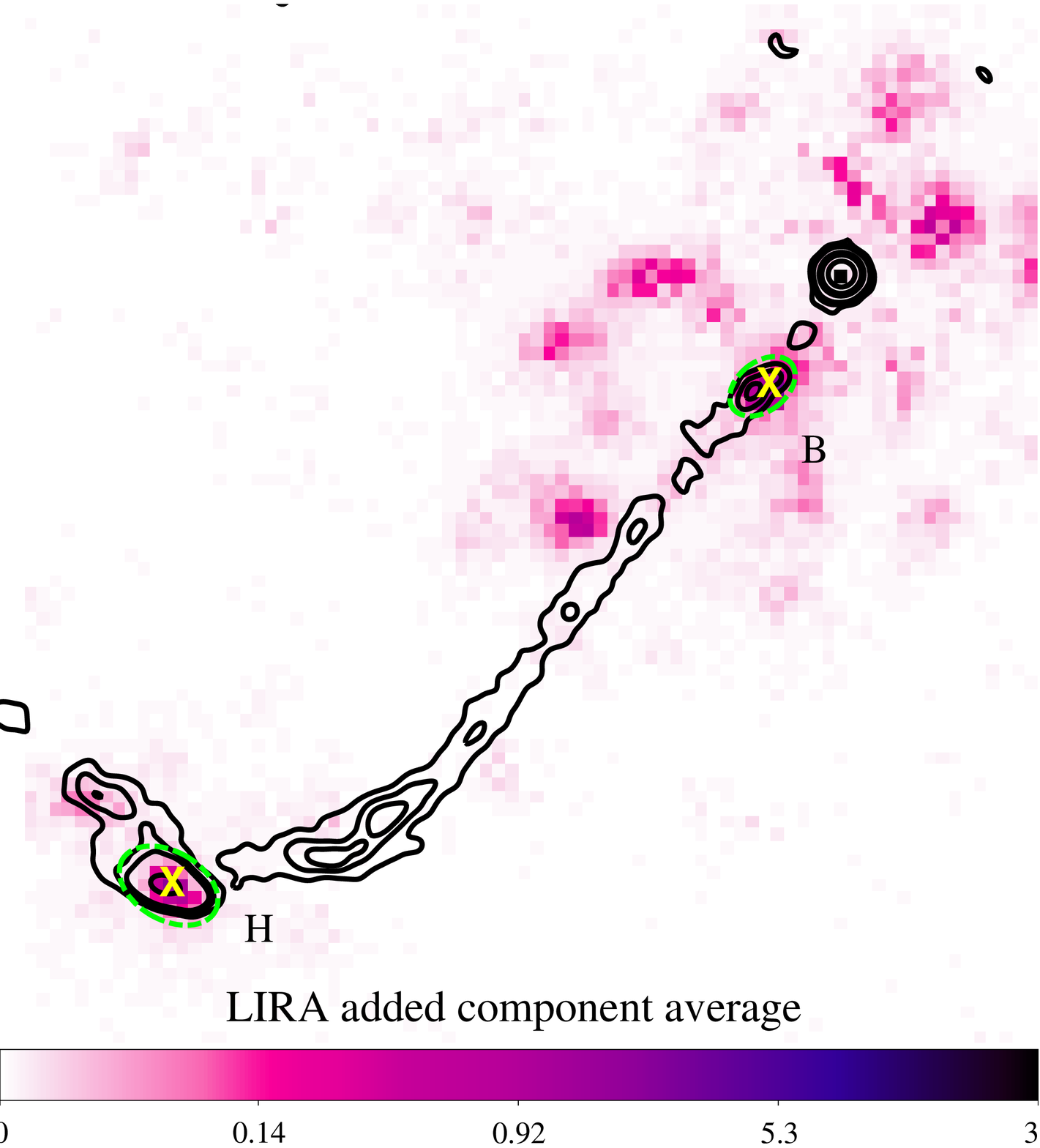}{0.5\textwidth}{(b)}
    }
    \gridline{
    \fig{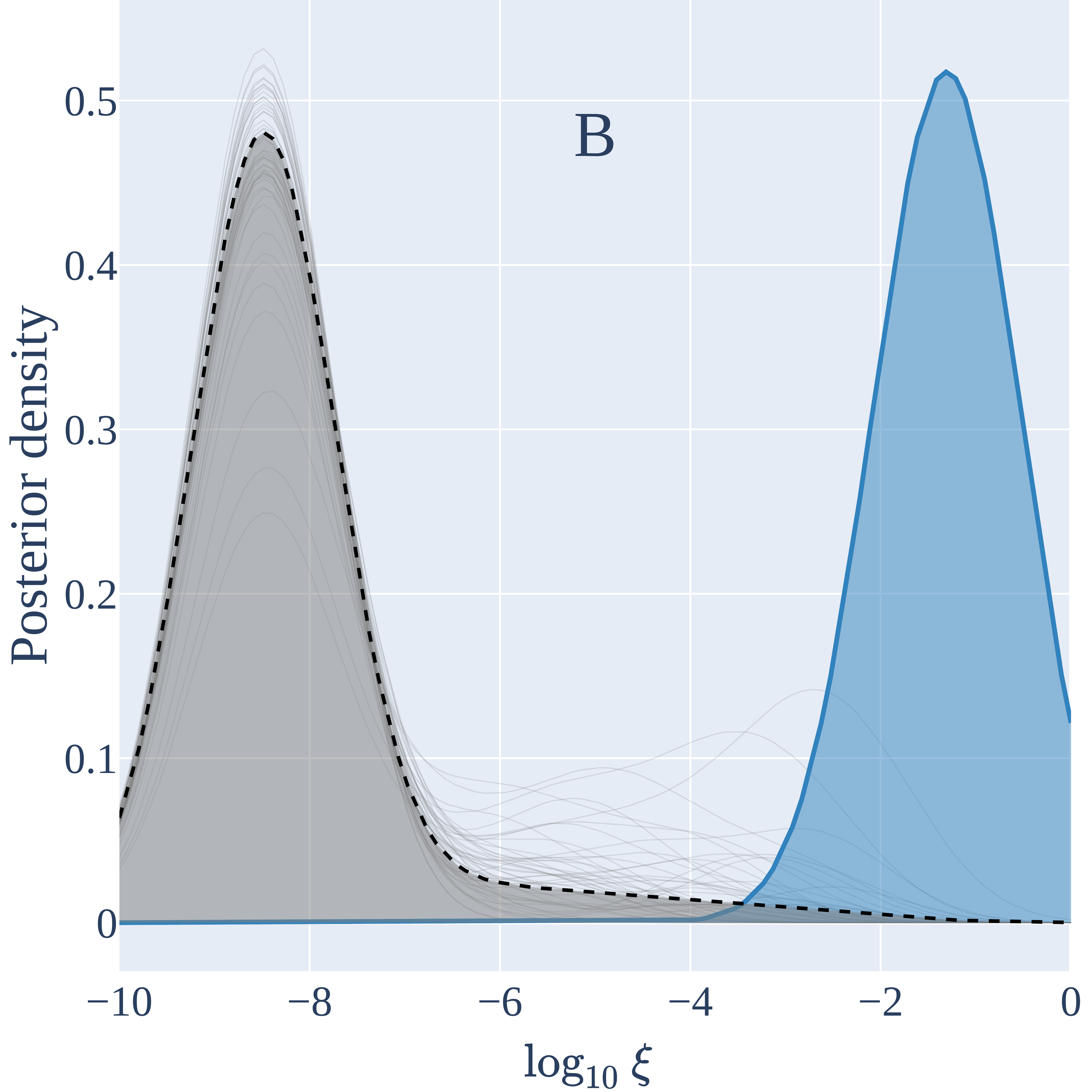}{0.33\textwidth}{(c)}
     \fig{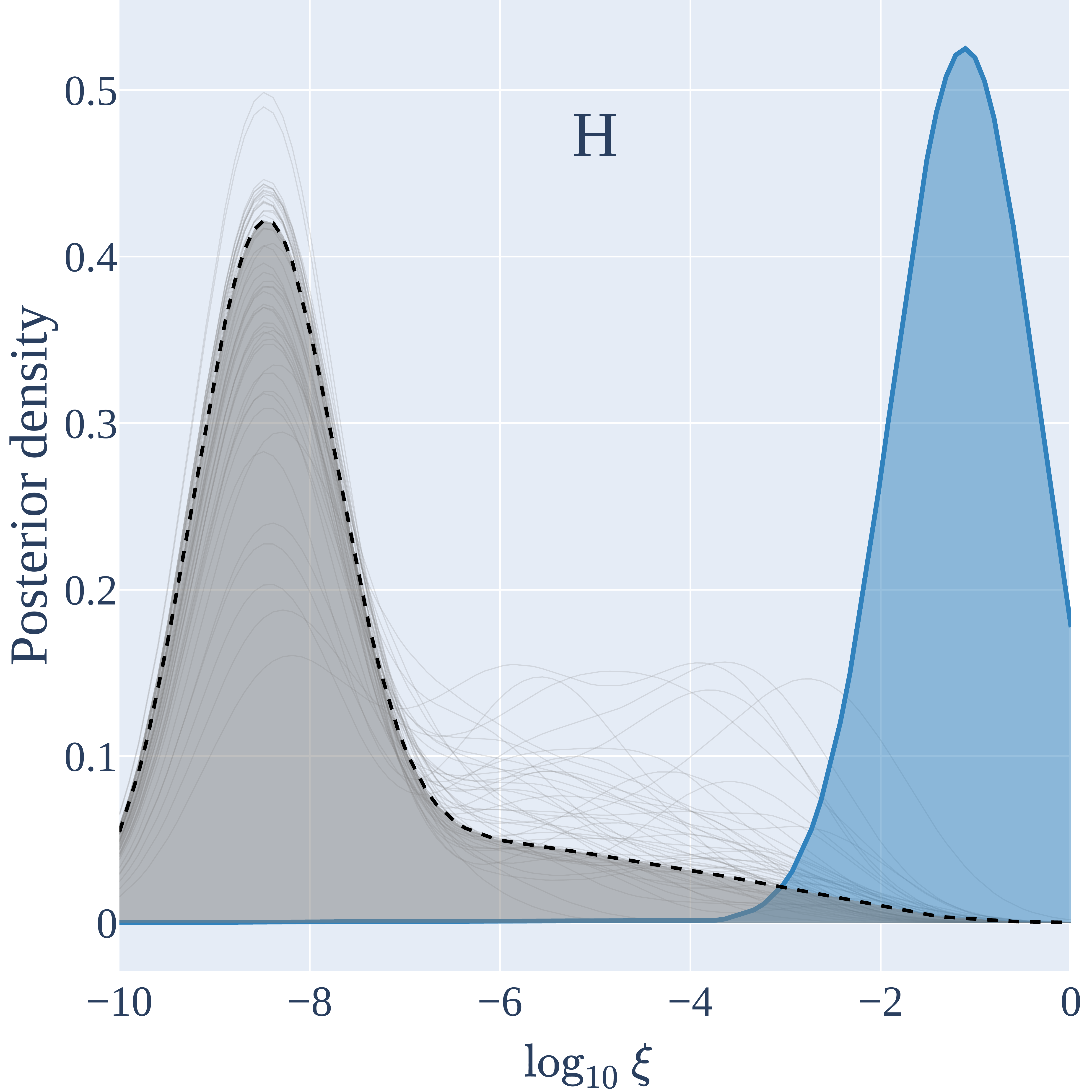}{0.33\textwidth}{(d)}
    }
    
\caption{Same as in Fig. \ref{fig:results_3C418}, with the VLA 4.8 GHz radio contours overlaid, for 3C 334, a quasar. \Radiocontours{0.15, 0.4, 0.7, 5, 10, 40}\label{fig:results_3C334}}
\end{figure*}
\textbf{3C 334} (Fig. \ref{fig:results_3C334}): This is a quasar. Knot B shows an Xf-offset of 0.22\as~(1.46 kpc). In contrast, the X-ray centroid in H, where the jet makes a 90\degree~apparent bend to the northeast, lies within 0.15\as~(1 kpc) of its radio peak.

\textbf{3C 418} (Fig. \ref{fig:results_3C418}): This is a quasar. We newly detect jet-linked X-ray emission from knot B using LIRA. See section \ref{subsec:new_detections} for details.

\begin{figure*}[h]
    \gridline{
        \fig{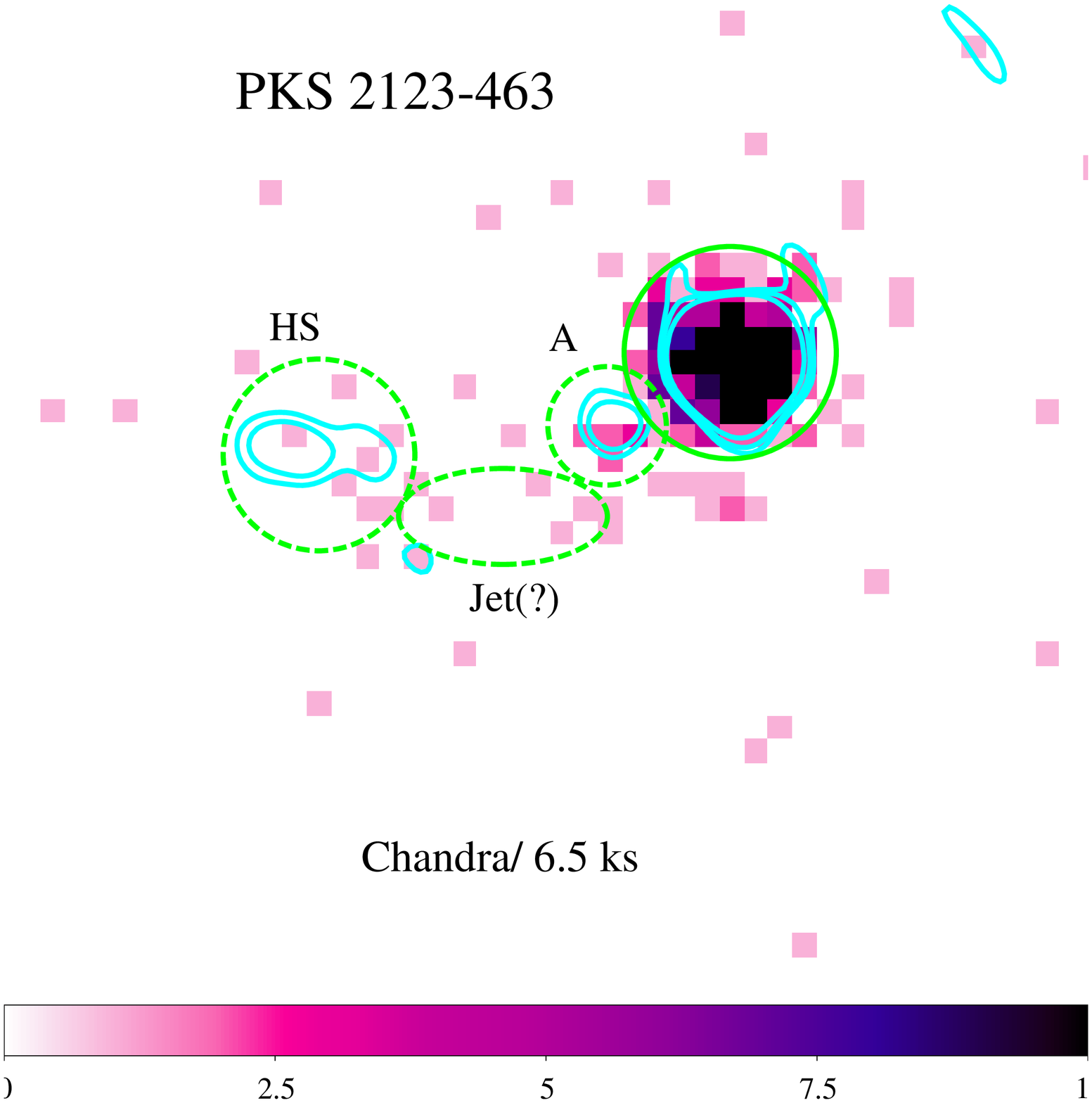}{0.5\textwidth}{(a)}
        \fig{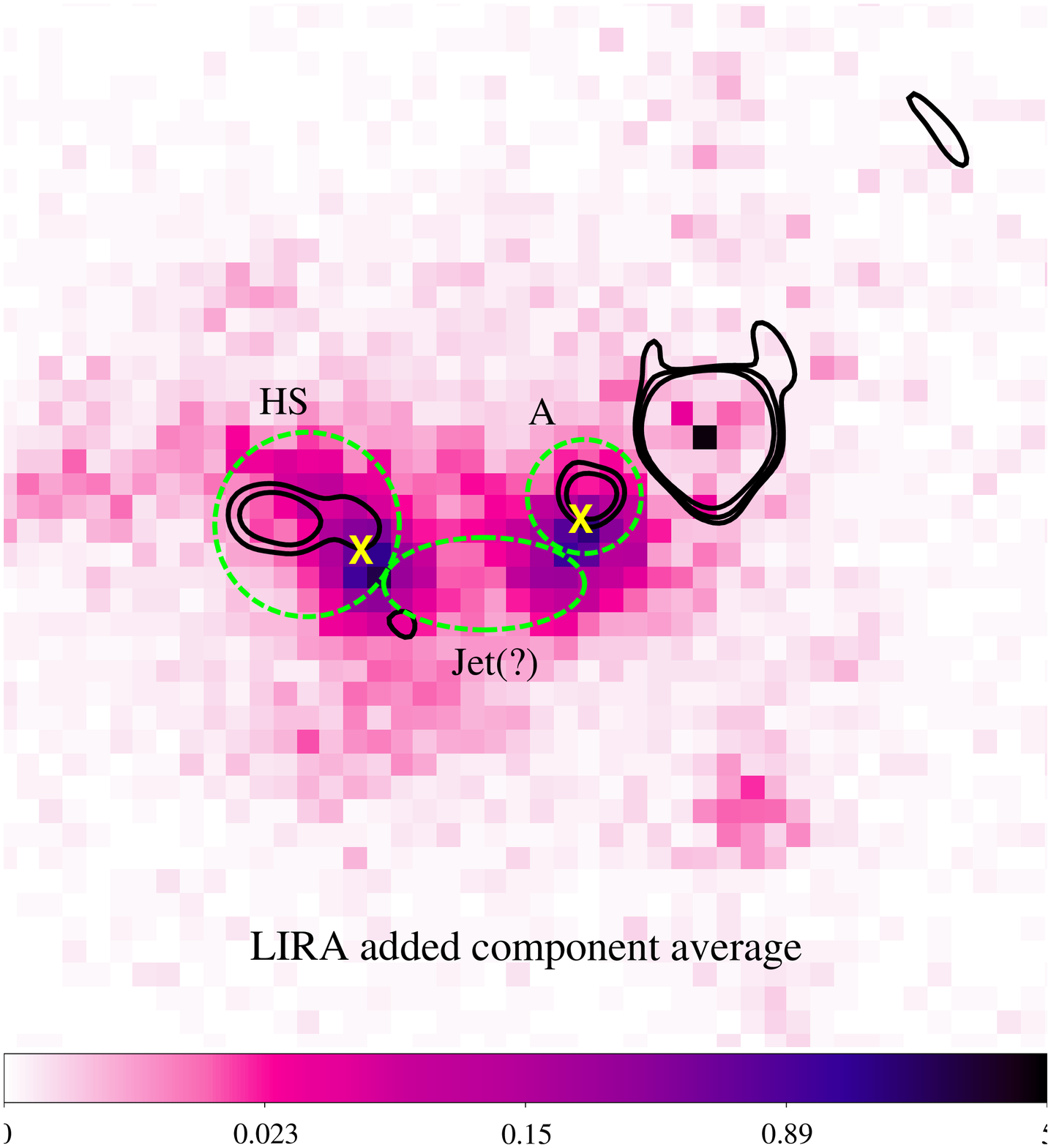}{0.5\textwidth}{(b)}
    }
    \gridline{
    \fig{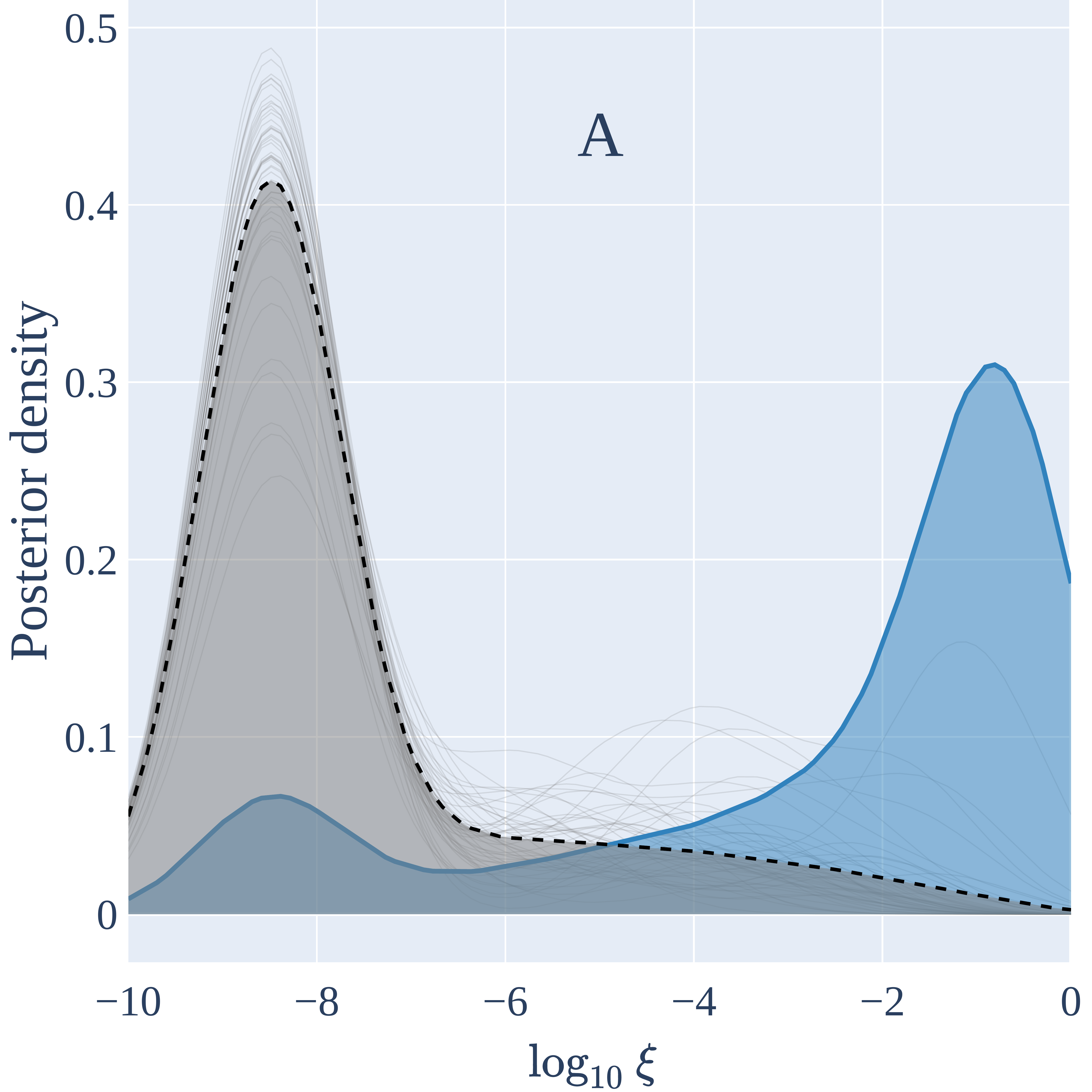}{0.33\textwidth}{(c)}
    \fig{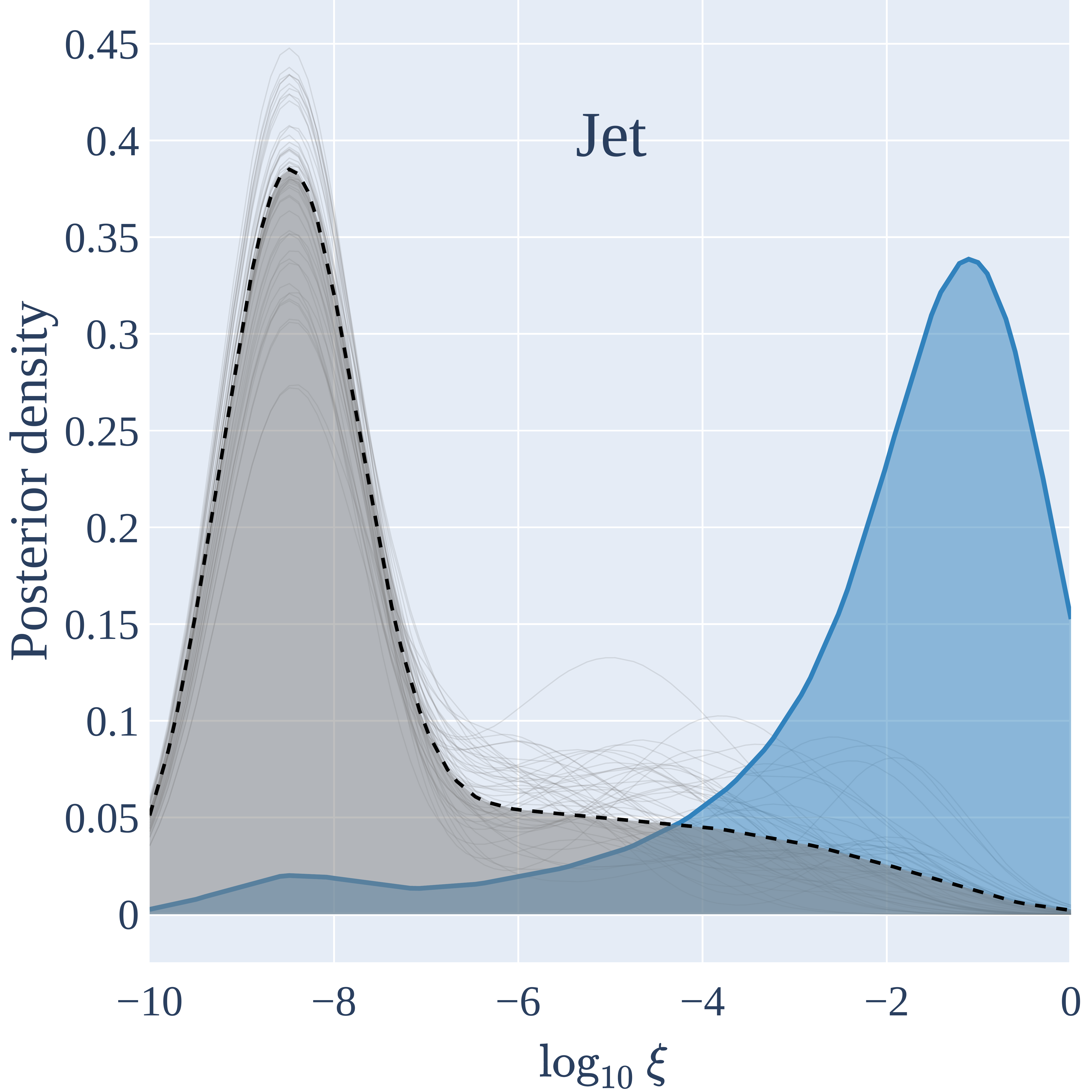}{0.33\textwidth}{(d)}
     \fig{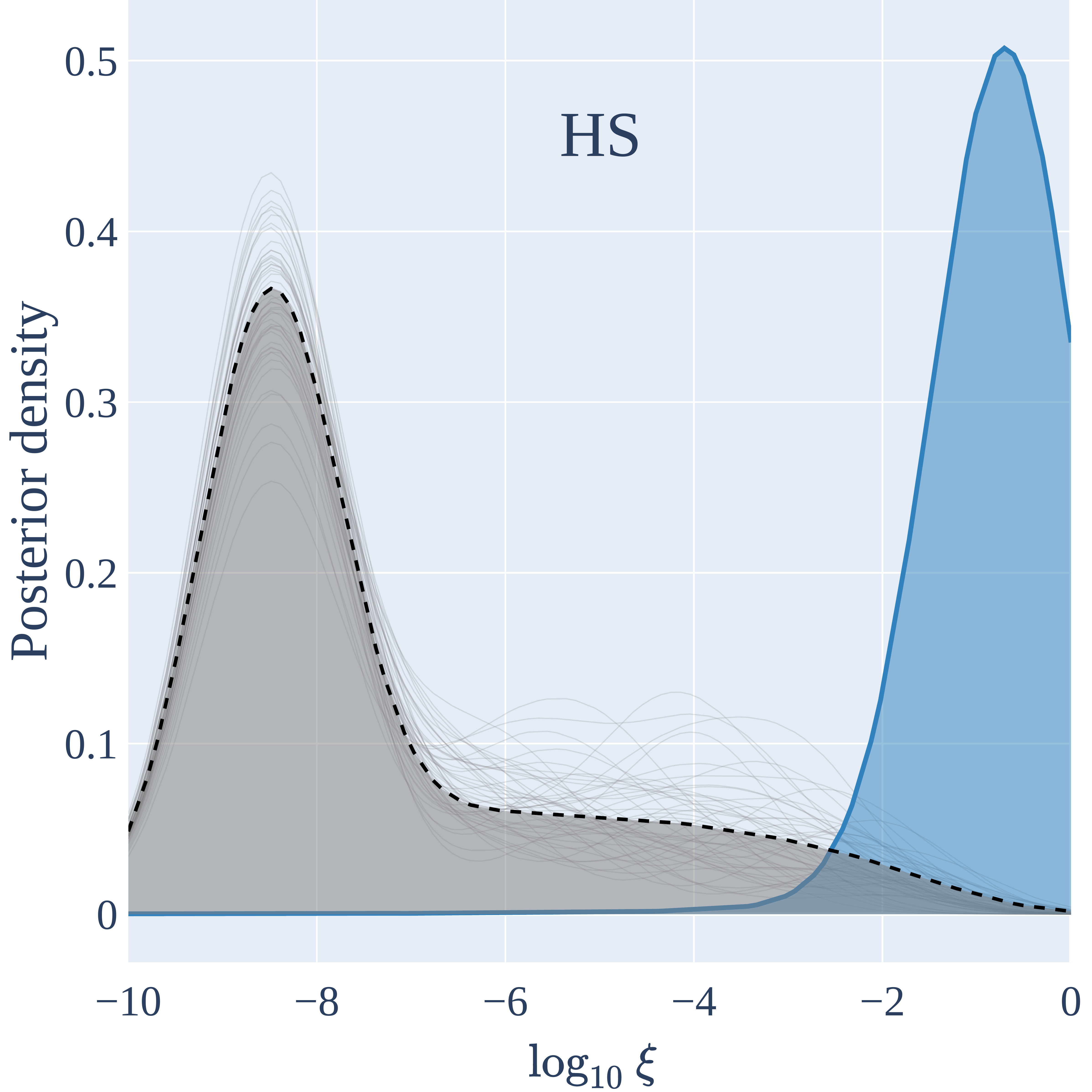}{0.33\textwidth}{(e)}
    }
    
\caption{Same as in Fig. \ref{fig:results_3C418}, for PKS 2123-463, with the ATCA 18 GHz radio contours overlaid, a high redshift quasar. \Radiocontours{1.3, 2, 4}\label{fig:results_pks2123-463}}
\end{figure*}
\textbf{PKS 2123-463} (Fig. \ref{fig:results_pks2123-463}): This is a core-dominated quasar. Knot A shows an Rf-type offset of 0.35\as~(3 kpc) while the hotspot shows an Xf-offset of 0.89\as~(7.62 kpc). We also detect significant X-ray emission along the putative direction of the jet between knot A and the hotspot, which, similar to 4C+25.21, suggests that IC/CMB presumably produces these X-rays (see section \ref{subsec:flux_ratio} for details).

\begin{figure*}[h]
    \gridline{
        \fig{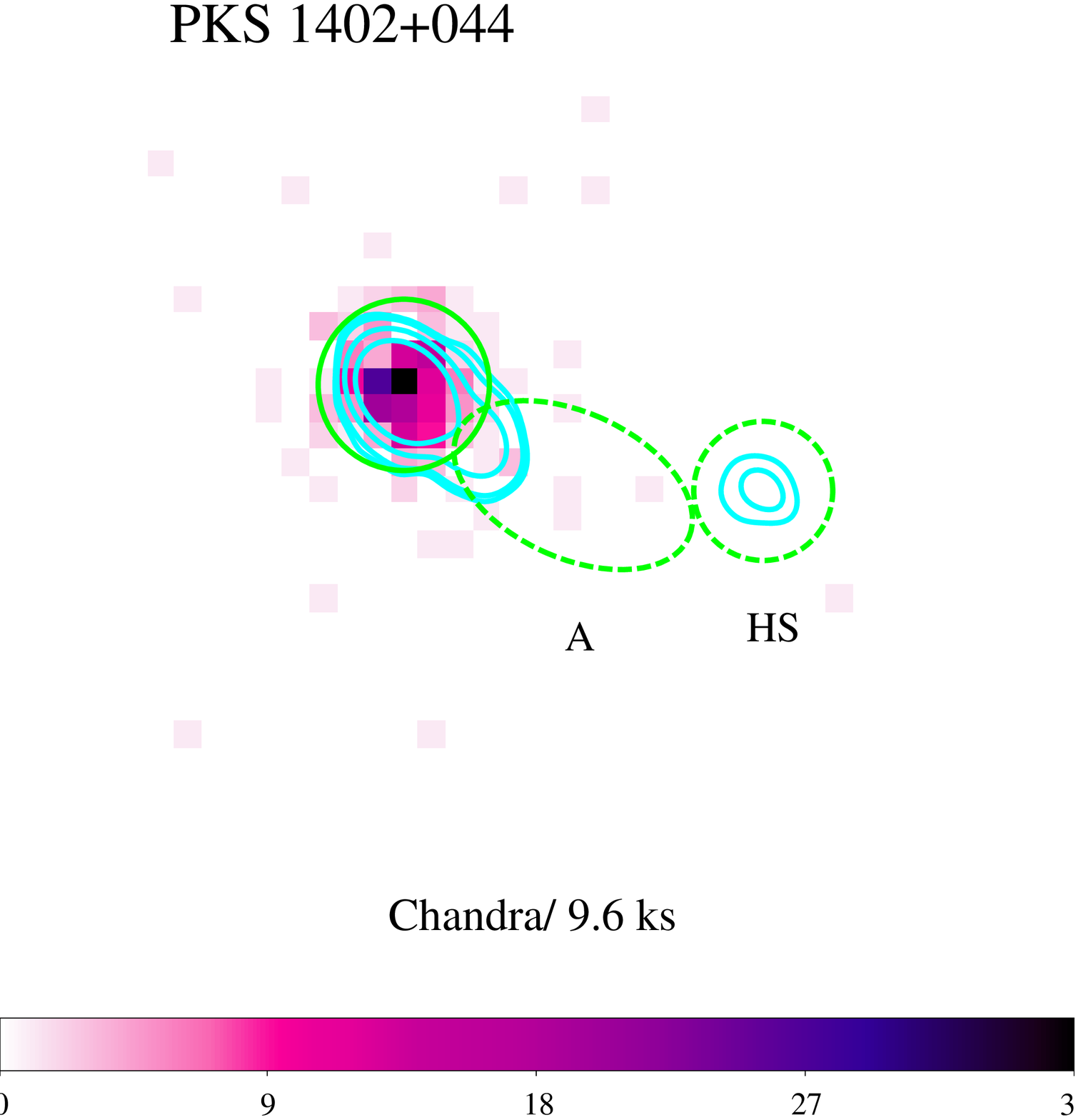}{0.5\textwidth}{(a)}
        \fig{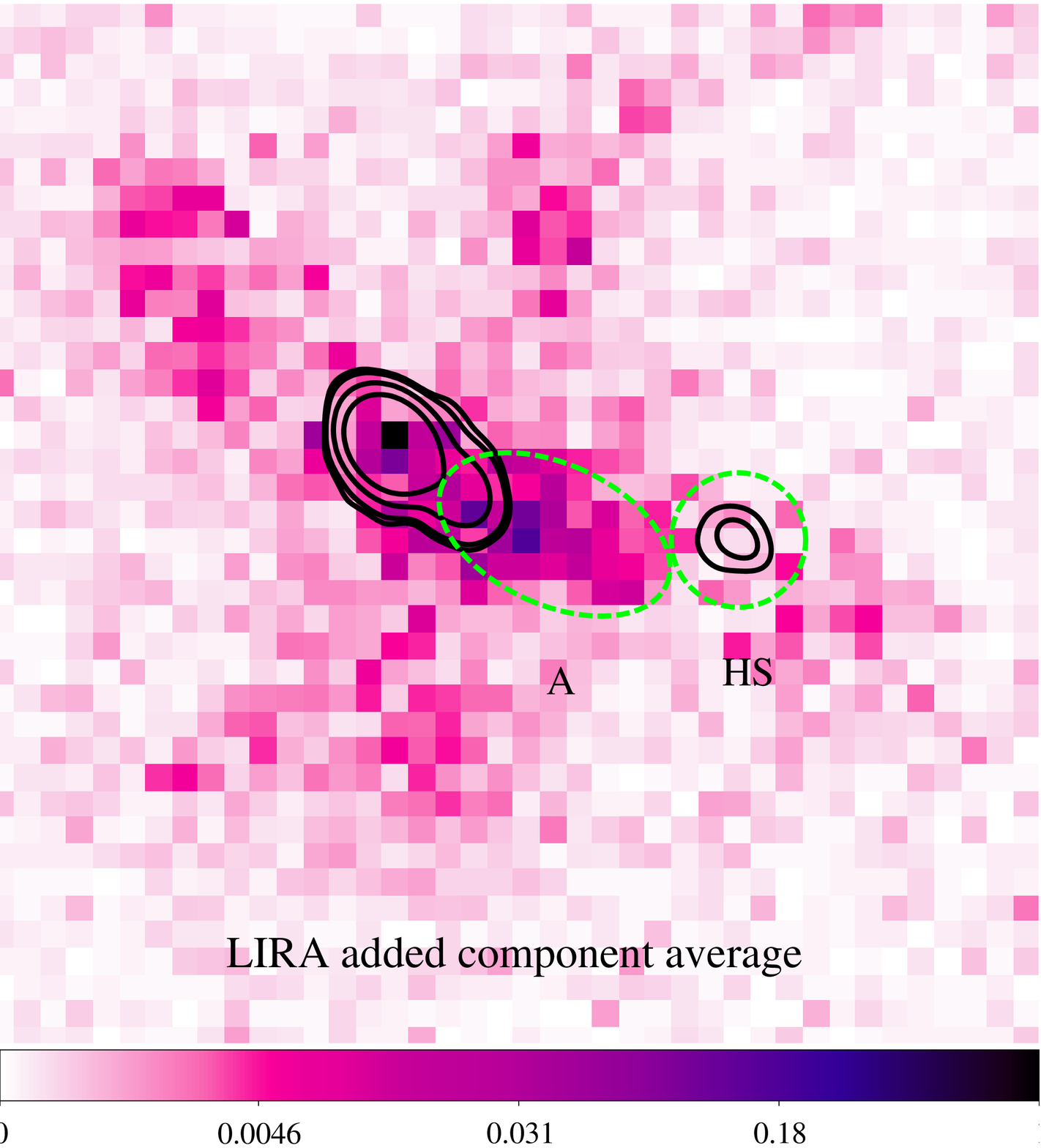}{0.5\textwidth}{(b)}
    }
    \gridline{
    \fig{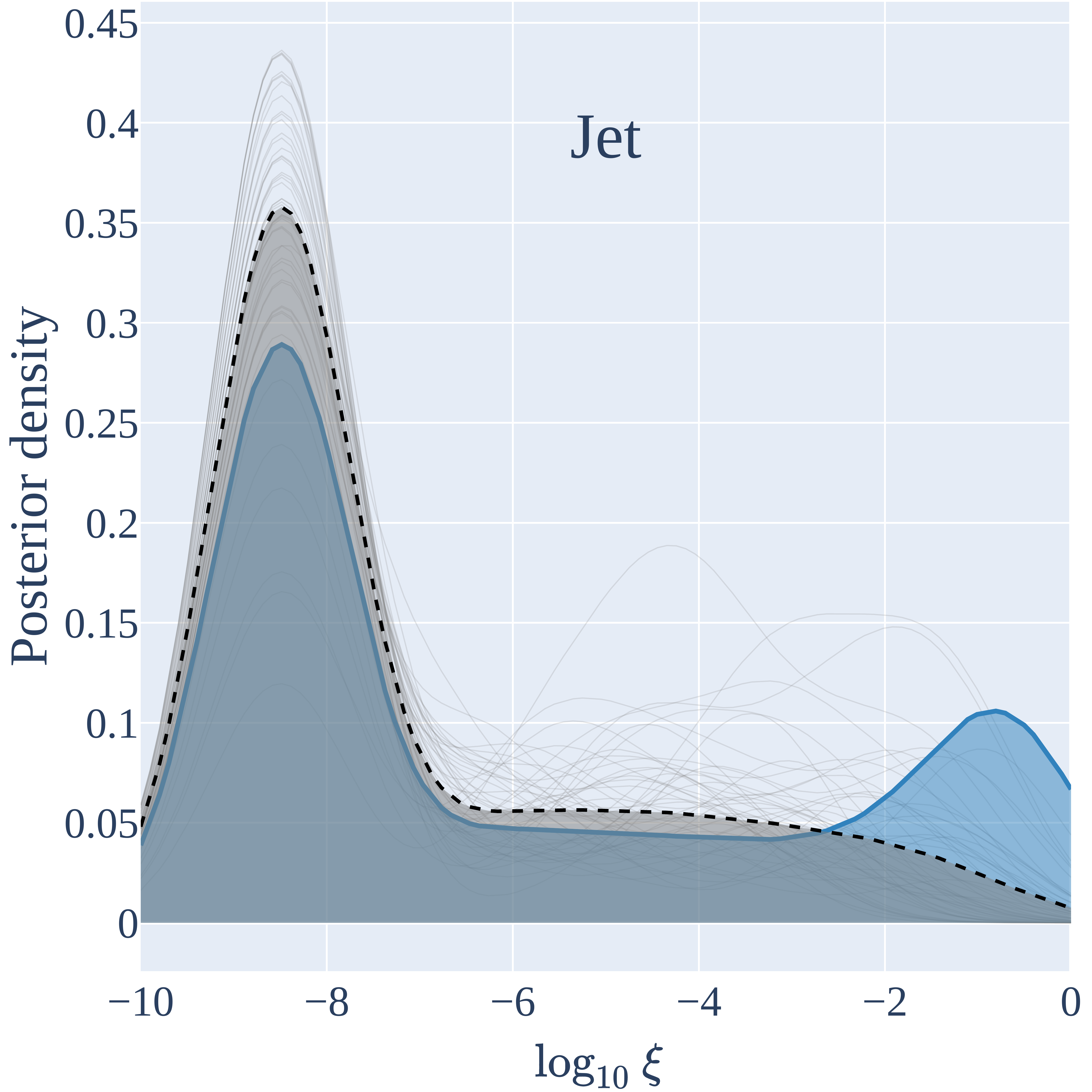}{0.33\textwidth}{(c)}
     \fig{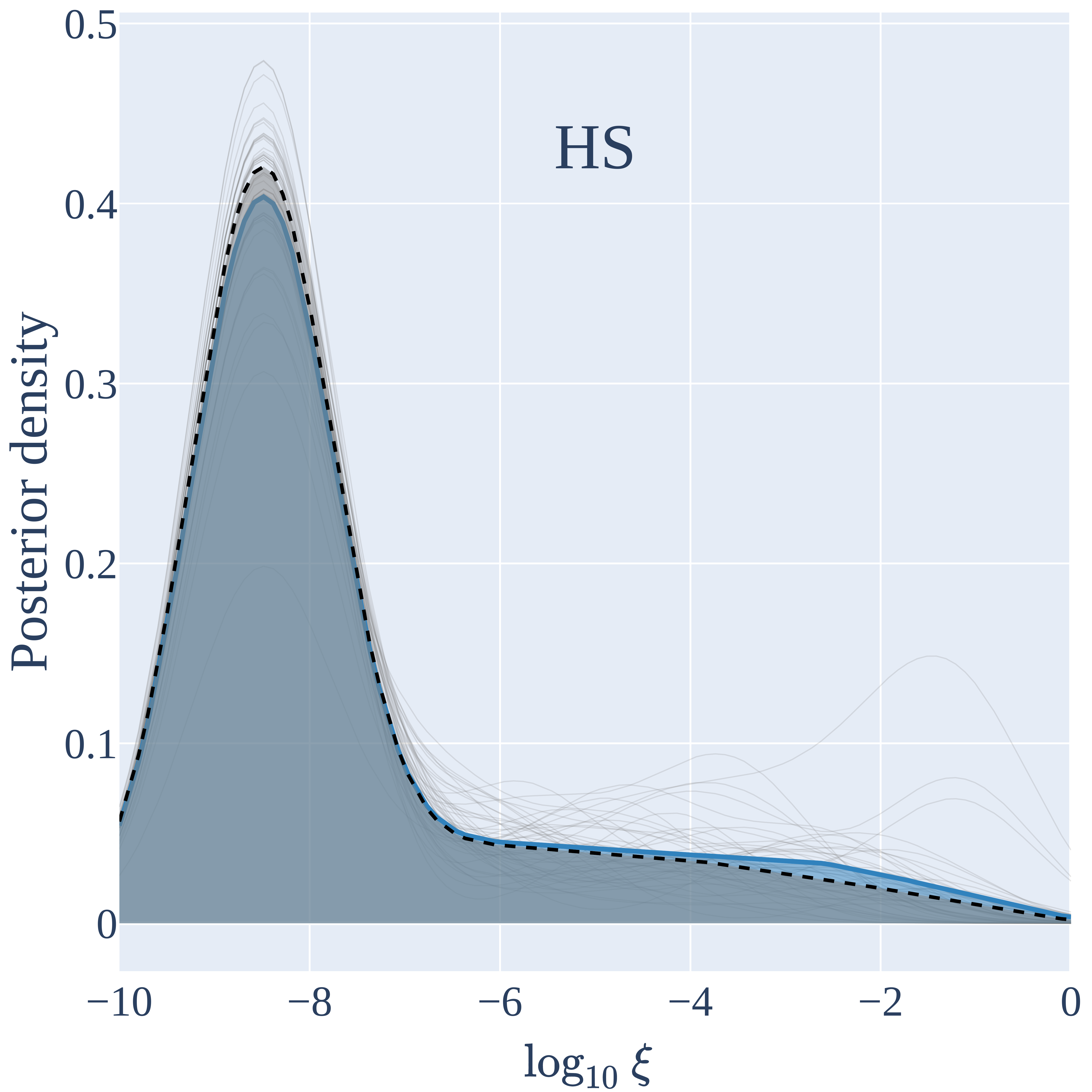}{0.33\textwidth}{(d)}
    }
    
\caption{Same as in Fig. \ref{fig:results_3C418}, with the VLA 4.8 GHz radio contours overlaid, for PKS 1402+044, a high-redshift quasar. \Radiocontours{0.8, 1.5, 8, 40}\label{fig:results_PKS1402+044}}
\end{figure*}
\textbf{PKS 1402+044} (Fig. \ref{fig:results_PKS1402+044}): This is a high-redshift quasar (z=3.2). We could not confirm (\textit{p}-value$\leq$0.04) any significant X-ray emission from the jet of this source, unlike the previous authors in \citet{schwartz2019relativistic}.

\begin{figure*}[h]
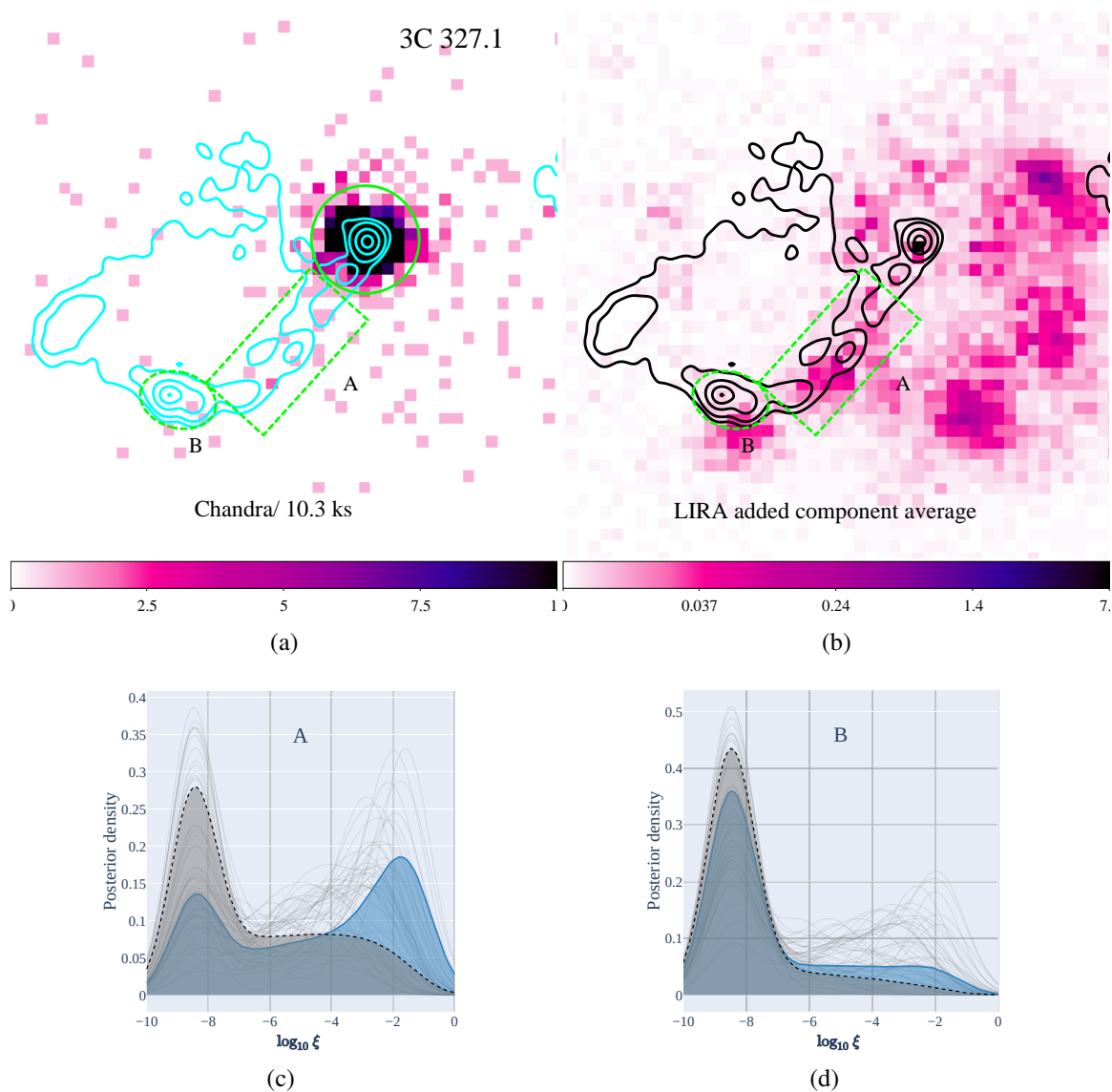

    \gridline{
        \fig{{3C327.1_lira_paper_count}.eps}{0.5\textwidth}{(a)}
        \fig{{3C327.1_lira_paper_add}.eps}{0.5\textwidth}{(b)}
    }
    \gridline{
    \fig{{3C327.1_A_lira_post}.pdf}{0.33\textwidth}{(c)}
     \fig{{3C327.1_B_lira_post}.pdf}{0.33\textwidth}{(d)}
    }
    
\caption{Same as in Fig. \ref{fig:results_3C418}, for 3C 327.1, a HERG. We did not measure the offset as there was no statistically significant emission from the chosen ROIs. \Radiocontours{0.5, 2, 10, 25, 40}\label{fig:results_3C327.1}}
\end{figure*}
\textbf{3C 327.1} (Fig. \ref{fig:results_3C327.1}): This is an FR-II source. Similar to PKS 1402+044, contrary to the previous claim in \citet{massaro2013chandra}, we could not confirm (\textit{p}-value$\leq$0.15)  the presence of any significant X-ray emission from its jet.

\section{Discussion\label{sec:discussion}}
\subsection{Offset Statistics}
In the majority (10 out of 13) of the clearly detected ($>$0.15$''$ separation) knot offsets, the X-ray centroid lies upstream of its corresponding radio peak (denoted as Xf or X-ray-first). Xf-offsets are also the majority (5 out of 6) amongst the tentative cases and are also the predominant type reported in the literature \citep[e.g.,][]{sambruna2004survey,2005MNRAS.360..926W,2007ApJ...657..145S,Clautice:2016zai,worrall2016x,Harris_2017,Marchenko_2017}. 
These offsets contradict an observational requirement of the commonly adopted IC/CMB models which expects the X-rays to either be coincident or extend beyond the radio \citep{2007RMxAC..27..188H,worrall2009x}.

In only three sources, Centaurus B, PKS~2123-463  and 4C+11.45, the X-ray centroid lies downstream of its radio peak (denoted as Rf or Radio-first), which are consistent with the IC/CMB interpretation. \citet{2011ApJ...729...26M} have previously noted such an offset in the jet of OJ 287, a BL Lac object. Recently,  \citet{Meyer_2019} have concluded that IC/CMB is its dominant X-ray emission mechanism. They measured the $\gamma$-ray flux from the jet and showed that it equals the level expected from the observed X-ray emission under IC/CMB \citep[see][]{georganopoulos2006quasar}. For Centaurus~B and PKS~2123-463, lack of observations in the optical prevents one from using the gamma-rays to test for steady IC/CMB from these source since the IR/optical is generally required to predict the gamma-ray flux level. However, the FR-II type morphology of PKS 2123-463 suggests that this is presumably an MSC-type jet and its Rf-offset may therefore indicate that IC/CMB is the dominant mechanism. We have also found significant extended X-ray emission between its inner knot and hotspot without any associated radio emission. This morphology combined with its high redshift also suggests that  IC/CMB dominates X-ray emission from the jet (see section \ref{subsec:iccmb} for details). In the case of 4C+11.45, the X-rays may instead be produced by a dense gas that deflects the jet while also getting shock heated by it (see Fig.~\ref{fig:results_4C+11.45}, where the extended nature of the jet-region emission is evident). Optical maps of 4C+11.45 presented in \citet{lehnert1999hubble} and \citet{heckman1991spatially} show large Ly$\alpha$ clouds around this offset and support the deflection scenario.  

Our sample contains five highly-superluminal sources with apparent jet speeds greater than 3.36, which are of interest because the high apparent jet speeds imply that the parsec-scale jet is at a small angle to the line-of-sight. In such sources, presuming no major change in orientation between the parsec and kpc scales, real offsets between the location of radio and X-ray emission might be more difficult to distinguish due to the extreme foreshortening of the jet. Of these five, we could not confirm the presence of X-ray emission from the jet in one source, PKS 1402+044. In the remaining sources we do detect an offset. Interestingly, three of them appear at an apparent bend in the jet (indicated with ``+Bnd" in column 5 of Table \ref{table:results}). In jets which are overall well-aligned, small deflections (changes in jet angle) can appear very extreme due to the effects of projection.
This may explain sources like TXS 0833+585 [$\beta_{app}=14.16$~\citet{britzen2008multi,2020arXiv200712661K}; Fig.~\ref{fig:results_TXS0833+575}], where the inner-most knot A in 
 shows only a small offset (0.07\as) while the outer knot B, which lies at an apparent bend, shows an Xf-offset (0.18\as). We can interpret these observations as a combined result of projection effects and limited instrumental resolution. The observed  $\beta_{app}$ limits the viewing angle of the pc-scale jet to at most \siml8\degree. Hence, for a small pc-to-kpc scale deflection, projection effects would shorten any offset in knot A by a factor of seven, and we may not resolve it. If both the knots contain offsets of similar magnitudes, then an offset in knot B suggests that the jet is more misaligned at this bend, thereby making the offset more apparent. 
 Two other scenarios are also possible: ({\sl i}) The offset in knot B is relatively higher (see section \ref{subsec:core-dist}), which may only require a small deflection in the jet. ({\sl ii}) Knot A may be intrinsically different from knot B and actually exhibit no offset. Distinguishing between these scenarios would require next-generation X-ray observatories with much higher resolution.

\begin{figure*}
    \gridline{
    \fig{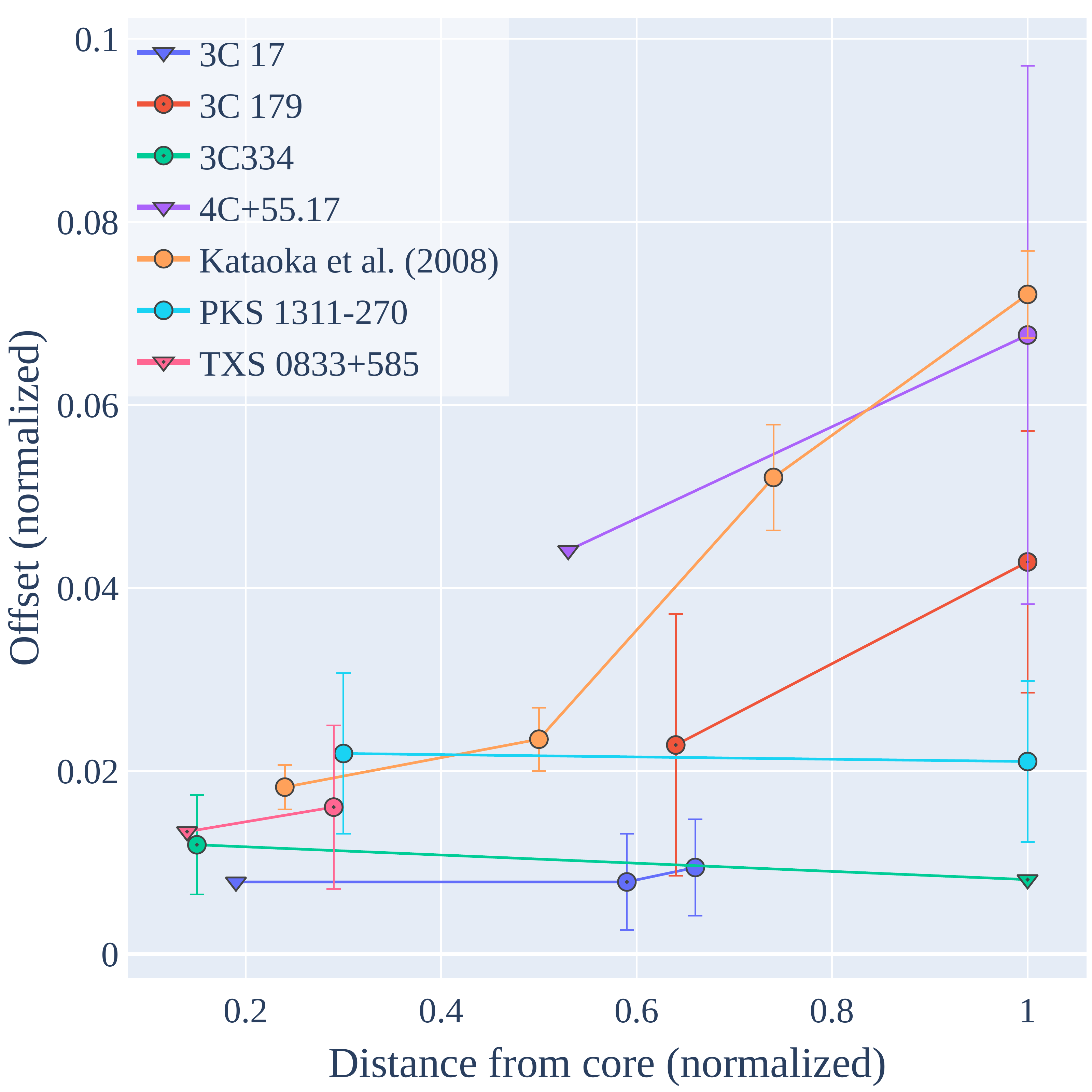}{0.5\textwidth}{(a)}
    \fig{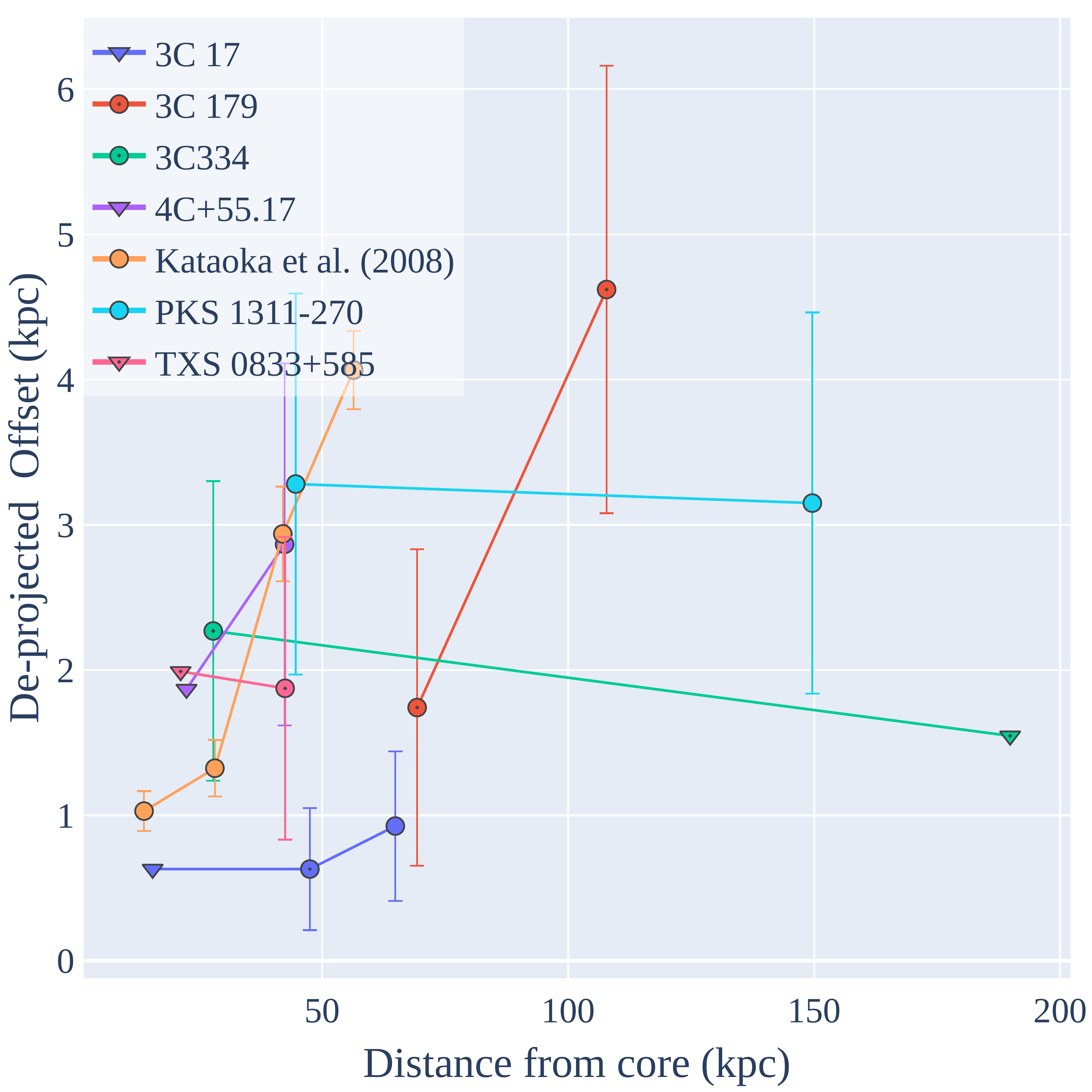}{0.5\textwidth}{(b)}
    }
    \gridline{
    \fig{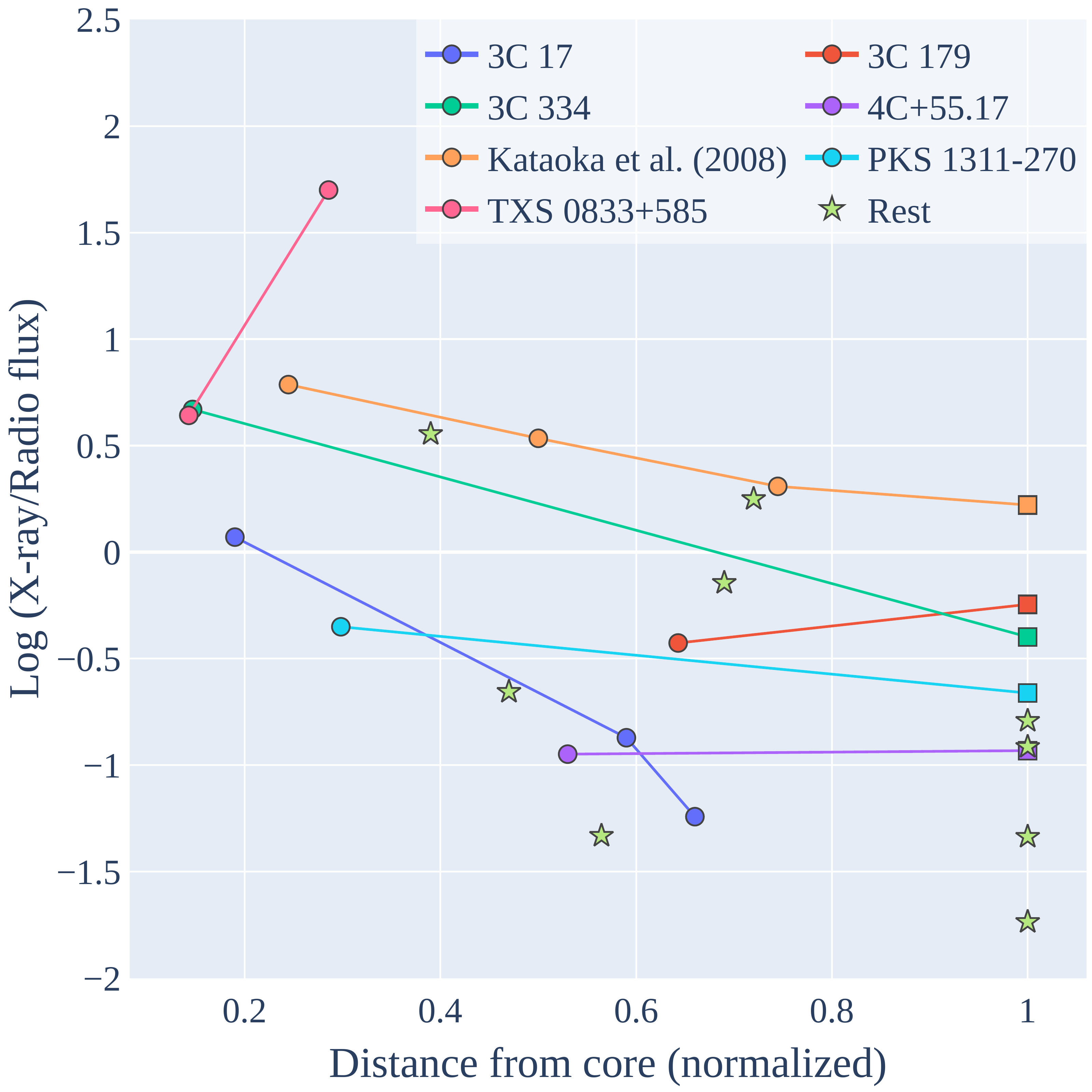}{0.5\textwidth}{(c)}
    \fig{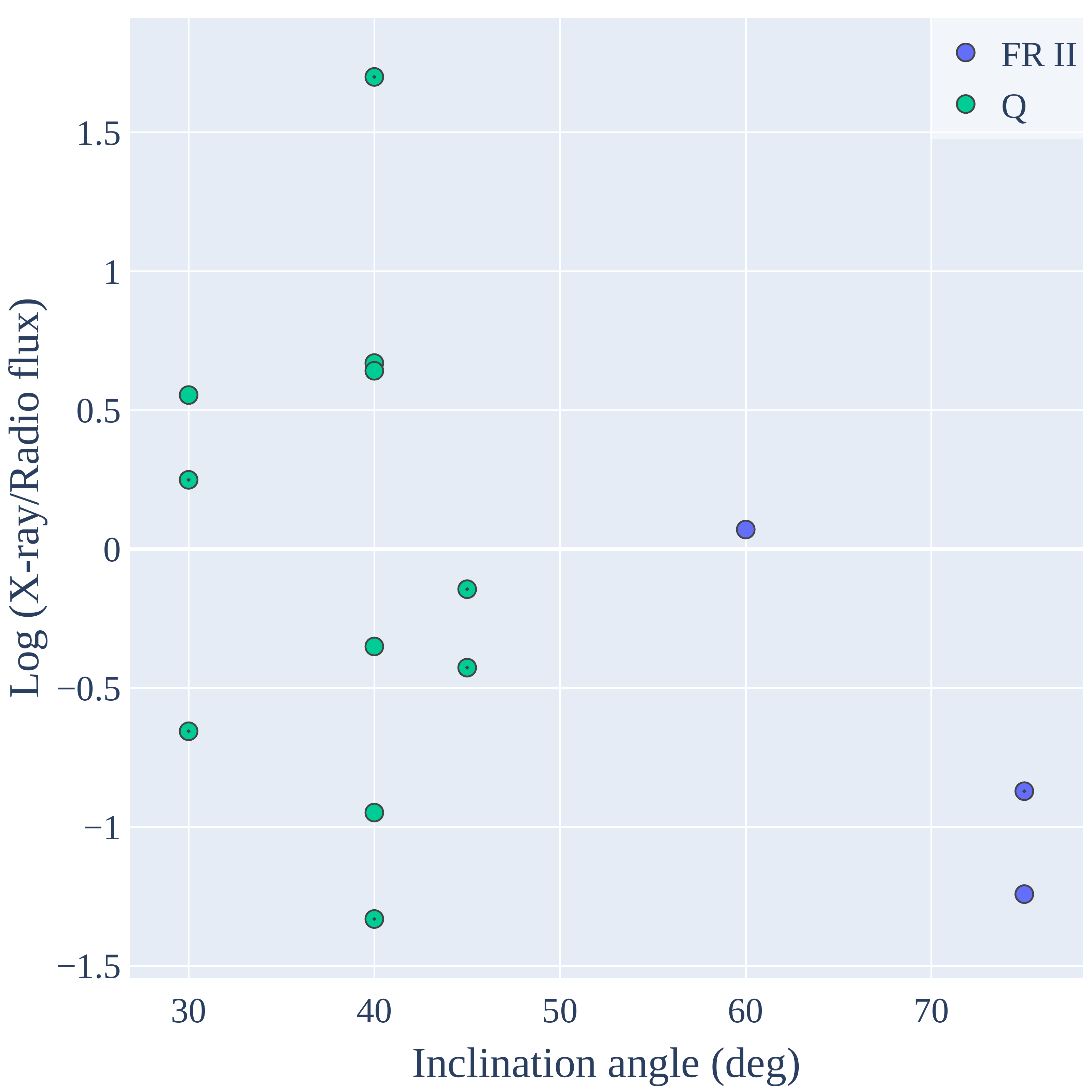}{0.5\textwidth}{(d)}
    }
    
    \caption{Offsets and flux ratios plotted against various parameters. (a) Normalized offset is plotted against normalized distance from core for selected jets. (b)  Same as (a) but with de-projected values. In (a) and (b) dot indicates a bent jet and the error bars are fixed at 0.1\as; Triangles denote knots with tentative offsets and are fixed at 0.15\as (upper limit). See the text for details on the angles used and associated uncertainties. (c) Flux ratio is plotted against normalized distance from core; squares indicate hotspots and circles knots, and lines indicate selected jets. Stars indicate features with only one X-ray feature in the jet. (d) Flux ratio is plotted against the inclination angle of the jet.}

    \label{fig:deproj_offsets}
\end{figure*}

\subsection{De-projecting the offsets}
Table \ref{table:results} provides lower-limits for the true offsets that are also projected onto the plane of the sky. Understanding the nature of these offsets would first require de-projecting them, which in turn requires the alignment angle of the jet. Although a few methods exist in the literature to estimate the alignment angle, for instance, from the ratio of core to lobe luminosities \citep[e.g.,][]{drouart2012jet,marin2016robust}, they only apply for specific redshift ranges and have large uncertainties for closely aligned sources. Because our sample covers a wide redshift range and contains superluminal sources, to start, we assigned representative upper-limits to the orientation angles based on their class. Adopting the mean angle estimates from the unification scheme proposed in \citet{urry1995unified}, we assigned an angle of 15\degree~ to the CDQ class, 30\degree~for LDQ, 60\degree~for BLRG and 90\degree~for NLRG. The choice of 60\degree~ for BLRG type sources is consistent with the recent estimates of the half-opening angle of the torus  \citep{2015MNRAS.454.1202S,marin2016robust}. When no sub-classification was available for a quasar, we assigned it a value of 40\degree~. For jets with apparent bends larger than 20\degree, we assumed an additional 15\degree~ increment.  Assuming the maximum possible non-disruptive bend in a relativistic jet is $\approx 49$\degree~ \citep{mendoza2002}, even small orientation effects are sufficient to produce large projected bends. Hence this increment may over-estimate the angle for closely aligned jets (e.g., quasars).

\subsection{De-projected offset versus distance from core\label{subsec:core-dist} }
We can also examine if and how the offsets are related to other properties of the jet. In a detailed study on 3C 353, a nearby FR-II jet, \citep{kataoka2008chandra} have found that the offsets in knots and hotspots increased with distance from the core. They also argued that 3C 353 is a representative FR-II source, and one can expect such a progression in other similar X-ray jets.
We searched for similar trends in our sample by selecting jets with at least two X-ray components on one side of the jet. Besides using the de-projected values, we also normalized offsets and distances by dividing them with the total length of the jet (measured from core to hotspot, along the jet). The normalized distance is independent of the alignment angle of the jet. For bent jets, this normalization would only modify the rate of increase would  and not the increasing trend itself.

Figure \ref{fig:deproj_offsets}a shows normalized offset plotted against normalized distance from core and 
 Figure  \ref{fig:deproj_offsets}b shows its de-projected version. We also added data from \citet{kataoka2008chandra} to both the figures for comparison. Triangles denote knots with tentative Xf-offsets and are fixed at 0.15\as (upperlimit). Dotted markers indicate a bent jet. In three of the six sources, the offsets increase with distance from core, consistent with 3C 353. The offsets in two of the remaining sources stay approximately constant.
 Only in 3C 334, a knot in the inner jet shows an Xf-offset while a knot at a bend that disrupts the jet shows no offset. Given the errors in measuring the offsets,
 further studies on a larger sample are necessary to establish the presence or absence of any trend between offsets and distance from core.  

\subsection{X-ray-to-radio Flux ratio\label{subsec:flux_ratio}}
We measured the X-ray-to-radio flux ratio ($\nu_XF_X/\nu_RF_R$; flux ratio, hereafter)  by evaluating the X-ray flux at 1 keV. We chose 4.8 GHz for the radio based on its availability for most sources, except in PKS 1046-409 and 3C 179, where we used the fluxes at 19 GHz and 8.4 GHz, respectively.
The flux ratio is known to systematically decrease along the jet in many cases \citep[e.g.,][]{sambruna2004survey,2007ApJ...657..145S,kataoka2008chandra}. It is possible that, within the IC/CMB model, the jet decelerates on kiloparsec-megaparsec scales via mass entrainment \citep{georganopoulos2004witnessing} and produces a declining flux ratio. However, \citet{hardcastle2006testing} found no significant trend for decelerating jets and noted that the cold matter required for entrainment is sometimes unreasonably high. \citet{kataoka2008chandra} have speculated that, within the synchrotron model, changing plasma conditions along the jet instead causes a gradual decline in the X-ray emission.
 
In our sample, we also notice a slight trend where the flux ratio tends to decrease with normalized distance from core (Fig. \ref{fig:deproj_offsets}c), except in TXS 0833+585, where the opposite happens (see section \ref{subsec:iccmb} for details). Correlation tests revealed a significant correlation between the two variables (for the entire sample), with Pearson's $R$=-0.47 (\textit{p}-value=0.018) and Spearman's $\rho$=-0.48 (\textit{p}-value=0.015). Although this trend is consistent with IC/CMB in a decelerating jet, it does not necessarily imply that IC/CMB dominates X-ray emissions from all the jets. We might instead, for instance, speculate that a faster spine up-scatters radio emission from a slower enclosing-sheath to X-ray wavelengths, and when combined with a decelerating jet, can also reproduce the observed trend. Furthermore, the IC/CMB model expects the flux ratio to decrease with increasing angle because the X-rays will be de-beamed faster than the radio. However, we did not find any significant evidence for such a trend (Fig.~\ref{fig:deproj_offsets}d) in our small sample, suggesting a synchrotron origin for the X-rays. This result is consistent with the findings in \citet{2011ApJS..197...24M} where the authors reported no difference between the flux-ratio distributions of knots in quasars (smaller alignment angles) and FR-IIs (larger alignment angles). We also tested for a correlation between offsets and flux ratio but found none.

\subsection{High redshift jets and IC/CMB\label{subsec:iccmb}}
While it seems increasingly unlikely that lower-redshift MSC jets have X-rays dominated by the IC/CMB process, this is not true of high-redshift jets, especially cases where X-rays have been observed with little to no corresponding radio emission \cite[e.g.,][]{yuan2003extended,Cheung_2012,Simionescu_2016,McKeough_2016,worrall2020inverse}. In light of this, we identified three high-redshift sources in our sample for further scrutiny.
In TXS 0833+588 (z=2.1), the flux ratio increases along the jet, contrary to the general decreasing trend. Moreover, of the two knots in this jet, the knot farther away from the core appears at an apparent bend. Because under the IC/CMB model, the X-rays are more beamed than the radio, stronger brightening in the X-rays at this point could be attributed to a small decrease in orientation angle at the bend. However, IC/CMB requires unreasonable kpc scale speeds to produce the observed X-rays in this knot. \citep{2018ApJ...856...66M}. Interestingly, the Xf-offset in this knot also argues against IC/CMB while the Rf-offset (tentative) in the inner knot is consistent with it. Hence, there may be multiple X-ray emission processes in the outer knot (e.g., an intrinsically brighter component), which, however, cannot be confirmed with the available data.

In the two other high-redshift jets, PKS 2123-463 (z=1.67; Fig. \ref{fig:results_pks2123-463}) and 4C+25.21 (z=2.69; Fig. \ref{fig:results_4C+25.21}), it appears that X-ray emission may be present along portions of the jet between the core and the hotspot that lacks significant radio emission. \citet{Simionescu_2016} have observed a similar morphology where the X-ray jet in B3 0727+409 (z=2.5) lacked a radio counterpart. They found that the existing data favours IC/CMB over the synchrotron interpretation. Therefore, although we only selected regions with well-defined radio emission to detect offsets, for PKS 2123-463 and 4C+25.21, we also analyzed regions without any radio emission and confirmed the presence of X-ray emission at a high significance. In both cases the X-ray emission in these regions appears to extend past the radio, a general characteristic of the IC/CMB interpretation \citep{2007RMxAC..27..188H,worrall2009x}. However, the situation is slightly ambiguous in the case of 4C+25.21 as the detected X-ray emission not only lies downstream of a bright inner jet but also upstream of a much fainter radio peak near the hotspot. These results further support the idea of a new class of high-redshift X-ray jets that lack radio emission \citep{Simionescu_2016}. However, it should be noted that such a morphology can be interpreted in an alternate way: instead of the jet, the lobes may also emit the observed X-rays via IC/CMB \citep[e.g.,][]{2008MNRAS.386.1774E,wu2017cmb}. In this case the low-frequency radio spectral index of the lobes and the X-ray emission are expected to match. However, it is impossible to distinguish between the two scenarios with the available data and deeper radio and X-ray observations are necessary.

\section{Summary\label{sec:conclusions}}
We have demonstrated a method to detect X-ray-to-radio offsets from low-count X-ray images using a powerful statistical tool called LIRA \citep{esch2004image,stein2015detecting}. Using a deep observation of PKS 0605--08, we showed that our method not only can accurately detect an offset under the low-count regime but also can provide an associated error.%
. Our analysis of 37 features in 22 jets detected by \emph{Chandra} in the low-count (\textless20) regime has revealed a significant number of `X-ray first' offsets -- 16 are clear detections ($>0.15''$ separation) while 5 are tenative ($0.05-0.15''$ separation). These greatly outnumber the number of Radio-first detections (3) or tentative detections (1), and only one jet/feature appears completely co-spatial.
We adopted a representative orientation scheme for the jets to de-project the offsets. In most of our sources, the offsets tend to increase with distance from core while the flux ratio decreases. We also identified three potential high-redshift jets where the X-ray emission may be dominated by the IC/CMB process. In two of these, we find evidence of X-ray emission without corresponding radio emission; in the third one, the flux ratio increases with distance from the core. Future work with a larger sample of X-ray detected jets will be needed to confirm the trends found here.

\acknowledgments
We thank Dr. Vinay L. Kashyap for valuable suggestions on computing the offsets using output images from LIRA. We acknowledge financial support from the National Science Foundation under Grant No. 1714380.

The scientific results reported in this article are based in part on observations made by the Chandra X-ray Observatory and data obtained from the Chandra Data Archive. This research has made use of software provided by the Chandra X-ray Center (CXC) in the application packages CIAO, ChIPS, and Sherpa. The National Radio Astronomy Observatory is a facility of the National Science Foundation operated under cooperative agreement by Associated Universities, Inc. The Australia Telescope Compact Array is part of the Australia Telescope National Facility which is funded by the Australian Government for operation as a National Facility managed by CSIRO. We acknowledge the Gomeroi people as the traditional owners of the Observatory site. This research is based on observations made with the NASA/ESA Hubble Space Telescope, obtained from the data archive at the Space Telescope Science Institute. STScI is operated by the Association of Universities for Research in Astronomy, Inc. under NASA contract NAS 5-26555. 

\facilities{VLA, EVLA, HST (STIS, WFPC2, WFPC3), CXO, ATCA}
\added{\software{Astrodrizzle (Hack et al. 2012), pyRAF (Science Software Branch at STScI 2012), CIAO (Fruscione et al. 2006), Sherpa (Freeman et al. 2001, Doe et al. 2007, Burke et al. 2020), ChiPS (Germain et al. 2006), MIRIAD (Sault et al. 1995), CASA (McMullin et al. 2007), ds9 (Joye et al. 2003), LIRA (Esch et al. 2004, Connors et al. 2007, Connors et al. 2011, Stein et al. 2015)}}

\clearpage

\clearpage
\appendix
\section{Optical/IR images and SEDs}
\begin{figure*}[h]
    \gridline{
        \fig{{3C6.1_WFPC2_PC1_F555W_lira_paper}.eps}{0.5\textwidth}{(a)}
        \fig{{3C6.1_NHS+SHS_SED_lira_paper}.pdf}{0.5\textwidth}{(b)}
    }
    \caption{Results for the optical and spectral analysis of 3C 6.1. (a) The HST Wide Field Planeraty Camera 2 (WFCP2) visible image at 5.5 $\times10^{14}$ GHz (5439 \AA) with the VLA 8.6 GHz radio contours overlaid. The left and right panes show NHS and SHS, respectively. Both the hotspots are not detected at this wavelength. (b) The broadband SEDs of NHS and SHS which indicate that X-rays require a second spectral component. \label{fig:hst_3C6.1}}
\end{figure*}

\begin{figure*}
    \gridline{
        \fig{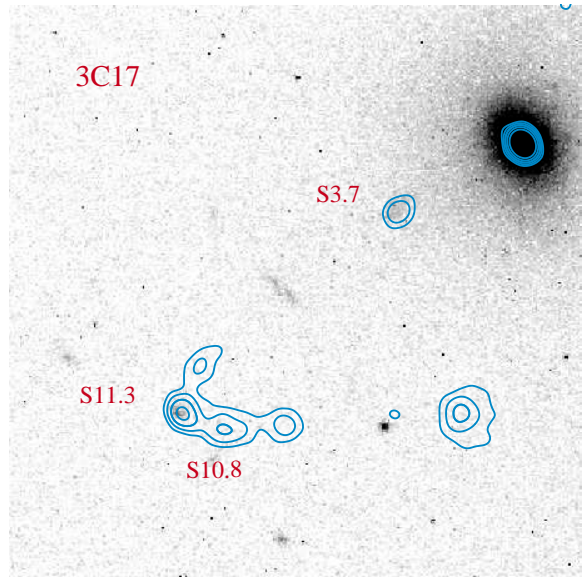}{0.5\textwidth}{(a)}
    }
    \caption{The HST Space Telescope Imaging Spectrograph (STIS) visible image at 4.16 $\times$ 10$^{14}$ Hz (7216 \AA) of 3C17 overlaid with the VLA 4.86 GHz radio contours. Knots S3.7 and S11.3 are detected in the optical.\label{fig:hst_3C17}}
\end{figure*} 

\begin{figure*}
    \gridline{
        \fig{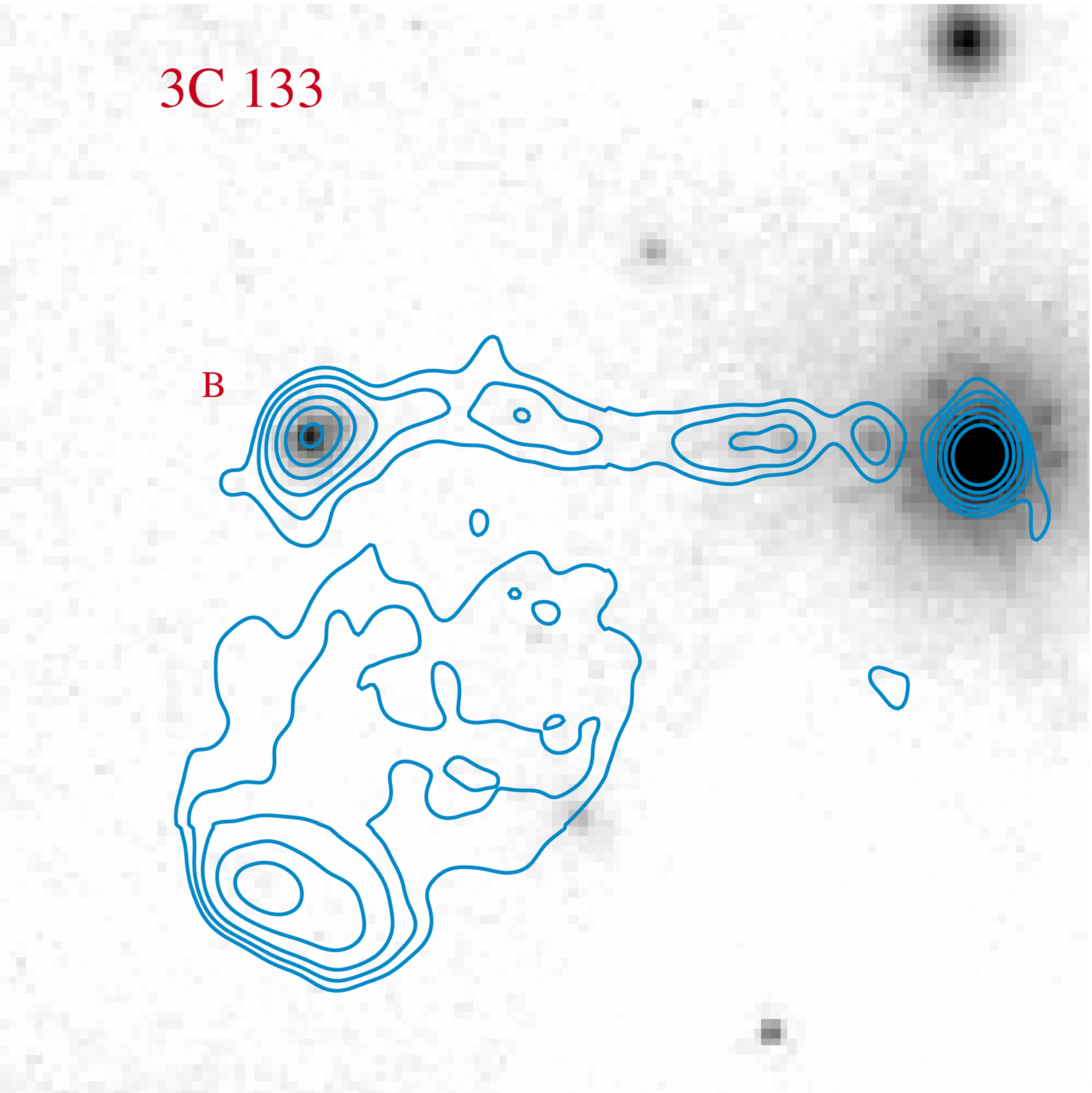}{0.5\textwidth}{(a)}
        \fig{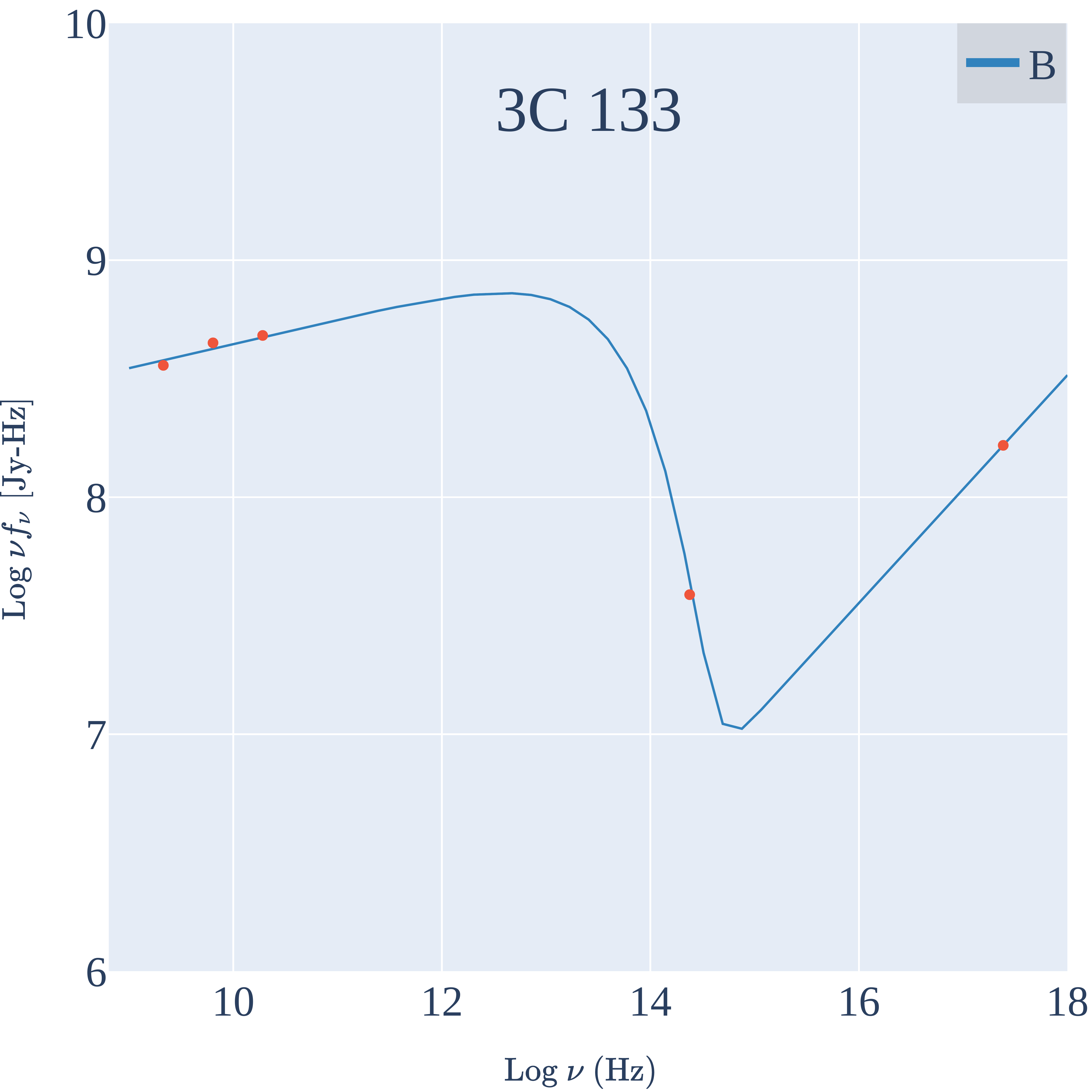}{0.5\textwidth}{(b)}
    }
    \caption{Results for the optical and spectral analysis of 3C 133. (a) The HST Near Infrared Camera and Multi-Object Spectrometer-2 (NICMOS-2) infrared image (IR) at 1.87 $\times10^{14}$ GHz (16030 \AA) with the VLA 4.86 GHz radio contours overlaid. (b) The broadband SED of knot B which indicates that the X-ray emission requires a separate spectral component.\label{ref:hst_3C133}}
\end{figure*} 

\begin{figure*}
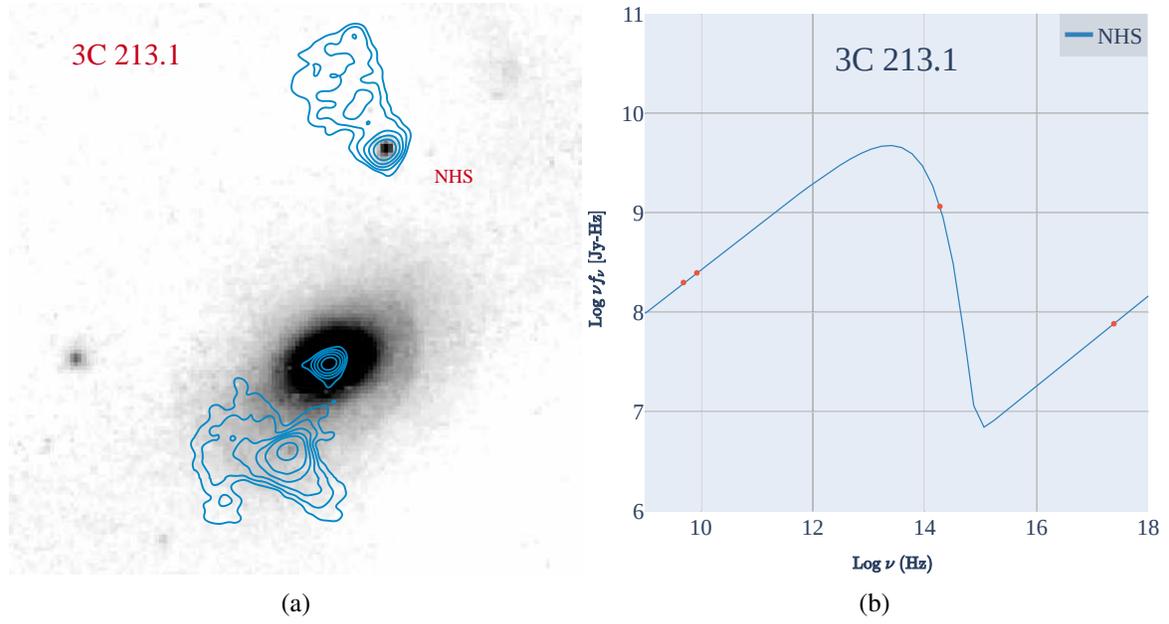

    \gridline{
        \fig{{3C213.1_hst_nic2_f1602w_lira_paper}.eps}{0.5\textwidth}{(a)}
        \fig{{3C213.1_NHS_SED_lira_paper}.pdf}{0.5\textwidth}{(b)}
    }
    \caption{Results for the optical and spectral analysis of 3C 213.1. (a)The HST NICMOS-2 infrared image at 1.87 $\times10^{14}$ GHz (16030 \AA) with the VLA 8.6 GHz radio contours overlaid. (b) The broadband SED of NHS which indicates that X-rays require a second spectral component.\label{fig:hst_3C213.1}}
\end{figure*} 

\begin{figure*}
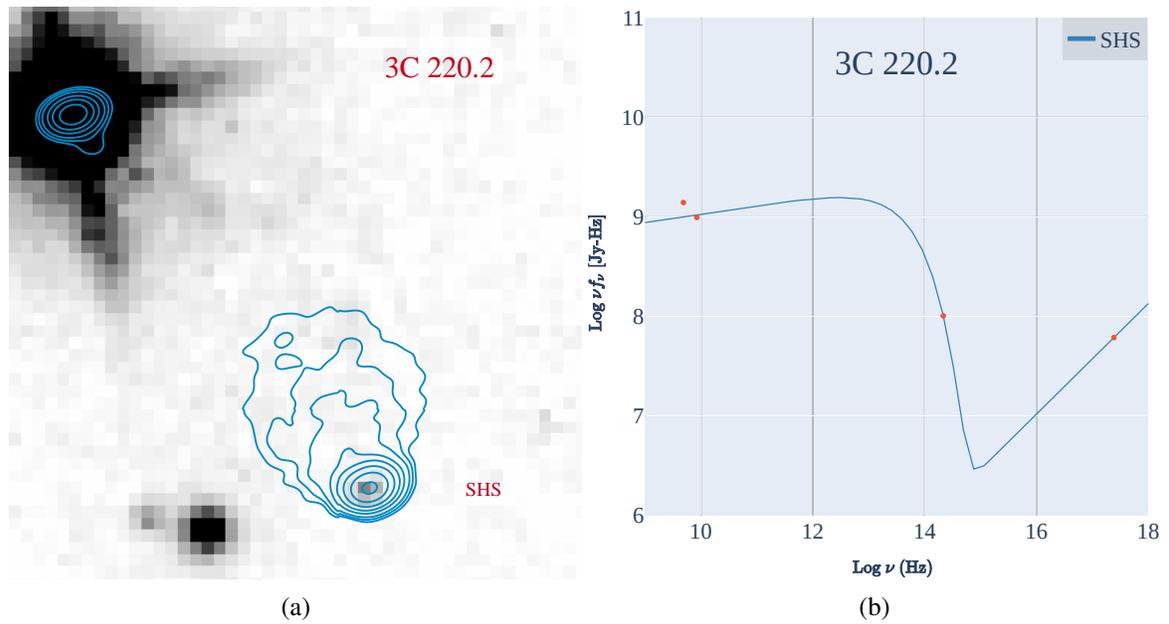

    \gridline{
        \fig{{3C220.2_WFC3_IR_F140W_lira_paper}.eps}{0.5\textwidth}{(a)}
        \fig{{3C220.2_SHS_SED_lira_paper}.pdf}{0.5\textwidth}{(b)}
    }
    \caption{Results for the optical and spectral analysis of 3C 220.2. (a) The HST Wide Field Camera-3 (WFC3) IR image at 2.15 $\times10^{14}$ GHz (13922 \AA) with the VLA 8.4 GHz radio contours overlaid. (b) The broadband SED of SHS which indicates that the X-rays require a separate spectral component.\label{fig:hst_3C220.2}}
\end{figure*} 

\begin{figure*}
    \gridline{
        \fig{{3C275.1_WFPC2_PC1_F675W_lira_paper}.eps}{0.5\textwidth}{(a)}
        \fig{{3C275.1_knotA_SED_lira_paper}.pdf}{0.5\textwidth}{(b)}
    }
    \caption{Results for the optical and spectral analysis of 3C 275.1. (a) The HST WFPC2 visible image at 4.46 $\times10^{14}$ GHz (6717 \AA) with the 8.4 GHz radio contours overlaid. Knot A is not detected at this wavelength. (b) The broadband SED of knot A which indicates that X-rays requires a second sepctral component.\label{fig:hst_3C275.1}}
\end{figure*} 

\begin{figure*}
    \gridline{
        \fig{{3C280.1_WFC3_F140W_lira_paper}.eps}{0.5\textwidth}{(a)}
        \fig{{3C280.1_D_SED_lira_paper}.pdf}{0.5\textwidth}{(b)}
    }
    \caption{Results for the optical and spectral analysis of 3C 280.1. (a) The HST WFPC3 IR at 2.15 $\times10^{14}$ GHz (13922 \AA) with the 4.8 GHz radio contours overlaid. Knot D is not detected at the IR wavelength. (b) The broadband SED of knot D where a two-component phenomenological fit was not possible. However, a single component is fit to the radio data which clearly cannot explain the observed X-ray emission. \label{fig:hst_3C280.1}}
\end{figure*} 

\begin{figure*}
    \gridline{
        \fig{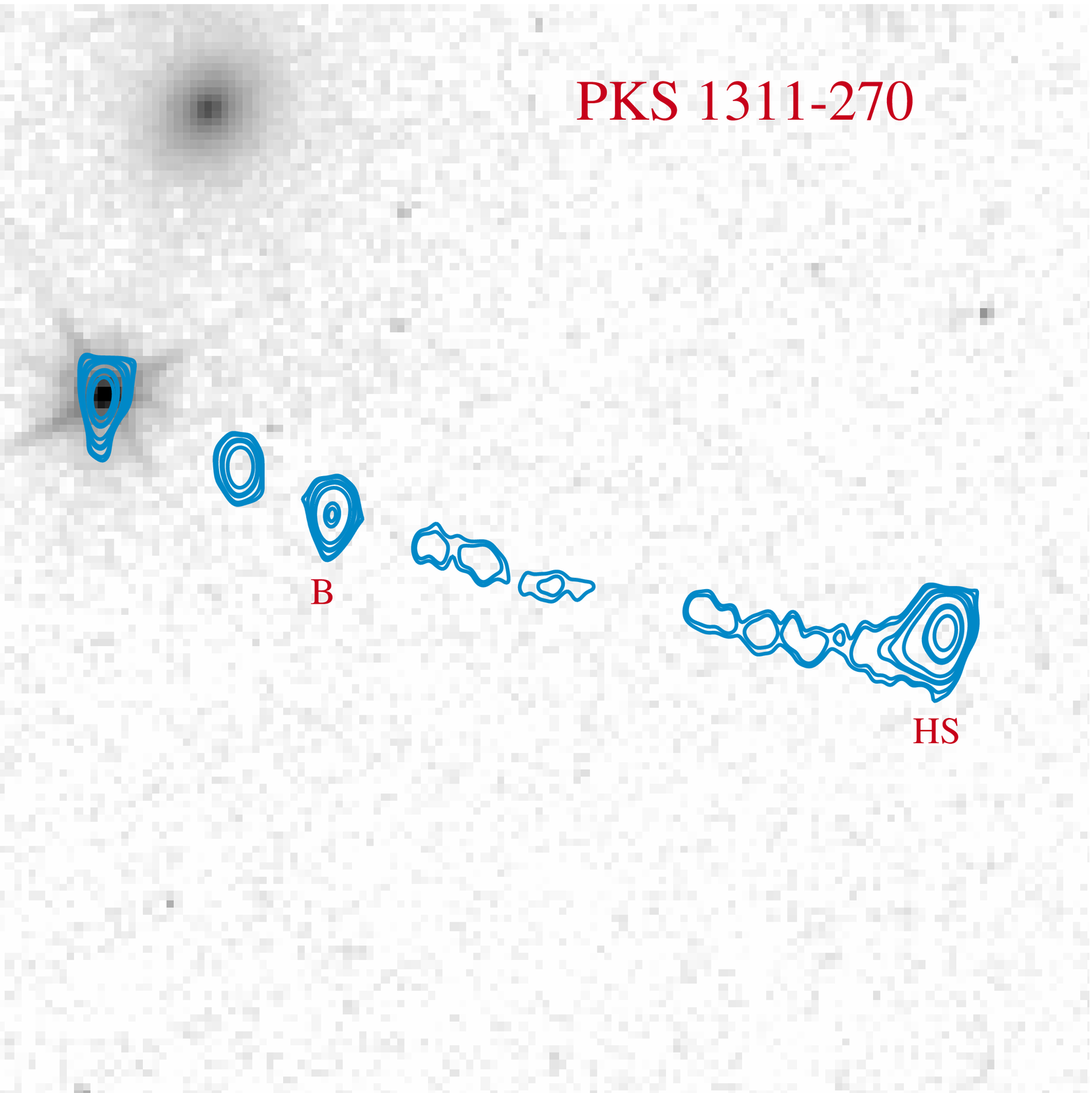}{0.5\textwidth}{(a)}
        \fig{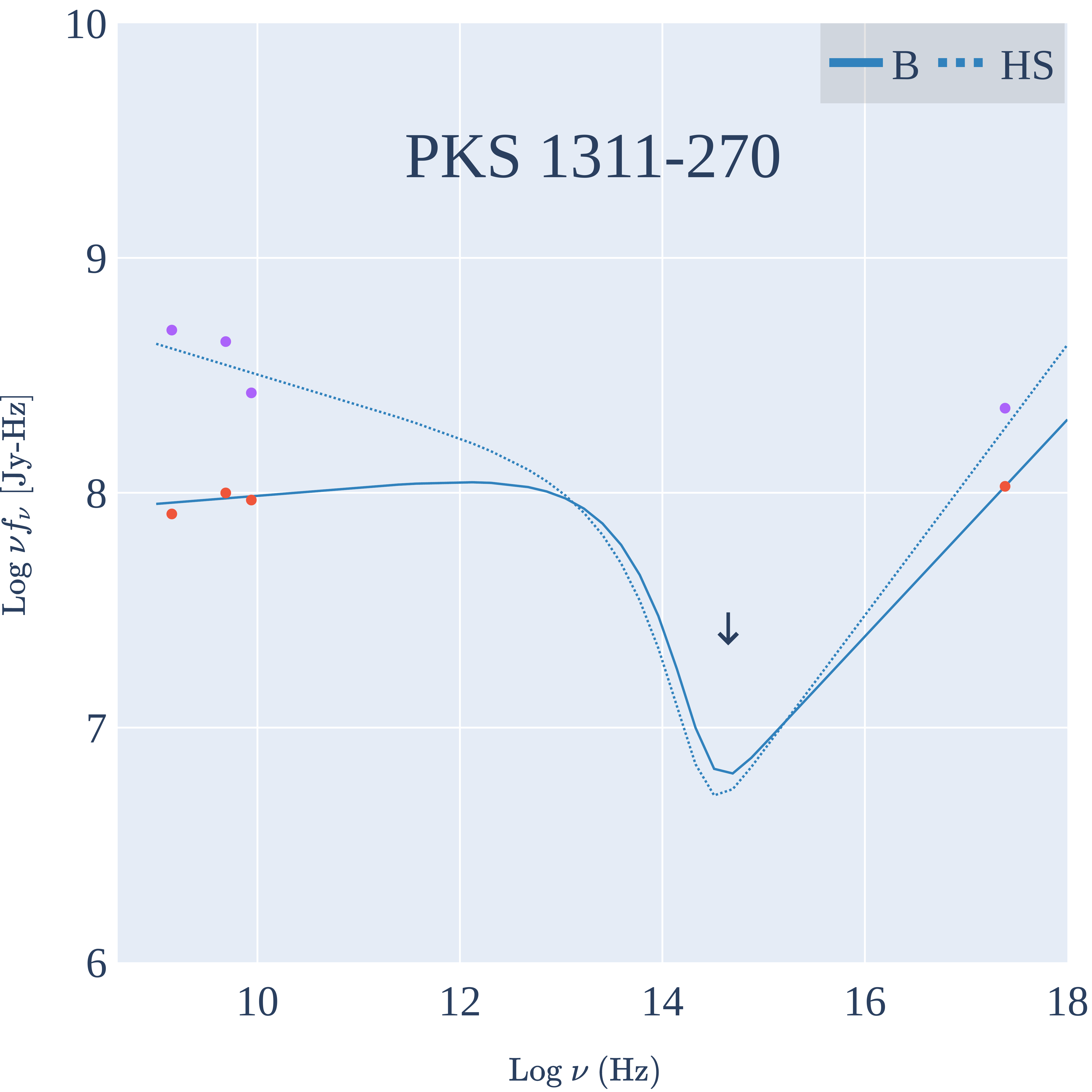}{0.5\textwidth}{(b)}
    }
    \caption{Results for the optical and spectral analysis of PKS 1311-270. (a) The HST WFPC2 visible image at 4.46 $\times10^{14}$ GHz (6717 \AA) with the 8.4 GHz radio contours overlaid. Knot B and HS are not detected at this wavelength. (b) The broadband SEDs of B and HS which indicate that the X-ray emission from both the components requires a second spectral component.\label{fig:hst_PKS1311-270}}
\end{figure*} 

\begin{figure*}
    \gridline{
        \fig{{4C+11.45_WFPC2_PC1_F555W_lira_paper}.eps}{0.5\textwidth}{(a)}
        \fig{{4C+11.45_knotC_SED_lira_paper}.pdf}{0.5\textwidth}{(b)}
    }
    \caption{Results for the optical and spectral analysis of 4C+11.45. (a) The HST WFPC2 visible image at 5.5 $\times10^{14}$ GHz (5439 \AA) with the VLA 8.4 GHz radio contours overlaid. Knot C is not detected at this wavelength. (b) The broadband SED of knot C which indicates that X-rays requires a second spectral component.\label{fig:hst_4C+11.45}}
\end{figure*} 

\begin{figure*}
    \gridline{
        \fig{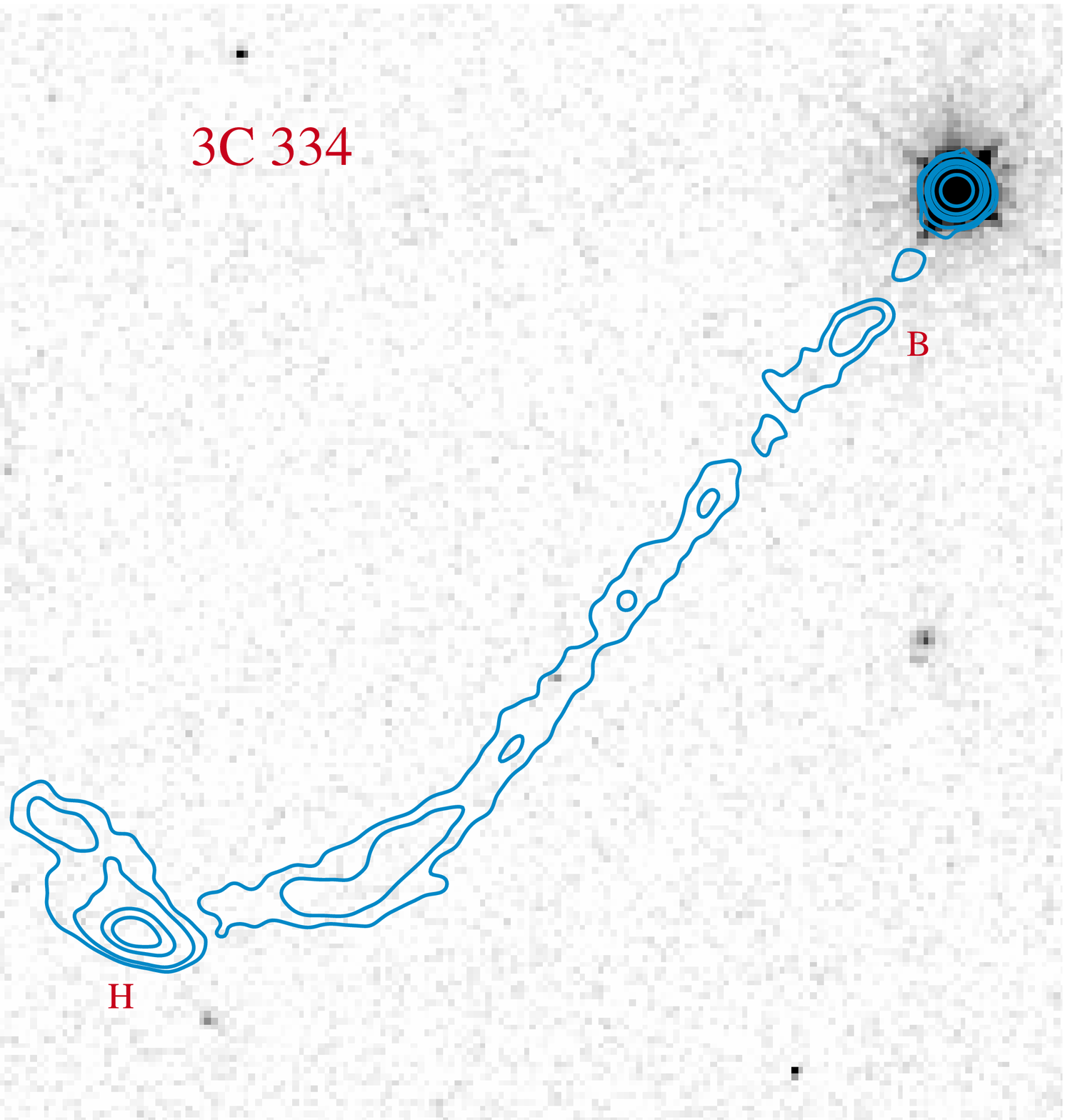}{0.5\textwidth}{(a)}
        \fig{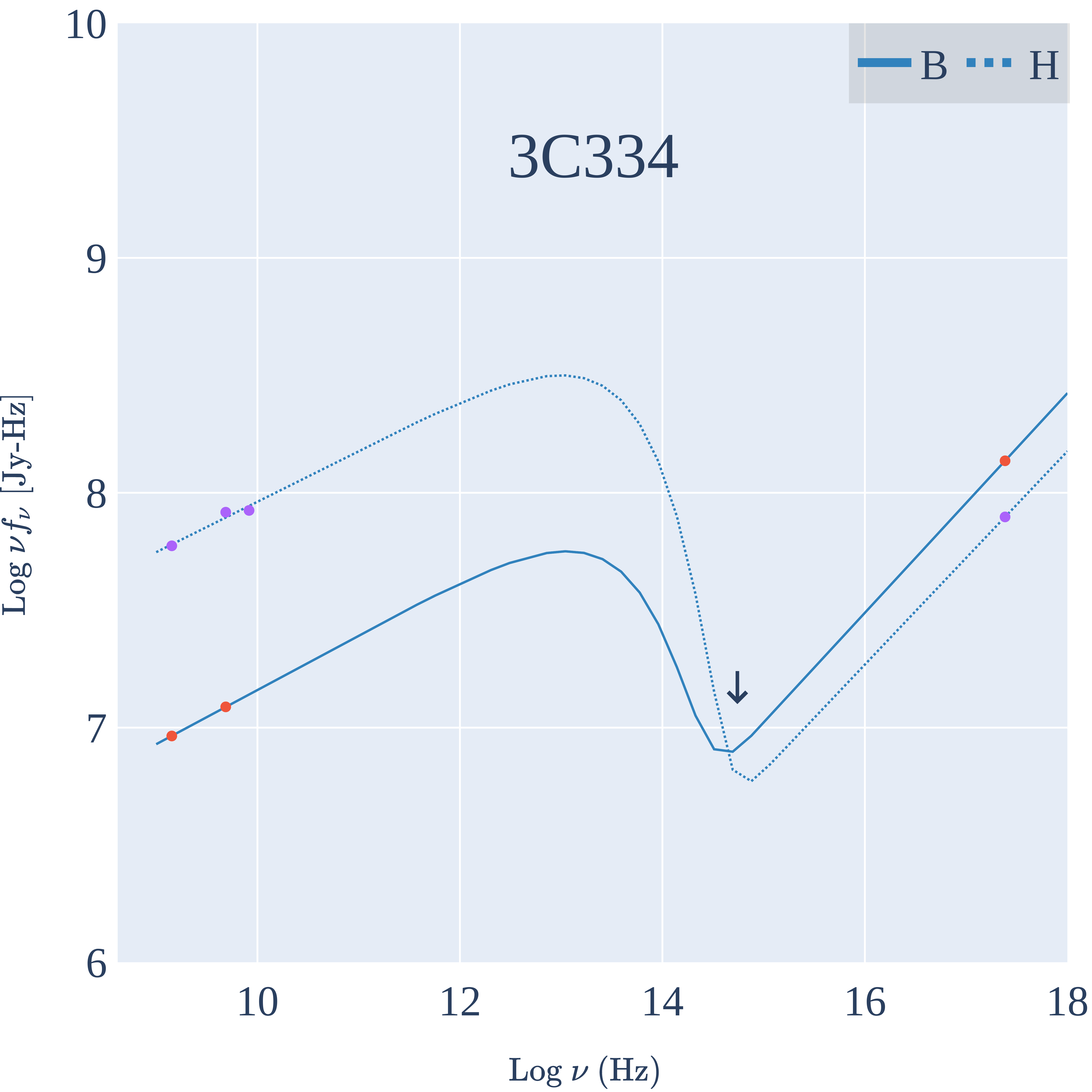}{0.5\textwidth}{(b)}
    }
    \caption{Results for the optical and spectral analysis of 3C 334. (a) The HST WFPC2 visible image at 5.5 $\times10^{14}$ GHz (5439 \AA) with the VLA 8.4 GHz radio contours overlaid. Knots B and H are not detected at this wavelength. (b) The broadband SEDs of knots B and H which indicates that X-rays from both the components requires a second spectral component.\label{fig:hst_3c334}}
\end{figure*}

\clearpage
\section{List of sources with low-count \textit{Chandra} observations}
\startlongtable
\begin{deluxetable*}{lcccchchll}

  \tablecaption{Summary of \textit{Chandra} detected low-count X-ray jets\label{table:sample}}
  \tabletypesize{\scriptsize}
  \tablehead{
    \colhead{Name\textsuperscript{a}} &
    \colhead{IAU Name\textsuperscript{b}} &
    \multicolumn{1}{p{1.5cm}}{\centering R.A. (J2000)\textsuperscript{c}\\hh:mm:ss} &
    \multicolumn{1}{p{1.5cm}}{\centering decl. (J2000)\textsuperscript{c}\\dd:mm:ss} &
    \colhead{z\textsuperscript{d}} &
    \nocolhead{$D_L$} &
    \multicolumn{1}{p{0.75cm}}{\centering Scale\textsuperscript{e}\\(kpc/\arcsec) } &
    \nocolhead{$\beta_{app}$\textsuperscript{e}}&
    \colhead{Class\textsuperscript{f}}&
    \multicolumn{1}{p{3cm}}{\centering Jet detection\textsuperscript{g}\\paper }
  }
  \tablecolumns{9}
  \startdata
  \cutinhead{Sources used for analysis}
  \object{3C 6.1} & 0013+790 & 00:16:31.147 & +79:16:49.88	 & 0.8400 & 5342 & 7.84 & &FR II (NLRG) &  \citet{2004ApJ...612..729H} \\
  \object{3C 17} & 0035-024 & 00:38:20.520	 & -02:07:40.72	 & 0.2200 & 1081 & 3.64 & &FR II (BLRG) &\citet{Massaro_2009} \\
  \object{3C 133} & 0459+252 & 05:02:58.480 & +25:16:25.11 & 0.2775 & 1460 & 4.34 & &FR II (HERG) &\citet{massaro2010chandra}\\
  \object{0529+075}\textsuperscript{\ddag} & 0529+075 & 05:32:38.998	 & +07:32:43.34 & 1.2540 & 8803 & 8.40 & 23.26 (1) &CDQ &\citet{2011ApJ...730...92H}\\
  \object{3C 179}\textsuperscript{\ddag} & 0723+679 & 07:28:10.896 & +67:48:47.03 & 0.8460 & 5386 & 7.70 & 8.98 (2)&LDQ &\citet{sambruna2004survey}\\
  \object{4C+25.21} & 0730+257 & 07:33:08.78 & +25:36:25.02 & 2.6860 &  & 8.135 & &Q & \citet{McKeough_2016}\\
  \object{TXS 0833+585}\textsuperscript{\ddag} & 0833+585 & 08:37:22.409 & +58:25:1.84 & 2.1010 & 16900 & 8.53 & 14.16 (3)&Q &\citet{McKeough_2016}\\
  \object{3C 213.1} & 0858+292 & 09:01:05.25 & +29:01:46.90 & 0.1939 & & 3.33 & &FR II (LERG) & \citet{massaro2010chandra}\\
  \object{3C 220.2} & 0927+362 & 09:30:33.473 &	+36:01:24.17 &	1.1574	& 814	& 8.47	& &Q&\citet{Stuardi_2018}\\
  \object{4C +55.17} & 0954+556 & 09:57:38.184 & +55:22:57.76 & 0.8996 & 5960 & 8.01 & & Q/CSO &\citet{tavecchio2007chandra}\\
  \object{PKS 1046-409} & 1046-409 & 10:48:38.270 & -41:14:00.11 & 0.6200 & 3668 & 6.78 & &CDQ &\citet{2005ApJS..156...13M}\\
  \object{3C 275.1} & 1241+266 & 12:43:57.649 & +16:22:53.39 & 0.5550 & 3204 & 6.40 & &LDQ &\citet{crawford2003extended}\\
  \object{3C 280.1} & 1258+404 & 13:00:33.364 & +40:09:07.28 & 1.6771 & & 8.62& & LDQ & \citet{jimenezgallardo2020textitchandra}\\
  \object{PKS 1311-270} & 1311-270 & 13:13:47.360 & -27:16:49.27 & 2.1860 & 18500 & 8.43 & &Q &\citet{McKeough_2016}\\
  \object{4C+11.45} & 1318+113 & 13:21:18.835 & +11:6:49.98 & 2.1792 & 17700 & 8.48 & &Q &\citet{McKeough_2016}\\
  \object{Centaurus B} & 1343-601 & 13:46:48.990 & -60:24:29.96 & 0.0130 & 55.5 & 0.26 & &FRI RG &\citet{2005ApJS..156...13M}\\
  \object{PKS 1402+044}\textsuperscript{*\ddag} & 1402+044 & 14:05:01.12 & +04:15:35.82	& 3.2091 & & 7.73 & 9.23 (1)&Q & \citet{schwartz2019relativistic}\\
  \object{B3 1428+422} & 1428+422 & 14:30:23.741 & +42:04:36.49 & 4.7150 & 44600 & 6.61 &  &Q&\citet{Cheung_2012}\\
  \object{3C 327.1}\textsuperscript{*} & 1602+014 & 16:04:45.320 & +01:17:51.02& 0.4620 & 2660 & 6.02 & &FR II (HERG)&\citet{massaro2013chandra}\\
  \object{3C 334}\textsuperscript{\ddag} & 1618+177 & 16:20:21.818 & +17:36:2.95 & 0.5551 & 3310 & 6.63 & 3.36 (2) &Q &\citet{2004ApJ...612..729H}\\
  \object{3C 418}\textsuperscript{\ddag} & 2037+511 & 20:38:37.042 & +51:19:12.43 & 1.6860 & & 8.61 & 6.86 (4)& Q & This work\textsuperscript{\dag}\\
  \object{PKS 2123-463} & 2123-463 & 21:26:30.704 & -46:05:47.90 & 1.6700 & 12583 & 8.56 & &CDQ&\citet{marshall2011x}\\
  \cutinhead{Sources excluded from analysis}
  \object{3C 13} & 0031+391 & 00:34:14.556 & 39:24:16.65 & 1.3510 & & 8.63 & & FR II (HERG) & \citet{wilkes2013revealing}\\
  \object{3C 16} & 0035+130 & 00:37:44.573 & +13:19:54.99 & 0.4050 & 2260 & 5.567 & &FR II (HERG) & \citet{massaro2013chandra}\\
  \object{3C 19} & 0038+328 & 0:40:55.0440 & +33:10:8.0200 & 0.4820 & 2780 & 6.15 & &FR II (LERG) &\citet{massaro2015chandra}\\
  \object{3C 52} & 0145+532 & 01:48:28.977 & +53:32:35.44 & 0.2900 & 1530 & 4.47 & &FR II (HERG) &\citet{massaro2010chandra}\\
  \object{3C 41} & 0123+239 & 01:26:44.300 & +33:13:11.00 & 0.7950	& 511	 & 7.70	& &FR II (HERG)&\citet{Massaro_2018}\\
  \object{3C 54} & 0152+435 & 01:55:30.162 & 	+43:45:55.43 & 0.8270 &	537 & 	7.81	&& FR II (HERG)&\citet{Massaro_2018}\\
  \object{3C 61.1} & 0210+860 & 02:22:35.046 & +86:19:06.17 & 0.1878 & 941.0 & 3.32 & &FR II (HERG) &\citet{massaro2010chandra}\\
  \object{3C 65} & 0220+397 & 02:23:43.191 & +40:00:52.45 & 1.1760 & 8290 & 8.49 &&FR II (HERG) &\citet{wilkes2013revealing}\\
  \object{3C 68.2} & 0231+313 & 02:34:23.856 & +31:34:17.46 & 1.5750 & 11900 & 8.69 & &FR II (HERG)&\citet{wilkes2013revealing}\\
  \object{0313-192} & 0313-192 & 03:15:52.100 & -19:06:44.30 & 0.0670 & 298 & 1.27 & &FR I RG&\citet{2006AJ....132.2233K}\\
  \object{3C 129} & 0445+449 & 04:49:09.064 & +45:00:39.40 & 0.0208 & 89 & 0.42 & & FR I RG&\citet{harris2002c}\\
  \object{3C 181} & 0725+147 & 07:28:10.305 & 14:37:36.24 & 1.3820 & 10100 & 8.64 & &Q&\citet{wilkes2013revealing}\\
  \object{3C 189} & 0755+379 & 07:58:28.108 & +37:47:11.80& 0.0428 & 187.0 & 0.83 & &FR I RG&\citet{worrall2001chandra}\\
  \object{3C 191} & 0802+013 & 08:04:47.972 & 10:15:23.69 & 1.9560 & 15500 & 8.60 & & Q&\citet{erlund2006extended}\\
  \object{4C+05.34} & 0805+046 & 08:07:57.54 & +04:32:34.53 & 2.8770 & & 7.99 & &Q & \citet{McKeough_2016}\\
  \object{3C 200} & 0824+294 & 08:27:25.397 & +29:18:44.90 & 0.4580 & 2630 & 5.99 & &FR II (LERG)&\citet{2004ApJ...612..729H}\\
  \object{3C 210} & 0855+280 & 08:58:09.961 & +27:50:51.57 & 1.1690 & 8240 & 8.48 & &FR II (HERG)&\citet{gilmour2009distribution}\\
  \object{PKS 0903-573} & 0903-573 & 09:04:53.179 & -57:35:05.80 & 0.6950 & 4221 & 7.12 & &CDQ&\citet{2005ApJS..156...13M}\\
  \object{4C+39.24} & 0905+399 & 09:08:16.900 & +39:43:26.00 & 1.882 & 14800 & 8.63 & &FR II (HERG)&\citet{2008MNRAS.386.1774E}\\
  \object{3C 219} & 0917+458 & 09:21:08.626 & +45:38:57.34 & 0.1740 & 831 & 2.90& &FR II RG (HERG)&\citet{2003MNRAS.340L..52C}\\
  \object{3C 225B} & 0939+139 & 09:42:15.387& 	+13:45:50.52 & 	0.5800 & 	349	 & 6.77 & &	FR II (HERG)&\citet{Massaro_2018}\\
  \object{3C 238} & 1008+066 & 10:11:00.379 &	+06:24:39.72 & 	1.4050	 & 10300 & 	8.65 &	 & FR II (HERG)&\citet{Stuardi_2018}\\
  \object{3C 239} & 1008+467 & 10:11:45.284 &	+46:28:18.79& 1.7810 & & 8.60 &  & FR II (HERG) & \citet{jimenezgallardo2020textitchandra}\\
  \object{3C 249} & 1059-009 & 11:02:03.774 &	-01:16:16.67 & 1.5540 & & 8.61 & & LDQ & \citet{jimenezgallardo2020textitchandra}\\
  \object{3C 257} & 1120+057 & 11:23:09.391 &	+05:30:18.50 & 2.4740 & & 8.25 & & FR II & \citet{jimenezgallardo2020textitchandra}\\
  \object{3C 268.1} & 1157+732 & 12:00:19.210	& +73:0:45.70 & 0.9700 & & 8.17 & & FR II (HERG) & \citet{massaro2015chandra}\\
  \object{3C 268.2} & 1158+318 & 12:00:59.110 & +31:33:27.90 & 0.3620 & 2000 & 5.21 & &FR II (HERG) &\citet{massaro2012chandra}\\
  \object{3C 280} & 1254+476 & 12:56:57.900 & +47:20:19.90 & 0.9960 & 6601 & 8.00 & &FR II (HERG)&\citet{donahue2003constraints}\\
  \object{3C 281} & 1305+069 & 13:07:53.929 & +06:42:14.30 & 0.6020 & 3538 & 6.70 & &LDQ&\citet{crawford2003extended}\\
  \object{4C+65.15} & 1323+655 & 13:25:29.702 & +65:15:13.293 & 1.6250 & 12163 & 8.60 & &LDQ &\citet{Miller_2009}\\
  \object{3C 293} & 1350+316 & 13:52:17.842& +31:26:46.50 & 0.0450 & 209 & 0.92 & &FR I (LERG) &\citet{massaro2010chandra}\\
  \object{3C 297} & 1414-037 & 14:17:23.999	&-04:00:47.54 & 	1.4061 & 	10300 & 	8.65 & 	&Q&\citet{Stuardi_2018}\\
  \object{3C 313} & 1508+080 & 15:11:00.010 & +07:51:50.00 & 0.4610 & 2650 & 6.01 & &FR II (HERG)&\citet{massaro2013chandra}\\
  \object{3C 322} & 1533+557 & 15:35:01.26 & +55:36:52.33 & 1.6810 & & 8.61 & & FR II & \citet{jimenezgallardo2020textitchandra}\\
  \object{3C 324} & 1547+215 & 15:49:48.900 & +21:25:38.10 & 1.2063 & 8387 & 8.40 & &FR II (HERG)&\citet{2004ApJ...612..729H}\\
  \object{3C 326.1} & 1553+202 & 15:56:10.170 &	+20:04:20.73 & 1.8250 & & 8.57 & & RG &
  \citet{jimenezgallardo2020textitchandra}\\
  \object{TXS 1607+183} & 1607+183 & 16:10:05.28 &  18:11:43.41 & 3.1180 & & 7.80 & &Q & \citet{schwartz2019relativistic}\\
  \object{3C 341} & 1626+738 & 16:28:4.050 & +27:41:43.00 & 0.4480 & 2560 & 5.91 & &FR II (HERG) & \citet{massaro2013chandra}\\
  \object{3C 349} & 1658+371 & 16:59:29.540 & +47:02:44.10 & 0.2050 & 1040 & 3.46 & &FR II (HERG)&\citet{massaro2010chandra}\\
  \object{3C 368} & 1802+110 & 18:05:06.454 & 	+11:01:35.06& 1.131 & & 8.44 & & FR II (HERG) & \citet{wilkes2013revealing}\\
  \object{3C 402} & 1940+504 & 19:41:46.050 & +50:35:48.40 & 0.0259 & 114 & 0.53 & &FR I&\citet{massaro2012chandra}\\
  \object{3C 436} & 2141+279 & 21:44:11.708 &	+28:10:19.030 & 0.2145 & & 8.68 & & FR II (HERG) & \citet{massaro2010chandra}\\
  \object{3C 437} & 2145+151 & 21:47:25.106 & +15:20:37.49 & 1.4800 & 11000 & 8.68 & &FR II&\citet{massaro2015chandra}\\
  \object{PKS 2155-152}\textsuperscript{\ddag} & 2155-152 & 21:58:06.282	 & -15:01:09.32	 & 0.6720 & 4049 & 7.02 & 21.46 (1) &CDQ &\citet{2011ApJ...730...92H}\\
  \object{PKS 2255-282}\textsuperscript{\ddag} & 2255-282 & 22:58:05.962 & -27:58:21.30 & 0.9260 & 6027 & 7.88 & 6.14 (5) &CDQ&\citet{marshall2011x}\\
  \object{3C 458} & 2310+050 & 23:12:54.408 & +05:16:45.98 & 0.2890 & 1530 & 4.46 & &FR II (HERG)&\citet{massaro2012chandra}\\
  \object{3C 470} & 2356+437 & 23:58:35.890 & +44:4:45.55 & 1.653. & 12600 & 8.69 & &FR II&\citet{2011ApJ...730...92H}\\
  \enddata
 \tablecomments{All the cosmological parameters are taken from NASA/IPAC Extragalctic Database(NED).}
\tablenotetext{a}{Commonly used name for this source.}
\tablenotetext{b}{IAU 1950 designation of the source.}
\tablenotetext{c}{The celestial positions of the source in J2000 coordinates taken from NED.}
\tablenotetext{d}{Redshift measurements taken from NED.}
\tablenotetext{e}{Angular scale of the source taken from NED.}
\tablenotetext{f}{The class column gives a radio jet morphology descriptor: Fanaroff-Riley I or II, CDQ(core dominated quasar), LDQ(Lobe dominated quasar), BL Lac, and an optical spectroscopic designation: LERG(low-excitation radio galaxy), HERG(high-excitation radio galaxy), NLRG(Narrow-line radio galaxy),BLRG(broad-line radio galaxy).  }
\tablenotetext{*}{Contrary to what is reported, we did not find significance evidence for any X-ray emission from the jet.}
\tablenotetext{\dag}{The \textit{Chandra} observation for this source was first published in \citet{jimenezgallardo2020textitchandra}. However, as the jet lied within the PSF of the core, no X-ray jet detection was indicated in their work.}
\tablenotetext{\ddag}{Sources with apparent superluminal motions in their parsec-scale jets \citep[see ][and the references therein]{2020arXiv200712661K}.}
\end{deluxetable*}

\bibliography{references}

\end{document}